\documentclass[ aps,twocolumn,groupedaddress,amsmath,amssymb]{revtex4-2}
\usepackage{graphicx}% Include figure files
\usepackage{dcolumn}% Align table columns on decimal point
\usepackage{bm}% bold math

\begin{document}

%%%   New Definitions
\newcommand{\eg}{{\it e.g.}}
\newcommand{\cf}{{\it cf.}}
\newcommand{\etal}{{\it et. al.}}
\newcommand{\ie}{{\it i.e.}}
\newcommand{\be}{\begin{equation}}
\newcommand{\ee}{\end{equation}}
\newcommand{\bea}{\begin{eqnarray}}
\newcommand{\eea}{\end{eqnarray}}
\newcommand{\bef}{\begin{figure}}
\newcommand{\eef}{\end{figure}}
\newcommand{\bce}{\begin{center}}
\newcommand{\ece}{\end{center}}
\newcommand{\red}{\textcolor{red}}

\newcommand{\dd}{\text{d}}
\newcommand{\ii}{\text{i}}
\newcommand{\lsim}{\lesssim}
\newcommand{\gsim}{\gtrsim}
\newcommand{\RAA}{R_{\rm AA}}
\newcommand{\sig}{\sigma_{\rm el}}
\newcommand{\rhoem}{\rho_{\rm em}}
\newcommand{\qcm}{q_{\textrm{CM}}}
\newcommand{\Ecm}{E_{\textrm{CM}}}

%\preprint{}

\title{Electric conductivity of hot pion matter}



\author{Joseph Atchison}
\email{jia08a@acu.edu}
\author{Ralf Rapp}
\affiliation{Cyclotron Institute and Department of Physics \& Astronomy, Texas A\&M University, College Station, TX 77843-3366, USA}
%\thanks{}


\date{\today}

\begin{abstract}
The determination of transport coefficients plays a central role in characterizing hot and dense nuclear matter. Currently, there are significant discrepancies between various calculations of the electric conductivity of hot hadronic matter. In the present work we calculate the electric conductivity of hot pion matter by extracting it from the  electromagnetic spectral function, via its zero energy limit at vanishing 3-momentum, within the Vector Dominance Model (VDM). 
Since within the VDM the photon couples to the hadronic currents primarily through the $\rho$ meson, we use hadronic many-body theory to calculate 
the $\rho$-meson's self-energy in hot pion matter, by dressing its pion cloud with thermal $\pi$-$\rho$ and $\pi$-$\sigma$ loops including vertex 
corrections to maintain gauge invariance. In particular, we analyze the low-energy transport peak of the spectral function, extract its behavior with 
temperature and compare to (the results of) existing approaches in the literature.
\end{abstract}

%\keywords{}

\maketitle


%%%%%%%%%%%%%%%%%%%%%%
\section{Introduction}
\label{sec:intro}
%%%%%%%%%%%%%%%%%%%%%%
A central goal of high-energy nuclear physics is to study and characterize hot and dense nuclear matter which can be created in heavy-ion collisions 
(HICs) over a large range of center-of-mass energies. A common way to characterize the long-wavelength properties of the medium are transport 
coefficients, which can be used to describe the transport of conserved charges through the fireball in nuclear collisions. One such coefficient is the 
electric conductivity, $\sig$, which will be the focus of this paper. In particular, we will take advantage of the close relation of $\sig$ to the thermal dilepton 
emission rate~\cite{Rapp:1999us}, as both quantities are directly proportional to the electromagnetic spectral function of the strongly interacting medium. 
On the one hand, this allows to establish connections between the processes that are widely implemented to describe low-mass dilepton and photon 
radiation observed in experiments at the SPS, RHIC, HADES and the LHC~\cite{LandoltBornstein,Salabura:2020tou,Tripolt:2022hhw}. On the other hand, 
renewed interest in the conductivity has recently been triggered by future plans to access the pertinent transport peak more directly in experiment through 
very low-mass and low-momentum dileptons, as being envisaged at the Schwer-Ionen Synchrotron (SIS), the Relativistuc Heavy-Ion Collider (RHIC) and 
the Large Hadron Collider (LHC)~\cite{HADES,UpStar,ALICE:2022wwr}. 
Thus far, available calculations of the electric conductivity of hot hadronic matter have utilized various formalisms and yielded results that vary considerably, 
by up to an order of magnitude or even more~\cite{Greif,Huot,Fraile,Finazzo,Aarts,Brandt2013,Brandt2016,Amato,Kadam,Ghosh,Ghosh2}. In addition, some
of the calculations appear to produce a conductivity that is below a conjectured quantum lower bound proposed in Ref.~\cite{Huot}. In this work, we 
seek to address the above questions by  performing a hadronic quantum many-body calculation of the conductivity, albeit constrained to a system
of hot pion matter. 

The dilepton emission rate is proportional to the imaginary part of the electromagnetic (EM) current-current correlation function, \ie, the EM spectral
function, $\rho_{\textrm{EM}}$. At the same time, the electric conductivity can be obtained from 
$\rho_{\textrm{EM}}$ as its low-energy limit at zero 3-momentum~\cite{LB16,LB18,Moore}. Within the vector dominance 
model (VDM) the EM correlator is proportional to the light vector meson propagators, $D_{V}$, most notably the $\rho$-meson's ($V$=$\rho$). 
In the vacuum and in a hadronic description, the latter is related to the $\rho$'s self-energy ($\Sigma_{\rho\pi\pi}$), which is governed by 
2-pion decays. In the medium, the dressing of the 
$\rho$'s pion cloud in a nuclear medium has been widely studied~\cite{Ko,Friman,ChanfrayGI,RappRho96,Urban:1999im}, and found to give a key
contribution to the low-mass dilepton enhancement observed in experiment.  The effects of thermal $\pi\pi$ scattering, generally believed to be less
important, have not been studied as widely~\cite{Rapp:1995fv,Song}. However, at relatively small temperatures and vanishing baryon chemical 
potential baryon-antibaryon excitations are suppressed and the effects of the lighter pions dominate. Since at very low dilepton masses one expects 
large contributions from the hadronic phase in nuclear collisions, and since at RHIC and the LHC the baryon-chemical potential is small (although
contributions from thermally excited baryon-antibaryons are not negligible), a reliable calculation of the conductivity of hot hadronic matter is likely to
require the inclusion of thermal pions. Toward this end we here calculate $\rho_{\textrm{EM}}$ at zero 3-momentum for hot 
pion matter and analyze the emerging conductivity, thereby serving as a first step in adding thermal $\pi\pi$ scattering to the baryonic effects 
calculated in previous works~\cite{Urban:1999im}. 

When dressing the pion cloud of the $\rho$-meson with medium excitations, \ie, pion self-energies, it is important to preserve EM gauge invariance, 
which requires the introduction of appropriate vertex corrections~\cite{HerrmannGI,ChanfrayGI,Urban:1999im}. The construction of these vertex 
corrections can be guided by Ward identities, which ensure that the resulting $\rho$-meson self-energy, and thus the EM correlation function, is 
4-dimensionally transverse, which is a necessary condition for gauge invariance.These vertex corrections were constructed, \eg, in 
Refs.~\cite{HerrmannGI,ChanfrayGI,Urban:1999im} for the case of nuclear matter and in Refs.~\cite{Song,Hohler:2015iba} for hot pion matter. 
The introduction of vertex corrections becomes even more challenging in the context of the conductivity, as to render it finite, a dressing of {\em all} 
in medium propagators with a finite width is required. In the present work, $\rho_{\textrm{EM}}$ and the electric conductivity 
will be calculated in hot pion matter both with and without vertex corrections.

This paper is organized as follows: in Sec.~\ref{sec:rho} we introduce our microscopic model for the  $\rho$-propagator and self-energy, first 
in the vacuum with constraints from scattering data, followed by the general form at finite temperature and a calculation of the conductivity for 
on-shell thermal pions that allows to recover the kinetic-theory result. In Sec.~\ref{sec:pion} we calculate the pion self-energy in a thermal pion 
gas, based on $S$- and $P$-wave scattering through sigma and rho resonances, again constrained by vacuum scattering data. Section~\ref{sec:vcorr} 
is dedicated to the construction of vertex corrections required to maintain gauge invariance at finite temperature. In Sec.\ref{sec:PIem} we discuss 
our numerical results for the in-medium EM spectral function, with focus on the low-energy transport peak and pertinent conductivity of hot pion matter, 
with and without vertex corrections, and put our results into the context of existing literature. In Sec.~\ref{sec:appl} we discuss two further applications,
namely a calculation of the charge susceptibility and a test of a current-conservation sum rule. Finally, we summarize and discuss future work 
in Sec.~\ref{sec:sum}.

%%%%%%%%%%%%%%%%%%%%%
\section{Hadronic Model for the rho meson}
\label{sec:rho}
%%%%%%%%%%%%%%%%%%%%%
In this section we discuss the connection between thermal dilepton rates and the conductivity via the EM spectral function and compute the latter within 
VDM (Sec.~\ref{ssec:EMspecVDM}) based on a hadronic model for the $\rho$ propagator. We  calculate the $\rho$ self-energy in vacuum 
(Sec.~\ref{ssec:rho-vac}) and at finite temperature (Sec.~\ref{ssec:rho-T}),  followed by a preliminary on-shell calculation of the conductivity to make 
contact with kinetic theory (Sec.~\ref{ssec:onshell}).

%%%%%%%%%%%%%%%%%%%%%%%%%%%%%%%%%%%%%%%%%%%%%
\subsection{EM spectral function in the vector dominance model}
\label{ssec:EMspecVDM}
%%%%%%%%%%%%%%%%%%%%%%%%%%%%%%%%%%%%%%%%%%%%%%%
The thermal dilepton emission rate can be expressed through the EM spectral function as~\cite{LB16,LB18}:
\begin{eqnarray}\label{eq2}
\frac{dR_{l+l-}}{d^4q}&=&\frac{\alpha_{\textrm{EM}}^2}{2\pi^3 M^2}f^B(q_0,T)\rho_{\textrm{EM}}(M,q,T,\mu_B) \ ,  
\end{eqnarray} 
where $f^B$ denotes the Bose-Einstein distribution, $M^{2}=q_{0}^{2}-\vec{q}\,^{2}$ is the dilepton's invariant mass, and 
$\alpha_{\textrm{EM}}=\frac{e^{2}}{4\pi}$ the fine-structure constant. The spectral function, in turn, is defined via the imaginary part of the EM 
current-current correlation in the strong-interaction medium, $\rho_{\rm EM}= -2 \ {\rm Im}\Pi_{\rm EM}$, with a polarization average implied as 
$\Pi_{\rm EM}\equiv g_{\mu\nu} \Pi_{\rm EM}^{\mu\nu}/3$.
The electric conductivity can be obtained from the spatial components of the spectral function at zero 3-momentum in the low-energy limit~\cite{Moore},
\begin{eqnarray}
\label{eq3}
\sig(T)&=&(-e^{2}/6) \lim_{q_0 \to 0} [\rho_{\textrm{EM}}^{ii}(q_0,\vec{q}=0,T)/q_{0} \ .
\end{eqnarray} 
Note that at finite $T$, the retarded spectral function goes linearly to zero with energy, and thus the division by $q_0$ leads to a finite result provided 
the imaginary part is non-vanishing for small $q_0$.

For invariant masses below approximately 1\,GeV, the EM correlator is well described by the VDM, represented by the 
so-called field-current identity between the vector meson fields, $V^\mu$ and the hadronic EM current:
\be
j^\mu_{\rm EM} = \frac{m_\rho^2}{g_\rho} \rho^\mu +  \frac{m_\omega^2}{g_\omega} \omega^\mu + \frac{m_\phi^2}{g_\phi} \phi^\mu  \  ,
\ee
where the $g_V$ denote pertinent hadronic couplings. Since $g_\omega \simeq 3g_\rho$, the primary contribution arises from the $\rho$ meson
leading to~\cite{LB19}:
\begin{eqnarray}
\label{VDM}
\textrm{Im}\Pi_{\textrm{\textrm{EM}}}^{\mu\nu}&\approx& \frac{m_{\rho}^{4}}{g_{\rho}^{2}}\textrm{Im}D_{\rho}^{\mu\nu} \ .
\end{eqnarray} 
The task of computing the EM spectral function is thus converted into calculating the $\rho$ spectral function which is given in terms of its 
self-energy, $\Sigma_{\rho}^{\mu\nu}$. In vacuum, the latter is generated by the $\rho$'s 2-pion cloud, including its 2-pion decay 
with a branching ratio of near 100\%. At finite temperature, the pions are subject to rescattering off thermal-medium particles, which we will 
approximate by a pion gas. We will discuss those two cases in the following sections.

%%%%%%%%%%%%%%%%%%%%%%%%%%%%
\subsection{Rho meson in vacuum}
\label{ssec:rho-vac}
\begin{figure}
\includegraphics[width=20pc]{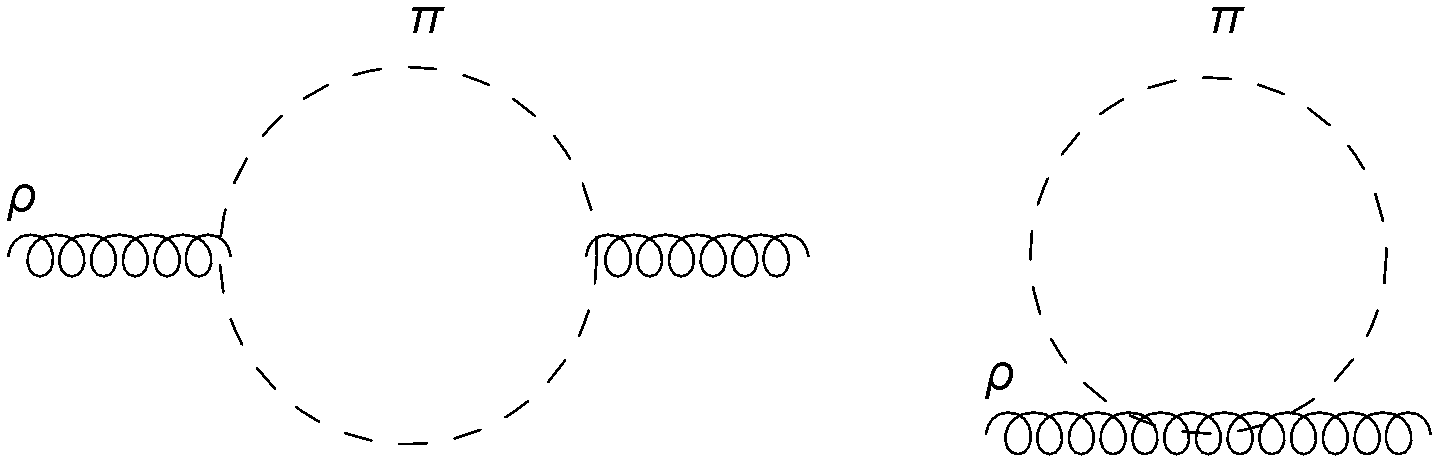}
\caption{Vacuum pion loops contributing to the $\rho$ self-energy to leading order in the coupling constant 
(left panel: $\pi\pi$-loop, right panel: tadpole loop).}
\label{figRhoSE}
\end{figure}

%%%%%%%%%%%%%%%%%%%%%%%%%%%%
When gauging the free pion Lagrangian with a $\rho$ meson using minimal substitution, and adding the $\rho$-meson mass and field-strength terms, 
one obtains the following effective Lagrangian for the $\pi$-$\rho$ system:
\begin{eqnarray}\label{eq6}
\mathcal{L}_{\pi}+\mathcal{L}_{\rho}&=&\frac{1}{2}\partial_{\mu}\vec{\phi} \cdot \partial^{\mu}\vec{\phi}-\frac{1}{2}m_{\pi}^{2}\vec{\phi} \cdot\vec{\phi}-\frac{1}{4}\rho_{\mu\nu}\rho^{\mu\nu}\nonumber\\
&&+\frac{1}{2}(m_{\rho}^{(0)})^{2}\rho_{\mu}\rho^{\mu},\\
\label{eq6a}
\mathcal{L}_{\pi\rho}&=&\frac{1}{2}ig_{\rho}\rho_{\mu}(T_{3}\vec{\phi}\cdot\partial^{\mu}\vec{\phi}+\partial^{\mu}\vec{\phi}\cdot T_{3}\vec{\phi})\nonumber\\
&&-\frac{1}{2}g_{\rho}^{2}\rho_{\mu}\rho^{\mu}T_{3} \vec{\phi}\cdot T_{3}\vec{\phi}.
\end{eqnarray}
Here, $\rho_{\mu\nu}=\partial_{\mu}\rho_{\nu}-\partial_{\nu}\rho_{\mu},T_{3}=-i\epsilon_{3ab}$, and $m_{\pi}=140 \, \textrm{MeV}$ is the pion 
mass. To leading order in the gauge coupling constant, $g_{\rho}$, the vacuum $\rho$ propagator, with a ``bare mass $m_\rho^{(0)}$, 
receives two contributions, diagrammatically depicted in Fig.~\ref{figRhoSE}. The first one, the $\pi\pi$-loop, gives rise to the vacuum $\rho\to\pi\pi$ 
decay,  while the purely real ``tadpole" diagram produces a constant shift in the $\rho$ mass. However, the tadpole is essential 
to ensure a 4-dimensionally transverse $\rho$ self-energy,  $q_\mu \Sigma_{\rho}^{\mu\nu} = 0$~\cite{HerrmannGI}.

\begin{figure}
\includegraphics[width=20pc]{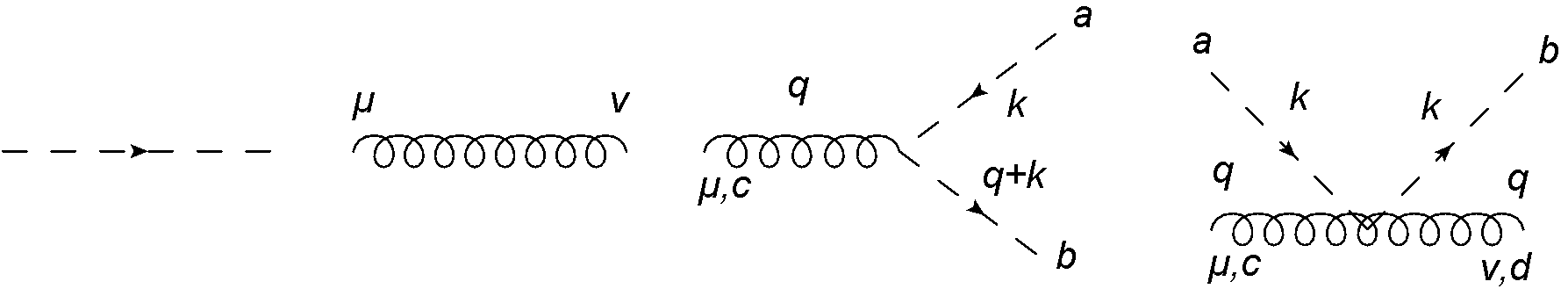}
\caption{\label{figB}From left to right: $\pi$-propagator, $\rho$-propagator, $\rho\pi\pi$ vertex, and $\pi\pi\rho\rho$ vertex.}
\end{figure}

From the  Lagrangian, the free pion propagator, $D_{\pi}$, $\rho\pi\pi$ vertex, $\Gamma_{\mu (ab)}^{(3)}$, and $\rho\rho\pi\pi$ vertex, 
$\Gamma_{\mu\nu (ab}^{(4)}$, follow as:
\begin{eqnarray}
\label{eq9a}
D_{\pi}(k)&=&\frac{1}{k^{2}-m_{\pi}^{2}+i\epsilon},\\
\label{eq9b}
\Gamma_{\mu\,abc}^{(3)}&=&g_{\rho}\epsilon_{cab}(2k+q)_{\mu},\\
\label{eq9c}
\Gamma_{\mu\nu\,abcd}^{(4)}&=&ig_{\rho}^{2}(2\delta_{ab}\delta_{cd}-\delta_{ac}\delta_{bd}-\delta_{ad}\delta_{bc})g_{\mu\nu} \ ,
\end{eqnarray}
where Greek (Roman) indices are used to denote Lorentz (isospin) space. Figure~\ref{figB} displays the propagators and vertices diagrammatically.
The $\rho$ self-energy energy diagrams can be evaluated as
\begin{widetext}
\begin{eqnarray}
\label{eq10x00x}
\Sigma_{\rho}^{\mu\nu}(q)&=&\frac{-i}{2}\int \frac{d^{4}k}{(2\pi)^{4}}D_{\pi}(k)D_{\pi}(q+k)\Gamma_{\mu\,3ab}^{(3)}(k,q)\Gamma_{\mu\,3ba}^{(3)}(q+k,-q) -\frac{1}{2}\int  \frac{d^{4}k}{(2\pi)^{4}}D_{\pi}(k)\Gamma_{\mu\nu\,aa33}^{(4)}(k,q) \ .
\end{eqnarray}
\end{widetext}
A symmetry factor of $\frac{1}{2}$ has been added to both terms to remove double counting of pion states, since the different pions are distinguishable in 
particle space (different charge states), but not in isospin space.
 
To regularize the loop integrals for  $\Sigma_{\rho}^{\mu\nu}$  we employ the Pauli-Villars scheme as a means to maintain gauge invariance~\cite{HerrmannGI,Urban1998}. In this scheme, $\rho$ self-energies are calculated by subtracting ``heavy-pion" propagators from the physical 
ones in the unregularized self-energy, yielding
\begin{eqnarray}
\label{eq11}
\Sigma_{\rho}'^{\mu\nu}(q)&=&\Sigma_{\rho}^{\mu\nu}(q,m_{\pi})-2\Sigma_{\rho}^{\mu\nu}(q,\sqrt{m_{\pi}^{2}+\Lambda_{0}^{2}})\nonumber\\
&&+\Sigma_{\rho}^{\mu\nu}(q,\sqrt{m_{\pi}^{2}+2\Lambda_{0}^{2}}) \ .
\end{eqnarray}
We adopt the values of $g_{\rho}=5.9$, $m_{\rho}^{(0)}=853 \, \textrm{MeV}$, and $\Lambda_{0}=1 \, \textrm{GeV}$, which have been previoulsy fitted 
to the $P$-wave $\pi\pi$ phase shifts and pion electromagnetic form factor~\cite{Urban:1999im}.

Finally, $\Sigma_{\rho}'^{\mu\nu}$ can be resummed into the vacuum $\rho$ propagator yielding~\cite{HerrmannGI}:
\begin{eqnarray}
\label{eq12a}
 D_{\rho}^{\mu\nu}(k)&=&D_{\rho}(k)(-g^{\mu\nu}+\frac{k^{\mu}k^{\nu}}{k^{2}})+\frac{k^{\mu}k^{\nu}}{(m_{\rho}^{(0)})^{2}k^{2}},\\
\label{eq12b}
D_{\rho}(k)&=&\frac{1}{k^{2}-(m_{\rho}^{0})^{2}-\Sigma_{\rho}'(k)},\\
\label{eq12c}
\Sigma_{\rho}'^{\mu\nu}(k)&=&(-g^{\mu\nu}+\frac{k^{\mu}k^{\nu}}{k^{2}})\Sigma_{\rho}'(k) \ .
\end{eqnarray}
In the follwoing, we drop the prime notation from the regukarized $\rho$ self-energy.
\\

%%%%%%%%%%%%%%%%%%%%%%%%%
\subsection{General form of $\rho$ self-energy at finite temperature}
\label{ssec:rho-T}
%%%%%%%%%%%%%%%%%%%%%%%%%
We calculate $\Sigma_{\rho}^{\mu\nu}$ at finite temperature within the imaginary-time formalism, using the methods outlined in 
Refs.~\cite{Fetter,Rapp:1996ym}; one obtains
\begin{widetext}
\begin{eqnarray}
\label{eq18}
\Sigma_{\rho}^{\mu\nu}(q)&=&g_\rho^2 \int \frac{d^3k}{(2\pi)^3} \int_{-\infty}^{\infty}\frac{dvdv'}{\pi^{2}}\frac{(2k+q)^{\mu} (2k+q)^{\nu}}{q_{0}+v-v'+i\epsilon}\textrm{Im}[D_{\pi}(v,\vec{k})]\textrm{Im}[D_{\pi}(v',\vec{k}+\vec{q})](f(v)-f(v'))+\textrm{I}_{\textrm{Tad}}  \ ,
\end{eqnarray}
\end{widetext}
where
\begin{eqnarray}\label{tad}
\textrm{I}_{\textrm{Tad}}&=&\frac{g_\rho^2g^{\mu\nu}}{4\pi^{4}} \int d^{3}k\int_{-\infty}^{\infty}dv\textrm{Im}[D_{\pi}(v,\vec{k})] f(v) \ .
\end{eqnarray}
The first term is complex, while the second is purely real. The imaginary part of $\Sigma_{\rho}^{\mu\nu}$ can be slightly simplified as
\begin{widetext}
\begin{eqnarray}\label{eq20}
\textrm{Im}\Sigma_{\rho}^{\mu\nu}(q)&=&g_\rho^2 \int \frac{d^3k}{(2\pi)^3} \int_{0}^{q_{0}}\frac{dv}{-\pi} \textrm{Im}[D_{\pi}(v,\vec{k})]\textrm{Im}[D_{\pi}(q_{0}-v,\vec{q}+\vec{k})](2k+q)^{\mu} (2k+q)^{\nu}(1+f(v)+f(q_{0}-v))\nonumber\\
&&+2g_\rho^2 \int \frac{d^3k}{(2\pi)^3} \int_{0}^{\infty}\frac{dv}{-\pi} \textrm{Im}[D_{\pi}(v,\vec{k})]\textrm{Im}[D_{\pi}(q_{0}+v,\vec{q}+\vec{k})](2k+q)^{\mu} (2k+q)^{\nu}(f(v)-f(q_{0}+v)) \  ,
\end{eqnarray}
\end{widetext}
where we have separated the self-energy into two "cuts". The first integral corresponds to the unitarity cut, which represents the vacuum 
$\rho\rightarrow \pi\pi$ decay and its Bose enhancement, while the second integral is referred to as Landau cut which represents $\rho\pi$ 
scattering through an intermediate pion state. 
The energy dependent part of the real part of $\Sigma_{\rho}^{\mu\nu}$ can be calculated from a dispersion relation, while the constant shift 
from the tadpole diagram can be calculated directly as
\begin{eqnarray}\label{eq22}
\textrm{Re}\Sigma_{\rho}^{\mu\nu}(q)&=&\frac{-1}{\pi}\textrm{p.v.}\int_{0}^{\infty} dv^{2}\frac{1}{q_{0}^{2}-v^{2}}\textrm{Im}\Sigma_{\rho}^{\mu\nu}(v,\vec{q})\nonumber\\
&&+\textrm{I}_{\textrm{Tad}} \ ,
\end{eqnarray}
where p.v. indicates the principal value of the integral.

At finite temperature the breaking of Lorentz invariance splits the $\rho$ propagator at finite 3-momemntum into transverse and longitudinal modes that
can be expressed in terms of projection operators $P_{T}^{\mu\nu}$ and $P_{L}^{\mu\nu}$ as~\cite{Gale}
\begin{eqnarray}\label{eq5a}
D_{\rho}^{\mu\nu}(q,T)&=&\frac{P_{T}^{\mu\nu}}{M^2-(m_{\rho}^{0})^{2}-\Sigma_{\rho}^{T}(q,T)}\nonumber\\
&&+\frac{P_{L}^{\mu\nu}}{M^2-(m_{\rho}^{0})^{2}-\Sigma_{\rho}^{L}(q,T)}\nonumber\\
&&+\frac{q^{\mu}q^{\nu}}{(m_{\rho}^{0})^{2}M^{2}} \ ,
\end{eqnarray} 
where $M^{2}=q_{0}^{2}-\vec{q}\,^{2}$ is the $\rho$ meson's invariant mass. The  projection operators are defined by
\begin{eqnarray}\label{eq5b}
P_{T}^{\mu\nu}&=&0, \; \textrm{for} \; \mu=0 \; \textrm{or} \; \nu=0,\nonumber\\
P_{T}^{\mu\nu}&=&\delta^{\mu\nu}-\frac{q^{\mu}q^{\nu}}{\vec{q}^{2}}, \; \textrm{for} \; \mu,\nu \in \{1,2,3\},\\
P_{L}^{\mu\nu}&=&\frac{q^{\mu}q^{\nu}}{M^{2}}-g^{\mu\nu}-P_{T}^{\mu\nu} \ .
\end{eqnarray}
The transverse and longitudinal components of the rho self-energy, $\Sigma_{\rho}^{T}$ and $\Sigma_{\rho}^{L}$, follow as 
\begin{eqnarray}\label{eq5c}
\Sigma_{\rho}^{\mu\nu}(q,T)&=&P_{T}^{\mu\nu}\Sigma_{\rho}^{T}(q,T)+P_{L}^{\mu\nu}\Sigma_{\rho}^{L}(q,T).
\end{eqnarray}
For $\vec{q}=0$, $\Sigma_{T}(q_0,T)=\Sigma_{L}(q_0,T)$, allowing one to write the conductivity in terms of only the transverse projection 
of the rho self-energy:
\be
\label{eq91x}
\sig=\frac{e^2}{g_{\rho}^{2}} \lim_{q_0 \to 0} \textrm{Im}\big[\frac{-(m^{0}_{\rho})^{4}}{q_{0}^2-(m_{\rho}^{0})^{2}-\Sigma_{\rho}^{T}(q_{0},\vec{q}=0)}\big] \ .
\ee

%%%%%%%%%%%%%%%%%%%%%%%%%%%%%%%%% 
\subsection{On-shell pion approximation}
\label{ssec:onshell}
%%%%%%%%%%%%%%%%%%%%%%%%%%%%%%%
For very small energies Im$D_{\rho}$ is approximately proportional to $\textrm{Im}\Sigma_{\rho}^{T}$, which is dominated by the Landau cut below $q_{0}\approx 2m_{\pi}$. Therefore the conductivity can be approximated  as
\begin{eqnarray}
\label{ApproxCond5x}
\sig&\approx&\frac{-2e^{2}}{3} \int \frac{d^3k}{(2\pi)^3} \int_{0}^{\infty}\frac{dv}{-\pi} [\textrm{Im}D_{\pi}(v,\vec{k})]^{2}\nonumber\\
&&\times\frac{4|\vec{k}|^{2}e^{\frac{v}{T}}}{T(-1+e^{\frac{v}{T}})^{2}} \ ,
\end{eqnarray}
where we have taken the limit $q_{0}\rightarrow0$ analytically. 
We see that one must introduce a finite pion width into $D_{\pi}$ in order to obtain a finite conductivity. This implies that we have to include at 
least two thermal-pion insertions in order to obtain a result beyond the ideal-gas limit. One can approximate the conductivity using a small, but finite, 
pion width, $\Gamma_{\pi}$. In this approximation the pion propagator becomes
\be
\label{ApproxDpia}
D_{\pi}(k)=\frac{1}{2\omega_{k}}\sum_{\pm}\frac{\pm1}{k_{0}\mp\omega_{k}+i\frac{\Gamma_{\pi}}{2}} \ ,
\ee
with $\omega_{k}=\sqrt{k^{2}+m_{\pi}^{2}}$. In the small-width limit $[\textrm{Im}D(k)]^{2}$ is approximately
\be
\label{ApproxDpi2d}
[\textrm{Im}D(k)]^{2} \approx \frac{\pi}{4\omega_{k}^{2}}\frac{1}{\Gamma_{\pi}}\delta(k_{0}-\omega_{k}) \ .
\ee
One can use Eqs.~(\ref{ApproxCond5x}) and (\ref{ApproxDpi2d}) to show that
\be
\label{ApproxCond11}
\sig=\frac{2 e^{2}}{3T} \int\frac{d^{3}\vec{k}}{(2\pi)^{3}}\frac{v_{k}^{2}}{\Gamma_{\pi}}f(\omega_{k})(1+f(\omega_{k})) \ ,
\ee
where $v_{k}=|\vec{k}|/\omega_{k}$ is the pion's velocity. This result agrees with similar calculations  where the conductivity is expressed in terms of 
the pion width~\cite{Fraile,Ghosh,Greif2}.

%%%%%%%%%%%%%%%%%%%%%%%%%%%%%%
\section{Thermal $\pi\pi$ scattering}
\label{sec:pion}
%%%%%%%%%%%%%%%%%%%%%%%%%%%%%%%%
In this section we calculate the pion self-energy due to thermal $S$- and $P$-wave $\pi\pi$ scattering. The self-energy can be resummed into a thermal 
pion propagator yielding
\be
\label{pionprop}
D_{\pi}(k_{0},\vec{k})=\frac{1}{k^{2}-m_{\pi}^{2}+\Sigma_{\pi}(k,T)} \ .
\ee
We first obtain the self-energy expression from the Matsubara formalism in terms of the $\pi\pi$ scattering amplitude (Sec.~\ref{ssec:sigpi}), 
discuss the implementation of phenomenological vertex form factors (Sec.~\ref{ssec:ff}) and constraints from vacuum $\pi\pi$ phase shifts and cross 
sections (Sec.~\ref{ssec:pipivac}),  and finally present our numerical results for the pion optical potentials in a hot pion gas (Sec.~\ref{ssec:optical}).

%%%%%%%%%%%%%%%%%%%
\subsection{Pion self-energy}
\label{ssec:sigpi}
\begin{figure}
\includegraphics[width=19pc]{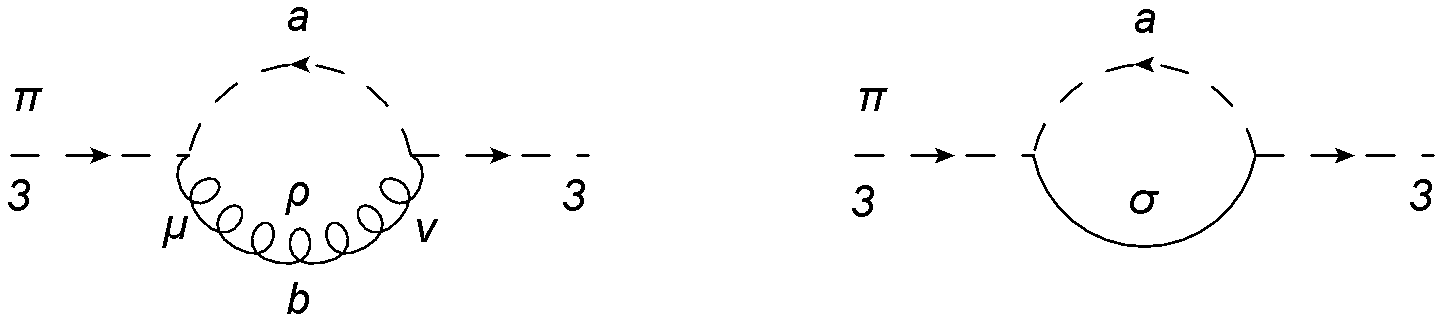}
\caption{\label{figD}The pion self-energy resulting from resonant scattering of a pion with a thermal pion through intermediate $\rho$ (left) and 
$\sigma$ (right) mesons.}
\end{figure}

%%%%%%%%%%%%%%%%%%%%%
Assuming the dominance of $s$-channel resonances we are able to describe $S$- and $P$-wave $\pi\pi$ interactions up to center-of-mass (CM) energies
 of about 1 GeV~\cite{Rapp:1995ir}, which is sufficient for the typical temperatures under consideration. Such scatterings can be fairly well approximated 
through $\sigma(500)$ and $\rho(770)$ resonances, whose pertinent diagrams are illustrated in Fig.~\ref{figD}.
We use the vertices and propagators established in Sec.~\ref{ssec:rho-vac} to construct the pion self-energy for $\pi\pi$ scattering through a 
$\rho$ resonance, while we follow the approach outlined in Refs.~\cite{Rapp:1995ir,Rapp:1996ym} to calculate scattering through an effective  
$\sigma(500)$ resonance. The in-medium pion self-energies ($R=\sigma,\rho$) are then obtained as
\begin{widetext}
\be
\label{sigma_pi}
\Sigma_\pi^{R}(k)=g_R^2\int \frac{d^3p}{(2\pi)^3}\int_{-\infty}^{\infty}\frac{dwdw'}{\pi^{2}}\Big[\textrm{Im}[D_{\pi}(w,\vec{p})]
\textrm{Im}[D_R(w',\vec{k}+\vec{p})]\frac{N_{R\pi\pi}[p,k,w']\textrm{FF}_R(\qcm^2)(f(w)-f(w'))}{k_{0}+w-w'+i\epsilon}\Big]_{p_{0}=w} 
\ee
\end{widetext}
with the vertex functions 
\be
\label{vrhopipi}
N_{\rho\pi\pi}[p,k,w']=-(k-p)^{2}+\frac{(k^{2}-p^{2})^{2}}{w'^{2}-(\vec{p}+\vec{k})^{2}} 
\ee
and
\be
\label{vsigpipi}
N_{\sigma\pi\pi}[p,k,w']=\frac{w'^{2}-(\vec{k}+\vec{p})^{2}-m_{\pi}^{2}}{2} \ . 
\ee
The vacuum $\rho$ propagator, $D_{\rho}^{\mu\nu}$, is taken from  Eq.~(\ref{eq12a}) and $D_{\sigma}$ is the vacuum $\sigma(500)$ propagator,
\begin{eqnarray}
\label{eqA}
D_{\sigma}(E)&=&\frac{1}{E^{2}-m_{\sigma}^{2}-\Sigma_{\sigma}(E)} \ ,
\end{eqnarray}
where $m_{\sigma}$ is the bare sigma mass and $\Sigma_{\sigma}$ its vacuum self-energy,
\begin{eqnarray}
\label{eq86}
\Sigma_{\sigma}(E)&=&3g_{\sigma}^{2}\int\frac{d|\vec{k}|\vec{k}^{2}}{(2\pi)^{2}}\frac{(E^{2}-m_{\pi}^{2})\textrm{FF}_{\sigma}(k)^{2}}{\omega_{k}(E^{2}-4\omega_{k}^{2}+i\epsilon)} \ .
\end{eqnarray}
Although we dress the intermediate $\rho(770)$ and $\sigma(500)$ propagators in $\Sigma_{\pi}$ with their vacuum self-energies, the pion propagators 
within $\Sigma_{\pi}$ are not dressed. These propagators will not induce an infinite conductivity, because they correspond to thermal pions, while only intermediate particle states transmit charge through the medium.

%%%%%%%%%%%%%%%%%%%%%%%%%%%
\subsection{Form factor}
\label{ssec:ff}
%%%%%%%%%%%%%%%%%%%%%%%%%%%
Phenomenological form factors $\textrm{FF}_R$ at each vertex are a standard tool to simulate the finite-size effects of hadrons while 
also regularizing vacuum loop integrals. For the purpose of the $\pi\pi$ scattering amplitude we do not employ the Pauli-Villars scheme (which is not 
necessary) but resort to a simpler form given by
\begin{eqnarray}
\label{eqAa}
\textrm{FF}_{\rho}(q_{\textrm{CM}})&=&\frac{\Lambda_{\rho}^{2}+m_{\rho}^{2}}{\Lambda_{\rho}^{2}+4(m_{\pi}^2+\qcm^2(p,k))}\\ 
\\
\label{FFSigmaR}
\textrm{FF}_{\sigma}(\qcm)&=&\frac{\Lambda_{\sigma}^{2}}{\Lambda_{\sigma}^{2}+4(m_{\pi}^2+\qcm^2(p,k))}
\\
\label{eqAc}
\qcm^2(p,k)&=&\frac{((p+k)^{2}-k^{2}-p^{2})^{2}-4k^{2}p^{2}}{4(p+k)^{2}}  \  ,
\end{eqnarray}
where $\qcm$ is pion momentum in the CM system of the collision. In the following section we constrain the parameters of the 
$\pi\pi$ amplitudes through vacuum scattering data.

%\begin{eqnarray}
%\label{eqAa}
%\textrm{FF}_{\rho}(\qcm)&=&\frac{\Lambda_{1\rho}^{2}}{\Lambda_{2\rho}^{2}+\qcm^2(p,k)} 
%\\
%\label{FFSigmaR}
%\textrm{FF}_{\rho}(\qcm)&=&\frac{\Lambda_{1\sigma}^{2}}{\Lambda_{2\sigma}^{2}+\qcm^2(p,k)}
%\\
%\label{eqAc}
%\qcm^2(p,k)&=&\frac{((p+k)^{2}-k^{2}-p^{2})^{2}-4k^{2}p^{2}}{4(p+k)^{2}},
%\end{eqnarray}

%%%%%%%%%%%%%%%%%%%%%
\subsection{Comparisons to vacuum data}
\label{ssec:pipivac}
\begin{figure}
\includegraphics[width=19pc]{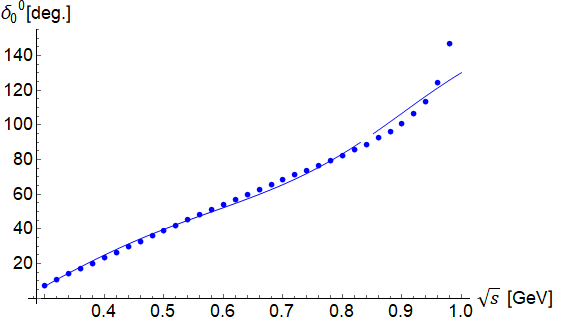}
\includegraphics[width=19pc]{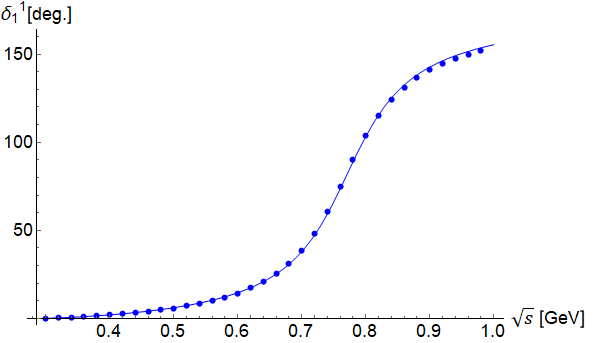}
\caption{Fits to the isoscalar $S$- (upper panel) and isovector $P$-wave (lower panel) $\pi\pi$ scattering phase shift via $\sigma$ and 
$\rho$ resonance scattering, respectively, compared to experimental data~\cite{Froggatt}.}
\label{fig_phaseshift}
\end{figure}

\begin{figure}
\includegraphics[width=19pc]{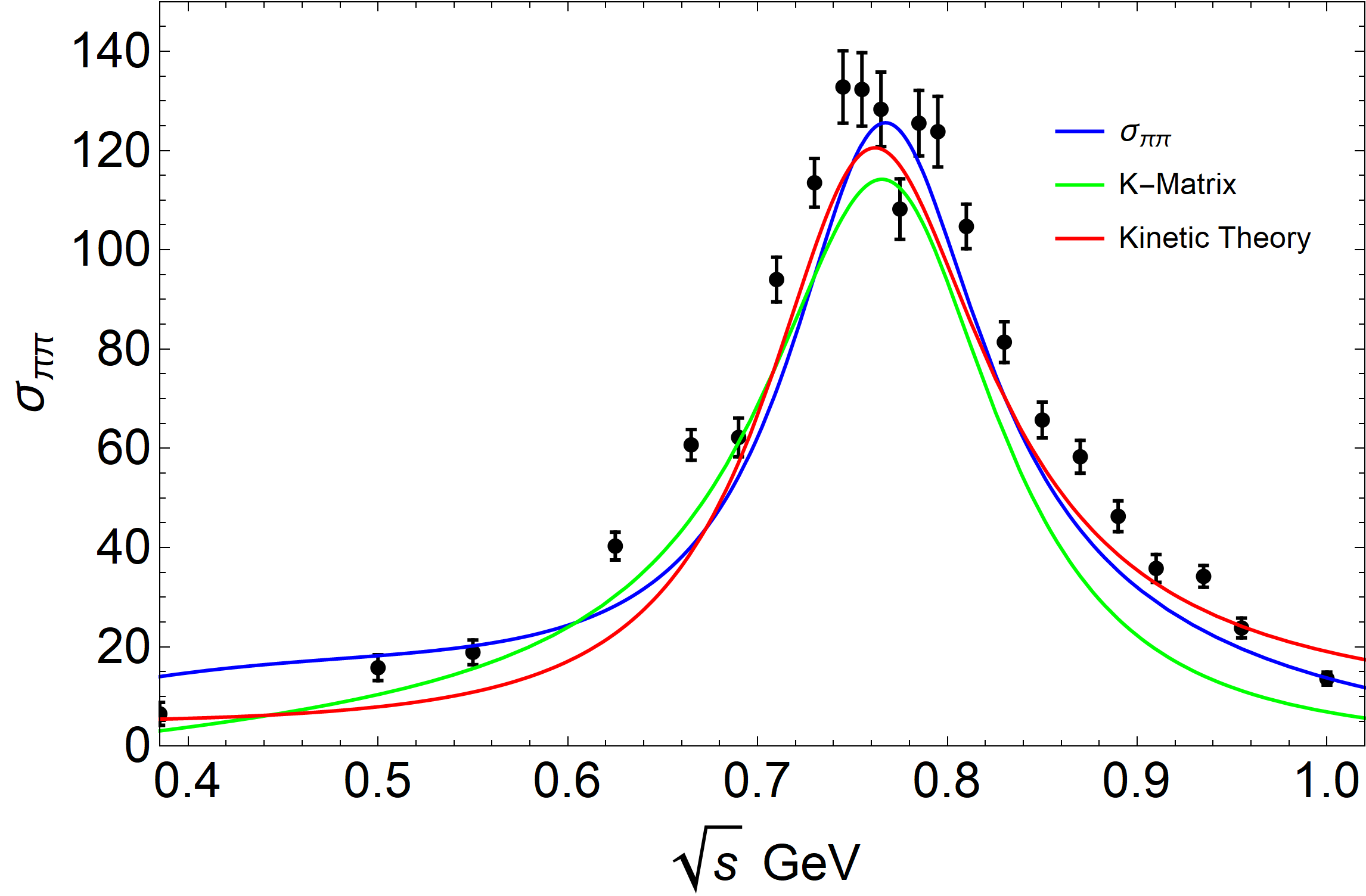}
\caption{Comparison of our $\pi\pi$ cross section (blue) to experimental data (black dots)~\cite{pipiCSdata} and two previous works with applications to the 
electric conductivity: Breit-Wigner ansatz for the $\rho$-resonance cross section (red)~\cite{Greif}, $K$-matrix formalism (green)~\cite{Kadam}. 
%For the K-matrix we plot only ref. \cite{Kadam}'s lowest lying resonance, the rho meson.
}
\label{fig_xsec}
\end{figure}

%%%%%%%%%%%%%%%%%%%
For comparisons to vacuum scattering data the pion energies, $k_{0}$ and $p_{0}$, are taken to be on-shell, while the CM energy, $\Ecm=\sqrt{s}$, is taken as an external variable. The CM momentum is then given by
\be
\label{qcm-on}
\qcm(s) =  \sqrt{\frac{s}{4}-m_\pi^2} \ .
\ee
From the vacuum resonance self-energies, the $\pi\pi$ scattering phase shift in the pertinent resonance channel ($R=\rho, \sigma$) can be obtained as
\be
\tan\delta_R(\Ecm) = -\frac{{\rm Im}\Sigma_R(\Ecm)}{{\rm Re} \Sigma_R(\Ecm)}  \ . 
\ee 
We utilize these data to fit our parameters resulting to describe the $S$- and $P$-wave $\pi\pi$ scattering phase shifts~\cite{Froggatt}. The 
fits are shown in Fig.~\ref{fig_phaseshift} based on the values: $g_{\sigma}=8.86$, $m_{\sigma}=0.934 \, \textrm{GeV}$, $\Lambda_{\rho}=1.85 \, \textrm{GeV}$, and $\Lambda_{\sigma}= 0.745\, \textrm{GeV}$.
%{\bf com: either MeVor GeV,but not both}\\
The fit quality is rather good up to CM energies of 1\,GeV which is sufficient for our application to a thermal pion gas 
for temperatures of up to 180\,MeV.
Finally, we calculate the resulting elastic $\pi\pi$ cross section by utilizing the optical theorem to express it through the imaginary part of the 
forward scattering amplitude $M_{\pi\pi}$, 
\be
\label{eq28a}
\sigma_{\pi\pi}(s)=- \frac{(1}{\sqrt{s(s-4m_{\pi}^{2})}}\textrm{Im}[M_{\pi\pi}(s)] \ ,
\ee
where the amplitude for a given spin-isospin channel can be related to the resonance propagator as
\begin{eqnarray}
\label{eq31c}
M_{\pi\pi}^{00}(s)&=&\frac{g_\sigma^{2}}{2}(s- m_{\pi}^{2})D_{\sigma}(s)\textrm{FF}_{\sigma}(\qcm^{2})\\
\label{eq31d}
M_{\pi\pi}^{11}(s)&=&g_\rho^{2}(s-4 m_{\pi}^{2})D_{\rho}(s)\textrm{FF}_{\rho}(\qcm^{2}).
\end{eqnarray}
In Fig.~\ref{fig_xsec} we compare our cross section to experimental data~\cite{pipiCSdata}, and to several other works where it was also used as an
input to calculating the electric conductivity~\cite{Kadam,Greif}.
The $\rho$ resonance peak in our cross section turns out to be somewhat narrower than experimental data, a feature that is apparently shared
by other calculations and possibly related to the absence of non-resonant "background" terms. 

%We intend to only dress $D_{\pi}$ with thermal interactions, therefore we drop all terms form eqs. \ref{eq25} and \ref{SigmaS1} that correspond to pion decays. Furthermore, we perform calculations in hadronic matter at temperatures between $100$ and $180 \; \textrm{MeV}$, where thermal $\rho$ and $\sigma$ massons are heavily suppressed, due to their large masses relative to the temperature. Thus, we do not include scattering with thermal $\rho$ or $\sigma$ mesons.

%%%%%%%%%%%%%%%%%%%%%%%%%%
\subsection{Optical potentials}
\label{ssec:optical}
%%%%%%%%%%%%%%%%%%%%%%%%%
To obtain the in-medium pion selfenergy we first compute the imaginary part directly from Eq.~(\ref{sigma_pi}) and subsequently the real part from a 
subtracted dispersion relation, 
\begin{eqnarray}
\label{eq36}
\textrm{Re}\Sigma_{\pi}(k)&=&\int_{-\infty}^{\infty}\frac{dw}{-\pi}\Big(\frac{\textrm{Im}\Sigma_{\pi}(w,\vec{k})}{k_{0}-w}-\frac{\textrm{Im}\Sigma_{\pi}(w,\vec{k})}{-w}\Big) \ .
\nonumber\\
\end{eqnarray}
The subtraction ensures that $\textrm{Re}\Sigma_{\pi}$ is zero for zero energy. This is motivated by the pion being a Goldstone boson, 
to approximately implement chiral constraints on the low-energy behavior of the amplitudes~\cite{Rapp:1995fv,Rapp:1995ir}.
A convenient and more intuitive way to represent the pion self-energy, whose dimension is quadratic in energy, is to define corresponding 
``optical potentials", which formally can be thought of as resulting from the leading-order corrections to the pion's dispersion relation. 
Their on-shell values take the form
\be
\label{eq35}
U_{\pi}(\vec{k}) = \Sigma_{\pi}(\omega_{k},\vec{k})/2\omega_{k} \ , 
\ee
representing the real and imaginary parts of the in-medium contributions to the on-shell pion self-energy (also note that $-2{\rm Im} U_\pi$ amounts
to the Breit-Wigner width, $\Gamma_\pi$ (or full-width-half-maximum), around the on-shell pole in the spectral function) 
\begin{figure}
\includegraphics[width=19pc]{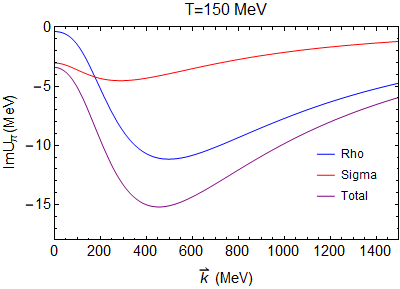}
\includegraphics[width=19pc]{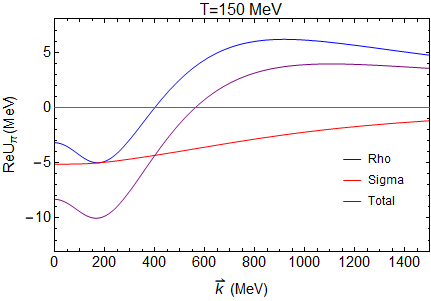}
\caption{Imaginary (upper panel) and real part (lower panel) of the on-shell pion optical potential in hot pion matter, plotted as a function of 3-momentum 
for $T=150 \, \textrm{MeV}$. Contributions from the rho (blue) and sigma (red) resonances are shown in addition to the total (purple).}
\label{fig_optpot}
\end{figure}
We graph the real and imaginary parts of the optical potential in Fig.~\ref{fig_optpot}; they generate a pion mass shifts of up to $\pm$10\,MeV, 
and pion widths of up to 30\,MeV. This is in rather good agreement with the results of 
Refs.~\cite{Rapp:1995fv,Rapp:1996ym} for the $\sigma$ and $\rho$ channels. However, for momenta of up to $\sim$\,1\,GeV the imaginary part 
of our total pion self-energy is a few MeV smaller than that obtained in the previous work, because the latter also contains non-resonant $S$-wave and 
$D$-wave contributions. As these contributions lead to additional complications with preserving gauge invariance, while their quantitative impact is only 
at the 10\% level, we neglect them for now.  
%We do not consider additional resonances here, because the $\rho$ and $\sigma$ provide the dominant contribution for low momentum, and the inclusion of further resonances would complicate efforts to maintain gauge invariance.

%%%%%%%%%%%%%%%%%%%%%%%%%%%%%%%%
\section{Gauge invariance in medium}
\label{sec:vcorr}
%%%%%%%%%%%%%%%%%%%%%%%%%%%%%%%%%%
In this section we scrutinize the issue of gauge invariance, that is violated when introducing in-medium pion self-energies and therefore needs to be
restored by accounting for pertinent vertex corrections. Toward this end, we first introduce in Sec.~\ref{ssec:ward} the Ward identities between 
2-, 3- and 4-point functions in our model that provide a sufficient criterion for a gauge invariant set of diagrams. In Sec.~\ref{ssec:vcorr} we lay out the 
general strategy of how the pertinent vertex functions can be constructed. In Sec.~\ref{ssec:inter-dress} we address the complications associated with 
having to dress intermediate propagators inside the vertex corrections which is dictated by rendering the conductivity finite, while in Sec.~\ref{ssec:vcorr-FF}
we elaborate on how to approximately maintain gauge invariance in the presence of form factors that figure in the vertex corrections. Finally, in 
Sec.~\ref{ssec:sigrhoT} we summarize our expression for the transverse $\rho$ self-energy in compact form at vanishing 3-momentum that will be 
used for the calculation of the conductivity.

%%%%%%%%%%%%%%%%%%%%%%%%%%%%%
\subsection{Ward identities}
\label{ssec:ward}
%%%%%%%%%%%%%%%%%%%%%%%%%
In the VDM the $\rho$ meson couples to a conserved current, and thus must be four-dimensionally transverse, 
$q_{\mu}\Sigma_{\rho}^{ \mu\nu}=0$. Transversality is ensured if the $\rho\pi\pi$ and $\rho\rho\pi\pi$ vertices satisfy the Ward-Takahashi identities
between the 2-point and 3-point functions, as well as the 3-point and 4-point functions in the model, \ie, 
\begin{eqnarray}
\label{eq43}
q^{\mu}\Gamma_{\mu\,ab3}^{(3)}&=&g_\rho \epsilon_{3ab}(D^{-1}_{\pi}(k+q)-D^{-1}_{\pi}(k)) \ ,\\
\label{eq43a}
q^{\mu}\Gamma_{\mu\nu\, ab33}^{(4)}&=&ig_{\rho} (\epsilon_{3ca}\Gamma_{\nu\,bc3}^{(3)}(k,-q)
\nonumber\\
&&-\epsilon_{3bc}\Gamma_{\nu\,ca3}^{(3)}(k+q,-q)) \ .
\end{eqnarray}
The Ward-Takahashi identities are straightforwardly satisfied in vacuum, but are upset by the introduction of a thermal pion self-energy in $D_{\pi}$. 
This can be remedied by considering thermal corrections to the $\rho\pi\pi$ and $\rho\rho\pi\pi$ vertices~\cite{Song,Urban1998}. Following these
references, the in-medium vertices can be written as
\begin{eqnarray}
\label{WT2}
\Gamma_{\mu \, ab3}^{(3)}&=&g_{\rho}\epsilon_{3ab}(2k+q)_{\mu}+\Gamma_{\mu \, ab3}'^{(3)} \ ,
\\
\label{WT3}
\Gamma_{\mu\nu \, ab33}^{(4)}&=&2ig_{\rho}^{2}(\delta_{ab}-\delta_{3a}\delta_{3b})g_{\mu\nu}+\Gamma_{\mu\nu \, ab33}'^{(4)} \ ,
\end{eqnarray}
where $\Gamma_{\mu \, ab3}'^{(3)}$ and $\Gamma_{\mu\nu \, ab3}'^{(4)}$ are vertex corrections to the $\rho\pi\pi$ and $\rho\rho\pi\pi$ vertices. 
The Ward identities will hold if the vertex corrections satisfy
\begin{eqnarray}
\label{eq47}
q^\mu\Gamma_{\mu \, ab3}'^{(3)}&=&g_\rho \epsilon_{3ab}(\Sigma_\pi(k)-\Sigma_\pi(k+q)),\\
\label{eq48}
q^\mu\Gamma_{\mu\nu \, ab33}'^{(4)}&=&ig_\rho (\epsilon_{3ca}\Gamma_{\nu \, bc3}'^{(3)}(k,-q)\nonumber\\
&&-\epsilon_{3bc}\Gamma_{\nu \, ca3}'^{(3)}(k+q,-q)),
\end{eqnarray}
In the following, we will use these as our guiding principle.

%%%%%%%%%%%%%%%%%%%%%%%%%%%%
\subsection{Vertex corrections}
\label{ssec:vcorr}
\begin{figure}
\includegraphics[width=19pc]{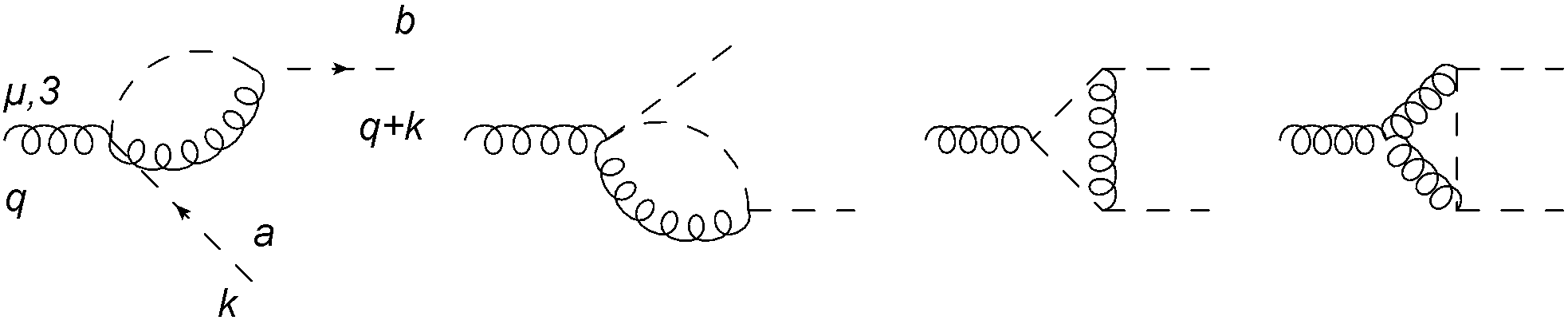}
\caption{\label{figVCS1}Corrections to the $\rho\pi\pi$ vertex due to $\Sigma_{\pi}^{(\rho)}$.}
\end{figure}

\begin{figure}
\includegraphics[width=19pc]{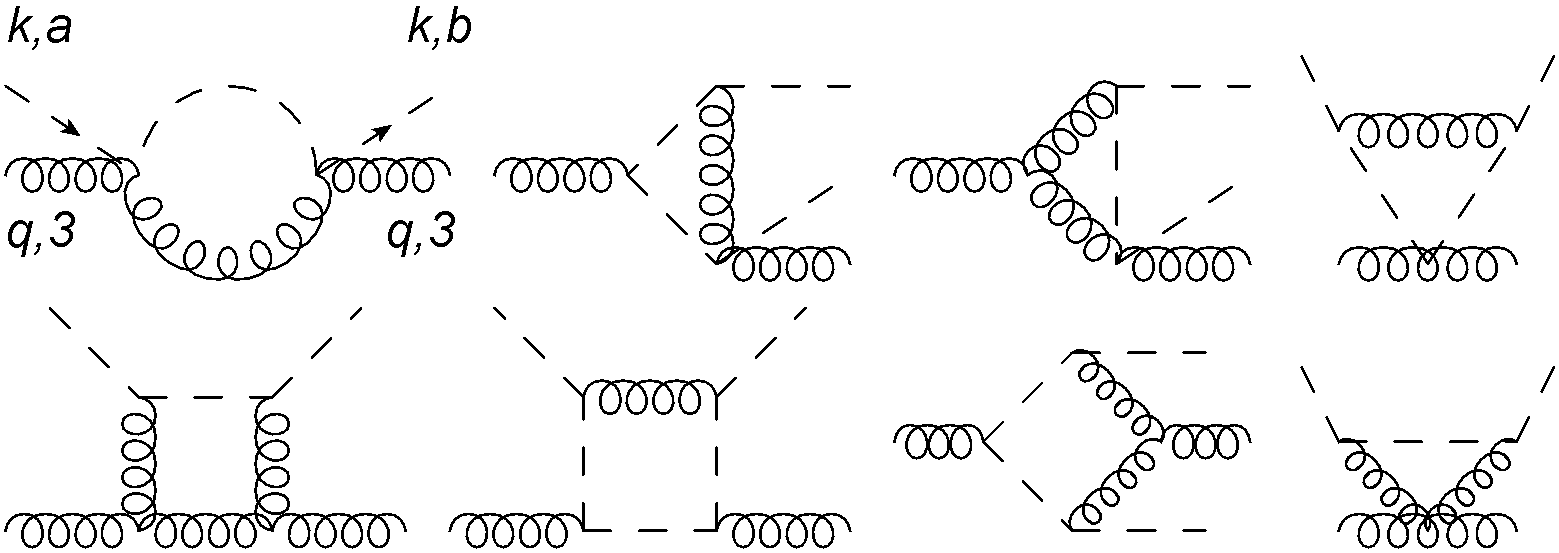}
\caption{\label{figVCS2}Corrections to the $\rho\rho\pi\pi$ vertex due to $\Sigma_{\pi}^{(\rho)}$.}
\end{figure}

\begin{figure}
\includegraphics[width=19pc]{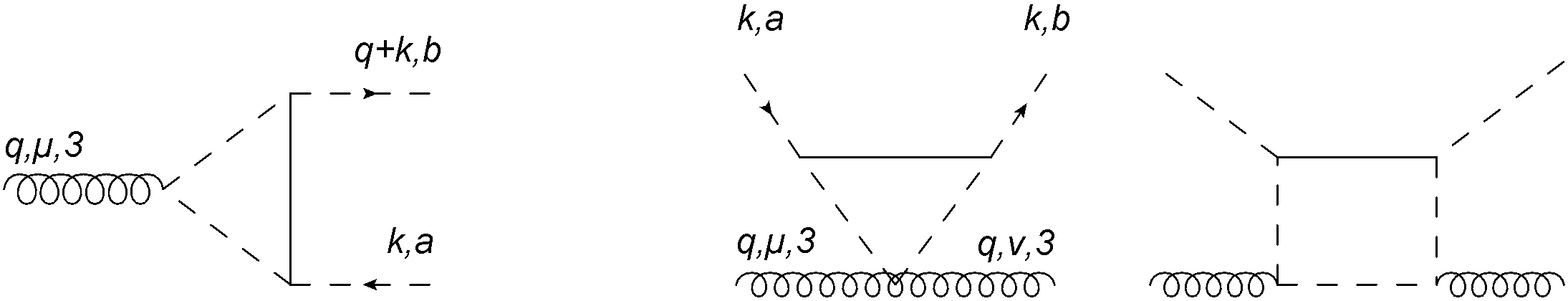}
\caption{\label{figVCS3}Left panel: Correction to the $\rho\pi\pi$ vertex due to $\Sigma_{\pi}^{(\sigma)}$. Right panels: Corrections to the $\rho\rho\pi\pi$ vertex due to $\Sigma_{\pi}^{(\sigma)}$.}
\end{figure}

%%%%%%%%%%%%%%%%%%%%%%%%%%%
The corrections to the $\rho\pi\pi$ vertex sufficient to satisfy the Ward identities can be generated by coupling a $\rho$ meson (or photon) to all 
possible charged-particle lines in $\Sigma_{\pi}$. Similarly, one can couple two $\rho$ mesons to $\Sigma_{\pi}$ in all possible configurations to obtain corrections to the $\rho\rho\pi\pi$ vertex~\cite{Urban1998}. We then analytically determine which vertex corrections are necessary to maintain the 
Ward identities when $D_{\pi}$ is dressed with only thermal $\pi\pi$ scattering. Figures \ref{figVCS1} through \ref{figVCS3} show the vertex 
corrections resulting from dressing $D_{\pi}$ with $\Sigma_{\pi}^{(\rho)}$ and $\Sigma_{\pi}^{(\sigma)}$. We show only the minimum diagrams necessary
to maintain gauge invariance. Consequently, there are fewer corrections for the $\sigma$ resonance interactions, because the $\sigma$ meson is neutral 
and the $\sigma\pi\pi$ vertex is a Lorentz scalar. Furthermore, the vertex corrections require the $\rho\rho\rho$, $\rho\rho\rho\rho$, and $\sigma\pi\pi$ vertices. Terms including the $\rho\rho\rho$ or $\rho\rho\rho\rho$ vertex contain multiple $\rho$ propagators, and are therefore suppressed 
by $1/m_{\rho}^{2}$. Thus we drop all corrections containing these vertices and do not write them here. 

%We have not established a $\sigma\pi\pi$ vertex. {\bf -- that sounds strange given our calculation of the sigma selfenergy; probably should drop this statement}. \\
In order to calculate vertex corrections involving a $\sigma$ meson we introduce an effective $\sigma\pi\pi$ vertex that corresponds to our calculation
of the vacuum self-energy of the $\sigma$ meson through $\pi\pi$ loops in Sec.~\ref{ssec:sigpi}, 
\begin{eqnarray}
\label{eq24}
\Gamma_{\sigma \, ab}^{(3)}&=&i\delta_{ab}g_{\sigma}\sqrt{s-m_{\pi}^{2}} \ , 
\end{eqnarray}
which reproduces our results for $\Sigma_{\sigma}$ and  $\Sigma_{\pi}^{(\sigma)}$. The expression for $\Gamma_{\sigma \, ab}^{(3)}$ can then 
be used to calculate vertex corrections involving the $\sigma\pi\pi$ vertex. We calculate all diagrams resulting from replacing a vacuum vertex with a corresponding vertex correction. For the four-point vertex corrections, we also include diagrams resulting from interchanging $\rho$-meson or pion 
propagators when they produce unique vertex corrections.

%%%%%%%%%%%%%%%%%%%%%%%%%%%%%%%%%%%%%%%
\subsection{Dressing intermediate particles}
\label{ssec:inter-dress}
%%%%%%%%%%%%%%%%%%%%%%%%%%%%%%%%%%%%%%%
The vertex corrections contain intermediate $\rho$, $\sigma$, and pion propagators. As we have seen for the Landau cut of $\Sigma_{\rho}^{\mu\nu}$, 
if these propagators are not dressed, the conductivity will diverge. Therefore, we dress the $\rho$ and $\sigma$ propagators with their vacuum 
self-energies and treat the intermediate-pion propagators self-consistently, \ie,
\be
\label{PionProp}
D_{\pi}(k)=\frac{1}{k^{2}-m_{\pi}^{2}-\Sigma_{\pi}(k)} \ .
\ee
However, thermal pions within vertex corrections and the pion self-energy are not dressed. The widths in the pion and $\rho$ propagators cause 
another layer of violation of gauge invariance. However, the violation due to dressing $D_{\rho}$ only occurs in vertex corrections containing multiple 
$\rho$ propagators, which are suppressed by powers of $1/m_\rho$. The violation due to dressing intermediate-pion propagators within vertex corrections 
can be corrected by dressing the $\rho\pi\pi$ vertices that couple to an external $\rho$ with three point-vertex corrections and dressing thermal-pion 
propagators, creating a self-consistency equation. These effects are expected to be small, due to the small pion width, thus we will not calculate these 
corrections in this work.

%%%%%%%%%%%%%%%%%%%%%%%%%%%%%%%%%%%
\subsection{Vertex correction form factors}
\label{ssec:vcorr-FF}
%%%%%%%%%%%%%%%%%%%%%%%%%%%%%%%%%
In Ref.~\cite{Urban1998} the in-medium dressing of the pions with nucleons and $\Delta(1232)$'s  was implemented in a non-relativistic approximation, 
so that the corresponding form factors depended only on the pion's three-momentum (the lab frame momentum). The form factors were generated by 
attaching heavy pions to the external pions in $\Sigma_{\pi}$ and the vertex corrections. These form factors violated the Ward identities, however, 
this violation was corrected by including further vertex corrections, derived by coupling $\rho$ mesons directly to heavy-pion propagators~\cite{Urban1998}. 
The additional vertex corrections could be expressed in terms of the pion self-energy and the original corrections to the $\rho\pi\pi$ vertex.
Two complications arise when applying the regularization procedure from Ref.~\cite{Urban1998} with the form factors introduced in Sec.~\ref{ssec:ff}. 
First, different form factors were introduced for the $S$- and $P$-wave contributions to the self-energy, and second, the CM momentum was used to 
as the variable. 

In order to extend the approach of Ref.~\cite{Urban1998} to multiple form factors we note that $\Sigma_{\pi}^{(\rho)}$ and its vertex corrections, 
$\Gamma'^{(3)\rho}_{\mu \, ab3}$ and $\Gamma'^{(4)\rho}_{\mu\nu \, ab33}$, independently satisfy Eqs.~(\ref{eq47}) and (\ref{eq48}), such that
\begin{eqnarray}
\label{eq47r}
q^{\mu}\Gamma'^{(3)\rho}_{\mu \, ab3}&=&g_{\rho} \epsilon_{3ab}(\Sigma_{\pi}^{(\rho)}(k)-\Sigma_{\pi}^{(\rho)}(k+q)) 
\\
\label{eq48r}
q^{\mu}\Gamma'^{(4)\rho}_{\mu\nu \, ab33}&=&ig_{\rho} (\epsilon_{3ca}\Gamma'^{(3)\rho}_{\nu \, bc3}(k,-q)
\nonumber\\
&&-\epsilon_{3bc}\Gamma'^{(3)\rho}_{\nu \, ca3}(k+q,-q)) \ .
\end{eqnarray}
Similarly, $\Sigma_{\pi}^{(\sigma)}$ and its vertex corrections, $\Gamma'^{(3)\sigma}_{\mu \, ab3}$ and $\Gamma'^{(4)\sigma}_{\mu\nu \, ab33}$, satisfy
\begin{eqnarray}
\label{eq47s}
q^{\mu}\Gamma'^{(3)\sigma}_{\mu \, ab3}&=&g_{\rho} \epsilon_{3ab}(\Sigma_{\pi}^{(\sigma)}(k)-\Sigma_{\pi}^{(\sigma)}(k+q))
\\
\label{eq48s}
q^{\mu}\Gamma'^{(4)\sigma}_{\mu\nu \, ab33}&=&ig_{\rho} (\epsilon_{3ca}\Gamma'^{(3)\sigma}_{\nu \, bc3}(k,-q)
\nonumber\\
&&-\epsilon_{3bc}\Gamma'^{(3)\sigma}_{\nu \, ca3}(k+q,-q)) \ .
\end{eqnarray}
One realizes that the $\sigma$ and $\rho$ resonances form subgroups that independently satisfy the Ward identities. Thus, the regularization procedure 
from Ref. \cite{Urban1998} can be applied separately to $S$- and $P$-wave scattering. The total regularized vertex corrections are then given by the 
sum of the corrections for each resonance.

Next we address the use of the CM momentum in the form factor. For $\pi\pi$ scattering we cannot apply the non-relativistic approximations which 
mandated to employ a form factor depending on the CM momentum, $\qcm$. Furthermore, the dependence on $\qcm$, rather than $\vec{k}$, 
prevents the introduction of spurious total-momentum dependencies. While the use of $\qcm$ renders the pion self-energy more robust, it is not clear 
how to satisfy the Ward identities in this framework. However, the violation of the Ward identities is proportional to the difference in the $\qcm^{2}$ 
of the two pions in the $\rho$ self-energy $\pi\pi$ loop over $\Lambda_{R}^{2}$. Since we are chiefly interested in $\sig$, which involves the 
zero-momentum low-energy limit of the EM spectral function, the violation is parametrically suppressed, because $\Lambda_{R}^{2}$ is on the order 
of several hundred MeV, while $\qcm$ is on the order of a few tens of MeV. This low energy suppression is expected because the 
form factors are constructed to suppress high-momentum behavior, while minimally affecting the low-momentum regime.
Furthermore, to leading order the violation is proportional to $q_{0}^{2}$ (where $q_0$ is the $\rho$-meson's energy), while $\Lambda_{\rho}$ 
is on the order of 1 GeV for the $\rho$ resonance. Thus, the violation due to the form factors in $\Sigma_{\pi(\rho)}$ should still be appreciably 
suppressed around the rho mass. Although $\Lambda_{\sigma} =0.745 \textrm{GeV})$ is of the same order as  $m_{\rho}$, the 
effects of $\pi\pi$-resonant scattering through a $\sigma$ resonance on the EM spectral function are suppressed for large $q_{0}$. Therefore, 
the violation of gauge invariance due to using $\qcm$ in the form factor is expected to be small even for $q_{0}$ around the rho mass.

Additionally, in order to further analyze the effect of the violation on our results, and provide an approximate correction for the violation, we follow 
the approach of Ref.~\cite{Urban1998} and generate additional vertex corrections involving heavy-pion propagators. However, at the vertices where 
a heavy-pion propagator couples to a thermal loop we replace the lab frame momentum, $\vec{k}$, with the center of mass momentum, $\qcm$. 
In Sec.~\ref{sec:PIem} we will assess the effect of the additional vertex corrections on the EM spectral function, in order to quantify the effect 
this violation of gauge invariance has on our results.

The total vertex corrections to the $\rho\pi\pi$ vertex, including terms where the $\rho$ meson couples to a heavy-pion propagator, can then 
be written in terms of the original vertex corrections plus terms involving the pion self-energy. Similarly, the total corrections to the 
$\rho\rho\pi\pi$ vertex can be written in terms of the original corrections plus terms involving the pion self-energy or the original vertex corrections 
to the $\rho\pi\pi$ vertex. The regularized corrections to the $\rho\pi\pi$ and $\rho\rho\pi\pi$ vertices are written out in Appendix~\ref{app:vcorr}.
\\

%%%%%%%%%%%%%%%%%%%%%%%%%%%%%
\subsection{$\rho$ self-energy at finite temperature}
\label{ssec:sigrhoT}
%%%%%%%%%%%%%%%%%%%%%%%%%%%%%
Here we write the transverse projection of $\Sigma_{\rho}^{\mu\nu}$ for arbitrary $\rho\pi\pi$ and $\rho\rho\pi\pi$ vertices, $\Gamma_{\mu \, ab3}^{(3)}(k,q)$ and $\Gamma_{\mu\nu \, ab33}^{(4)}(k,q)$, at $\vec{q}=0$:
\begin{widetext}
\begin{eqnarray}
\label{eq102}
\Sigma_{\rho}^{T}(q)&=&\frac{4\pi T}{3}\sum_{n(even)}\int\frac{d|\vec{k}|\vec{k}^{2}}{(2\pi)^3}D_{\pi}(k)\Big[D_{\pi}(q+k)\big(g_{\rho}\epsilon_{3ab}(2|\vec{k}|)+\Gamma_{3 \, ab3}'^{(3)}(k,q)\big)\big(g_{\rho}\epsilon_{3ba}(2|\vec{k}|)+\Gamma_{3 \, ba3}'^{(3)}(k+q,-q)\big)\nonumber\\
&&-\big(\frac{3g_{\rho}^{2}}{\pi}+i\Gamma_{ii \, aa33}'^{(4)}(k,q)\big)\Big]_{k_{0}=i\omega_{n}},
\end{eqnarray}
\end{widetext}
where we have performed the angular integrations analytically, and $\omega_{n}$ are discrete Matsubara frequencies. The vertex corrections are complex and introduce nontrivial energy dependence into the vertices, and thus must be written with a spectral representation, before Matsubara sums are performed. In appendix 2 we establish the relevant spectral representations and carry out the summations in eq.~(\ref{eq102}).

%%%%%%%%%%%%%%%%%%%%%%%%%%%%%%%%%%%%%%%
\section{Electromagnetic spectral function in pion matter}
\label{sec:PIem}
%%%%%%%%%%%%%%%%%%%%%%%%%%%%%%%%%%%%%%%%
We are now in position to discuss our numerical results, starting with the in-medium $\rho$ self-energy in Sec.~\ref{ssec:sigrho},  followed by the 
EM spectral function in Sec.~\ref{ssec:EMspec} and the temperature dependence of the electric conductivity in Sec.~\ref{ssec:cond}.

%%%%%%%%%%%%%%%%%%%%%%%%%%%
\subsection{Rho self-energy}
\label{ssec:sigrho}
\begin{figure}
\includegraphics[width=19pc]{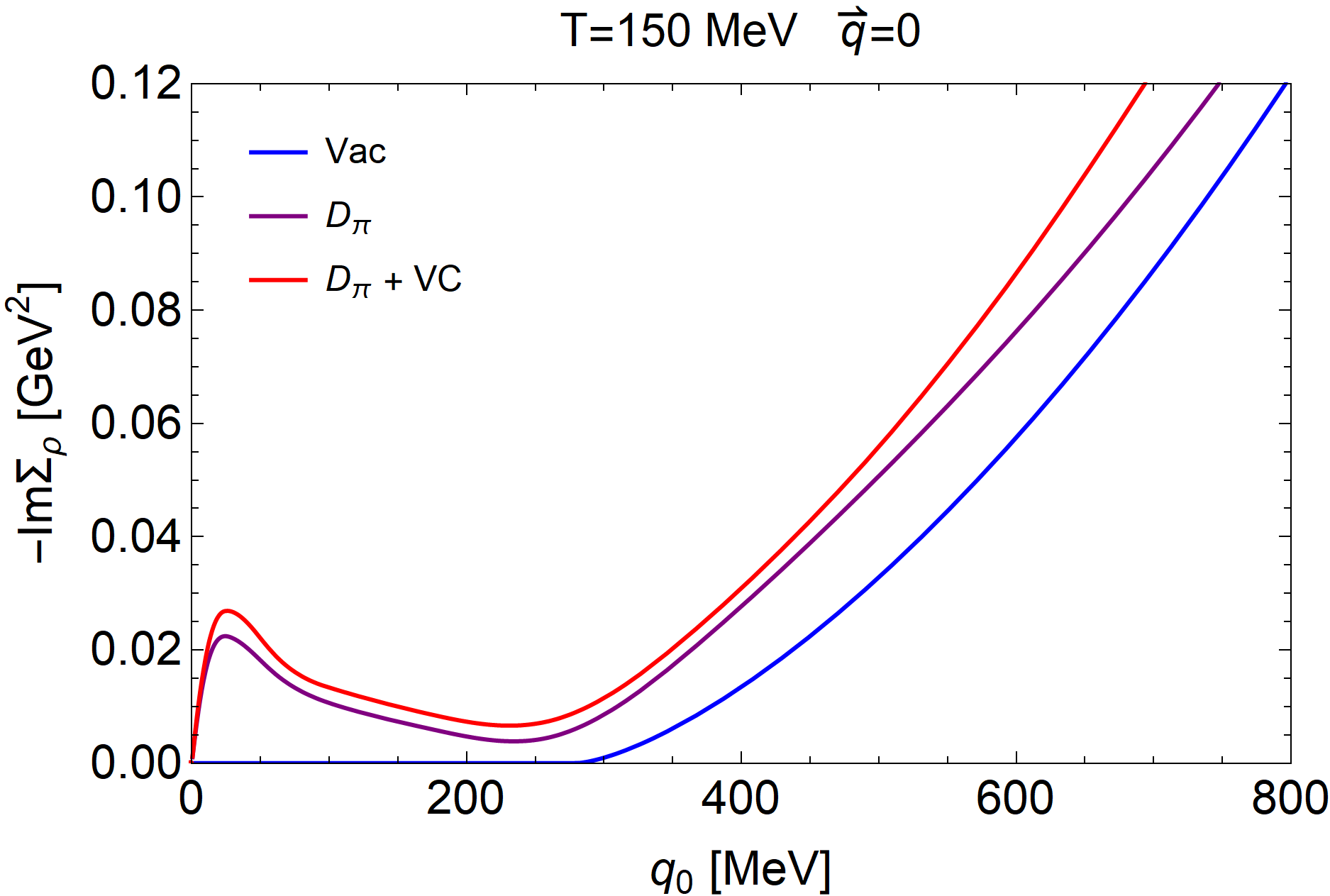}

\vspace{0.5cm}

\includegraphics[width=19pc]{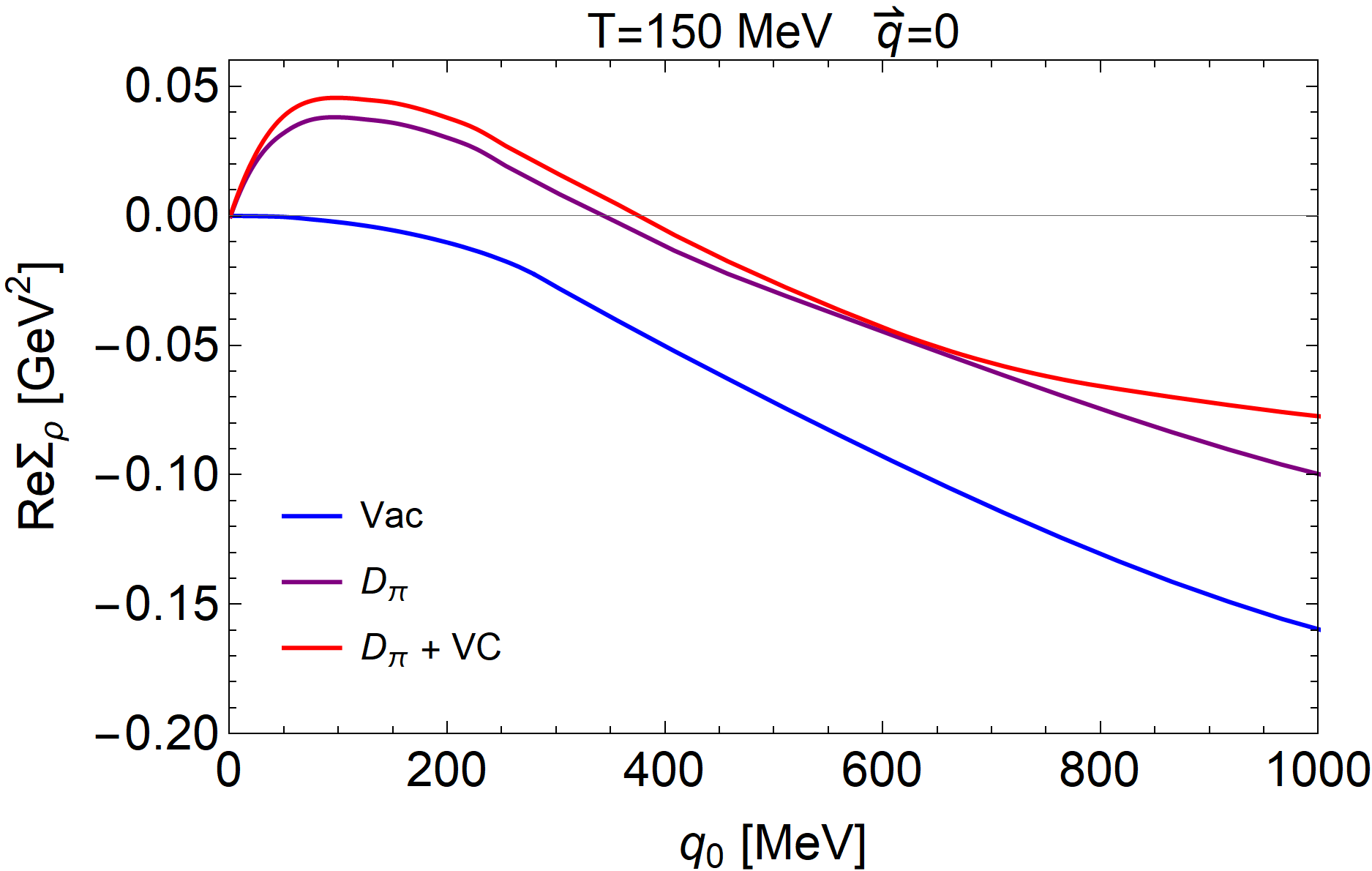}
\caption{Imaginary (upper panel) and real part (lower panel) of the transverse projection of the $\rho$ self-energy in vacuum (blue line) and  at $T$=150 MeV 
with (purple) and without vertex corrections (red). The results include both $\rho$ and $\sigma$ resonances in the $\pi\pi$ scattering amplitude.}
\label{fig:sigrho-vc}
\end{figure}

\begin{figure}
\includegraphics[width=19pc]{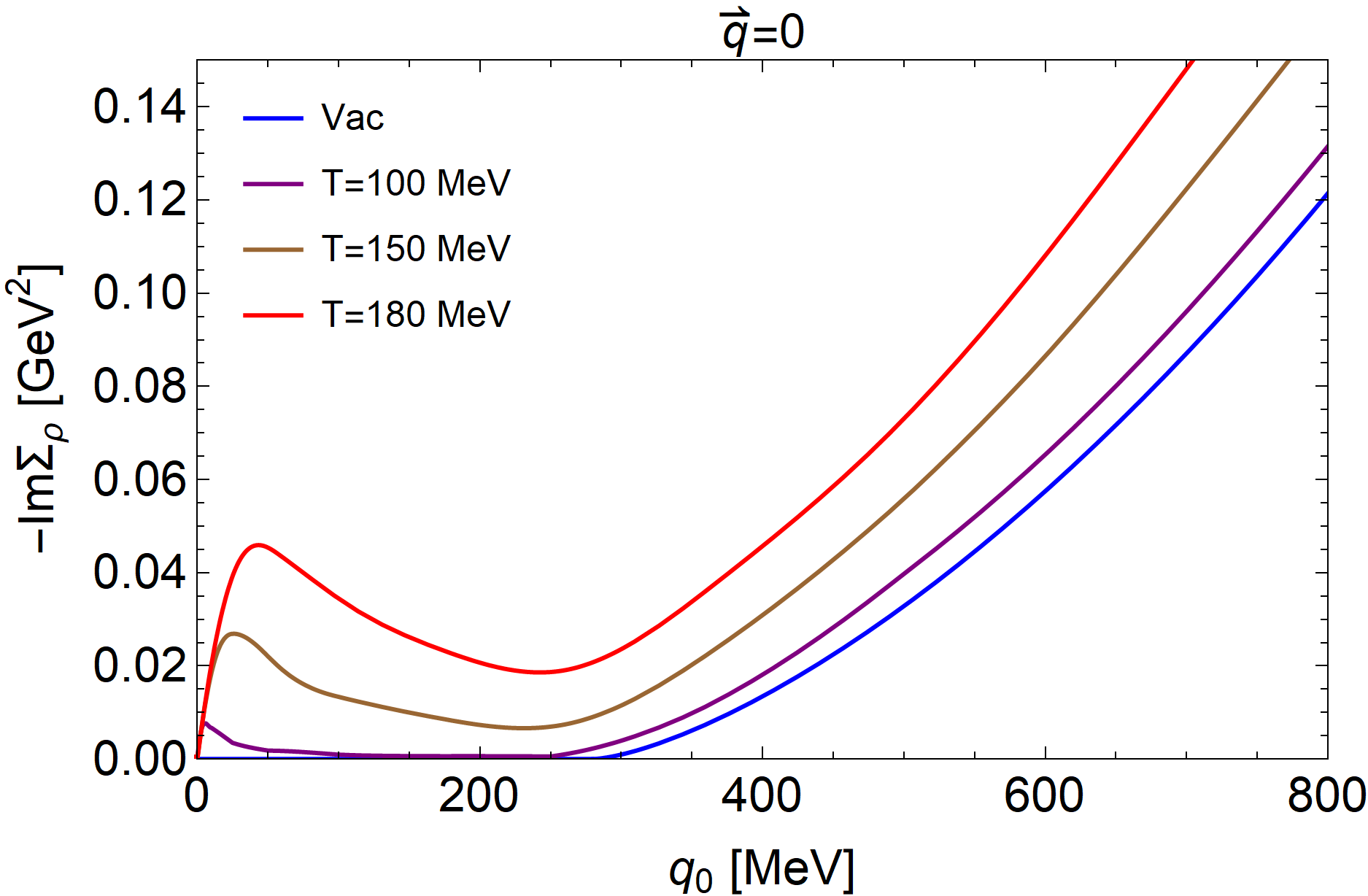}

\vspace{0.5cm}

\includegraphics[width=19pc]{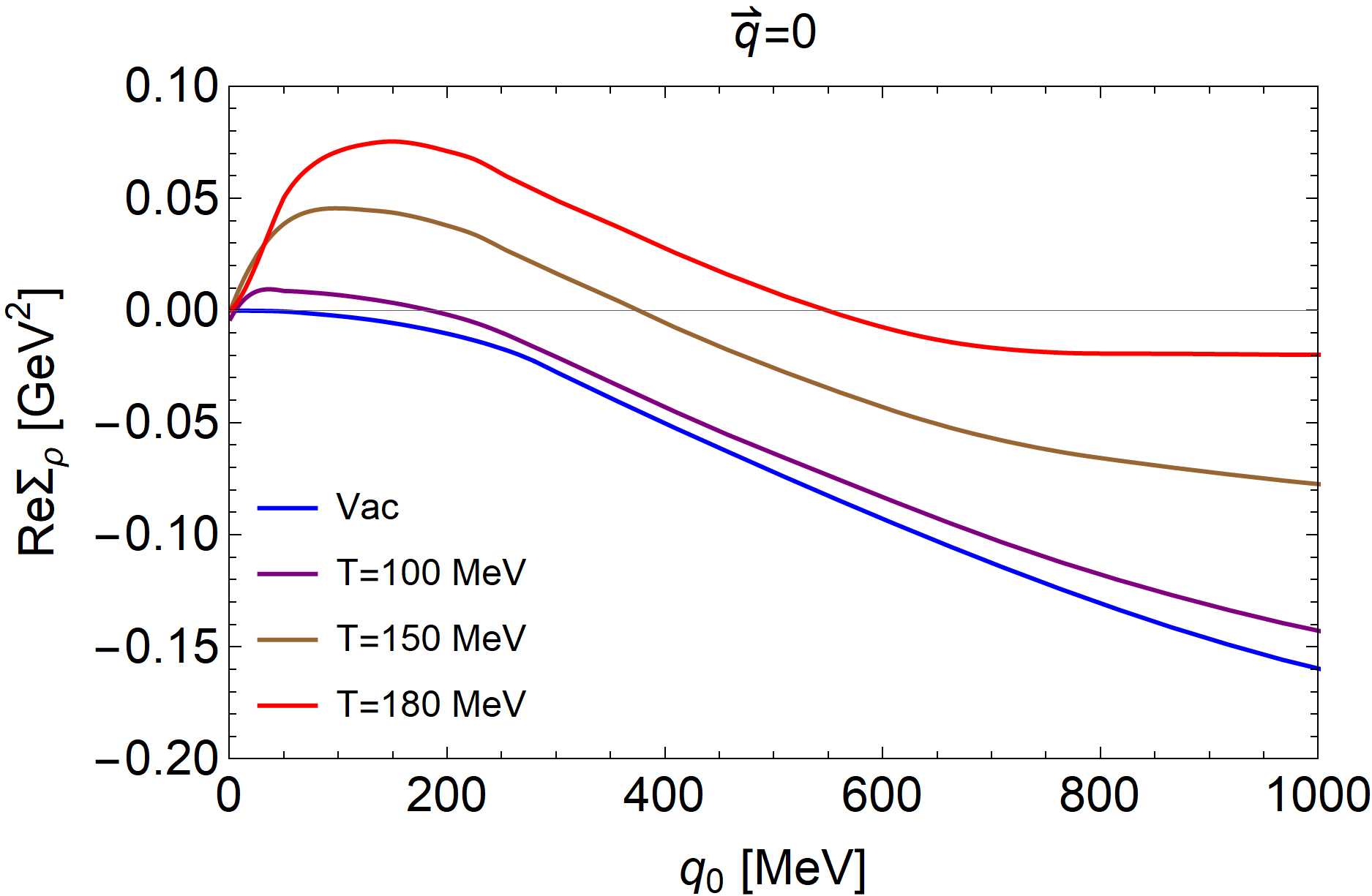}
\caption{Imaginary (upper panel) and real (lower panel) part of the transverse projection of the $\rho$ self-energy at $\vec{q}=0$ in vacuum (blue), and for various temperatures ($T$=100 MeV: purple, $T$=150 MeV: brown, and $T$=180 MeV: red), including vertex corrections. Results include the $\rho$ and $\sigma$ resonances in the $\pi\pi$ scattering amplitude.}
\label{fig:sigrhoT}
\end{figure}

%%%%%%%%%%%%%%%%%%%%%%%%%%
Let us first illustrate the role of the vertex corrections. In Fig.~\ref{fig:sigrho-vc} we display the imaginary and real parts of the transverse projection 
of the $\rho$ self-energy at $\vec{q}=0$, with and without vertex corrections, at a temperature of $T$=150 MeV. Compared to the vacuum result,
the use of in-medium pion propagators enhances the imaginary part considerably, most notably below the two-pion threshold where the vacuum 
result vanishes. At the same time the magnitude of the real part reduces, except below threshold, where the vacuum result is small and attractive 
while the in-medium one has turned significantly repulsive. The inclusion of vertex corrections increases the imaginary part, which is expected as they
create additional inelastic channels, again most notably in the
sub-threshold region, while the real part also shows a slight increase.  
Next, we turn to the temperature dependence of the $\rho$ self-energy, illustrated in Fig.~\ref{fig:sigrhoT} (which includes vertex corrections).
As we have seen before, the medium effects on $\textrm{Re}\Sigma_{\rho}^{T}$ are repulsive, causing the rho mass to increase. 
The increase of $\textrm{Im}\Sigma_{\rho}^{T}$ with temperature is again most pronounced in the low-energy regime, while the increase at higher
energies produces a broadening of the $\rho$ resonance peak. In particular, the Landau cut 
of the $\rho$ self-energy, corresponding to $\pi+\rho\to\pi$ scattering, generates a marked bump at very low energy, which, as we will see more clearly in 
the next section, corresponds to a broadening of the transport peak in $\rho_{\textrm{EM}}$. 

%%%%%%%%%%%%%%%%%%%%%%%%%%%%%
\subsection{EM spectral function}
\label{ssec:EMspec}
%%%%%%%%%%%%%%%%%%%%%%%%%%%%%%
In this section we will discuss our results for the EM spectral function, averaged over transverse polarization and divided by energy with a normalization
that yields the value of the electric conductivity at the intercept at zero energy, $e^2\rho_{\textrm{EM}}^{ii}/6q_{0}$, recall Eq.~(\ref{eq3}). In 
particular, the division by energy will clearly exhibit the transport peak near vanishing energy.

\begin{figure}
\includegraphics[width=19pc]{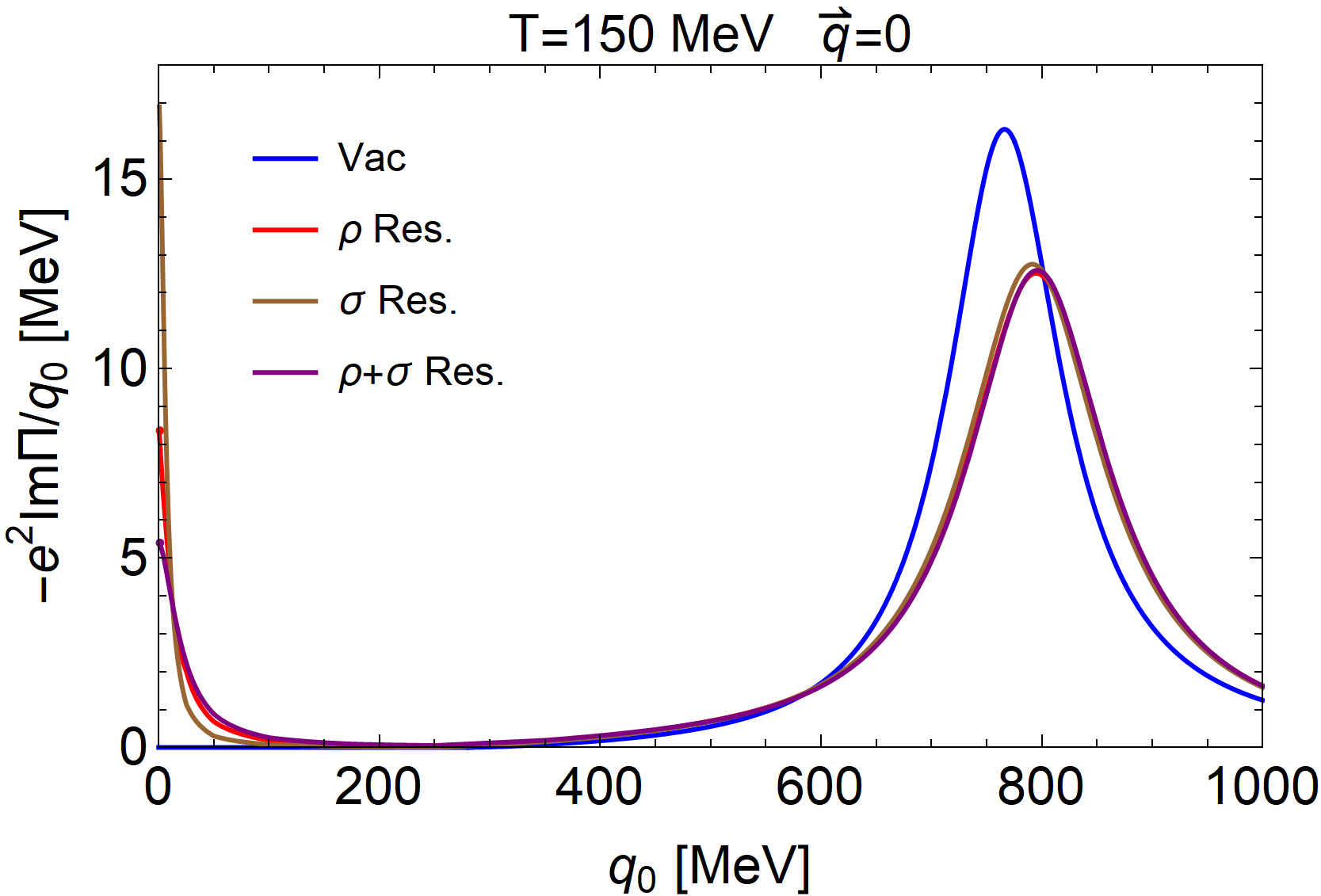}

\vspace{0.5cm}

\includegraphics[width=19pc]{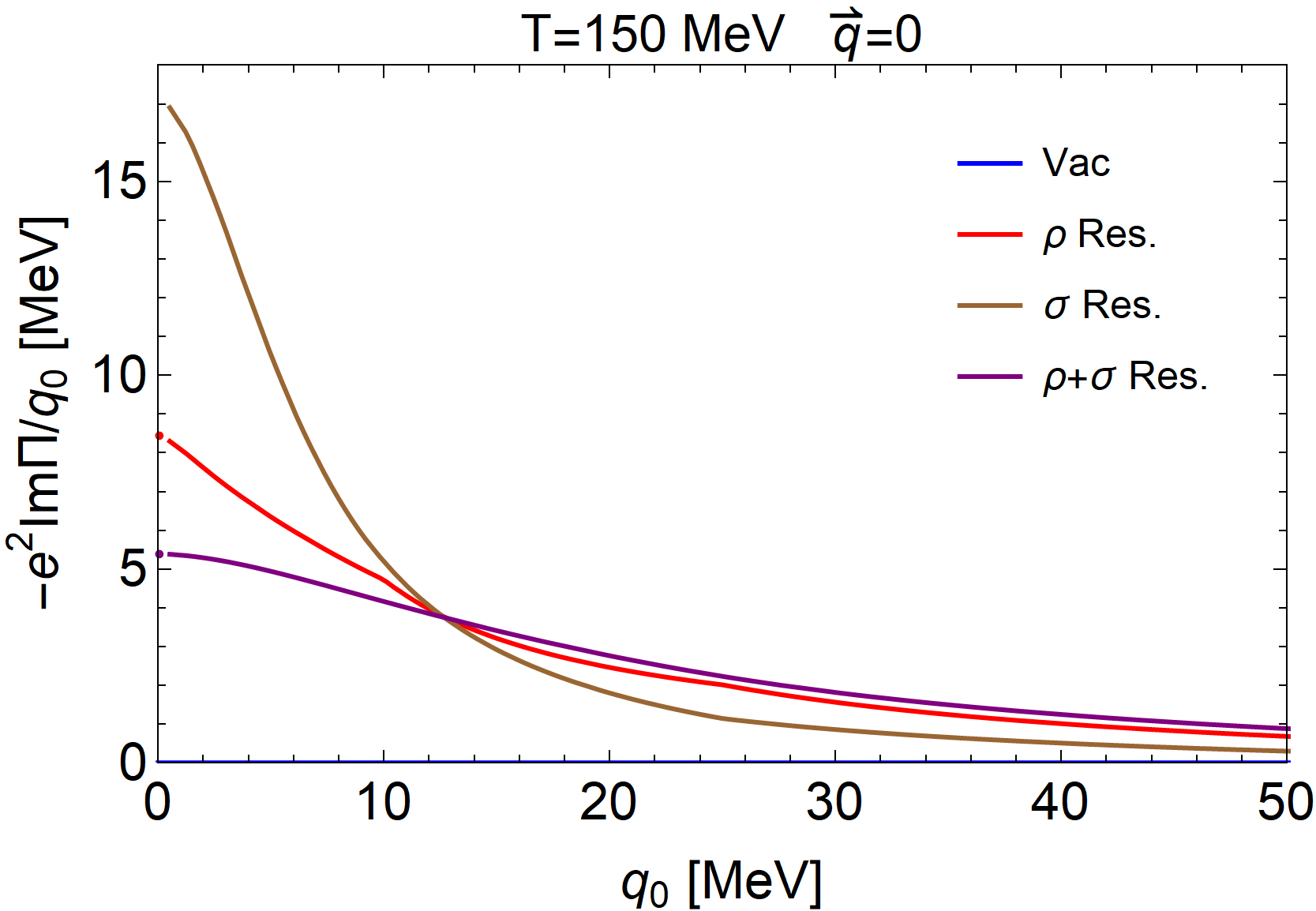}
\caption{Electromagnetic spectral function, scaled such that the zero-energy intercept corresponds to $\sig$, plotted as a function of energy at $\vec{q}=0$
(upper and lower panels differ by the selected energy range). Results are shown excluding vertex corrections for only $P$-wave scattering (red lines), 
only $S$-wave scattering (brown lines), and allowing $P$- and $S$-wave scattering (purple lines), and compared to the vacuum curve (blue line).}
\label{fig:piem-contr}
\end{figure}
We start by analyzing the different contributions from thermal $\pi\pi$ scattering, first focusing on the case without vertex corrections at a fixed temperature
of $T$=150\,MeV, cf.~Fig.~\ref{fig:piem-contr}. Maybe somewhat surprisingly, at high energies,  where the unitarity cut dominates, the rescattering 
contributions encoded in the pion self-energy have rather little impact on the medium effects on the $\rho$-meson resonance peak; its thermal broadening 
is almost entirely driven by the Bose enhancement of the intermediate pions in the pion gas; it also features a thermal mass shift that is probably a bit too 
large since our $\pi$-$\rho$  Lagrangian is not chirally symmetric (and therefore missing, \eg, an attractive contribution from $\rho+\pi\to a_1$ scattering; 
the simultaneous implementation of chiral symmetry and gauge invariance is beyond the scope of this work). In the low-energy region we now clearly 
see the development of the transport peak, essentially generated by the Landau cut. The pion width generated by scattering off thermal pions plays
a key role, with the  $\rho$ resonance providing the dominant contribution: when only including the $\rho$ contribution, the transport peak is much 
broader and the conductivity is about a factor 2 smaller compared to the case with only the $\sigma$ contribution.   

\begin{figure}
\vspace{0.3cm}

\includegraphics[width=20pc]{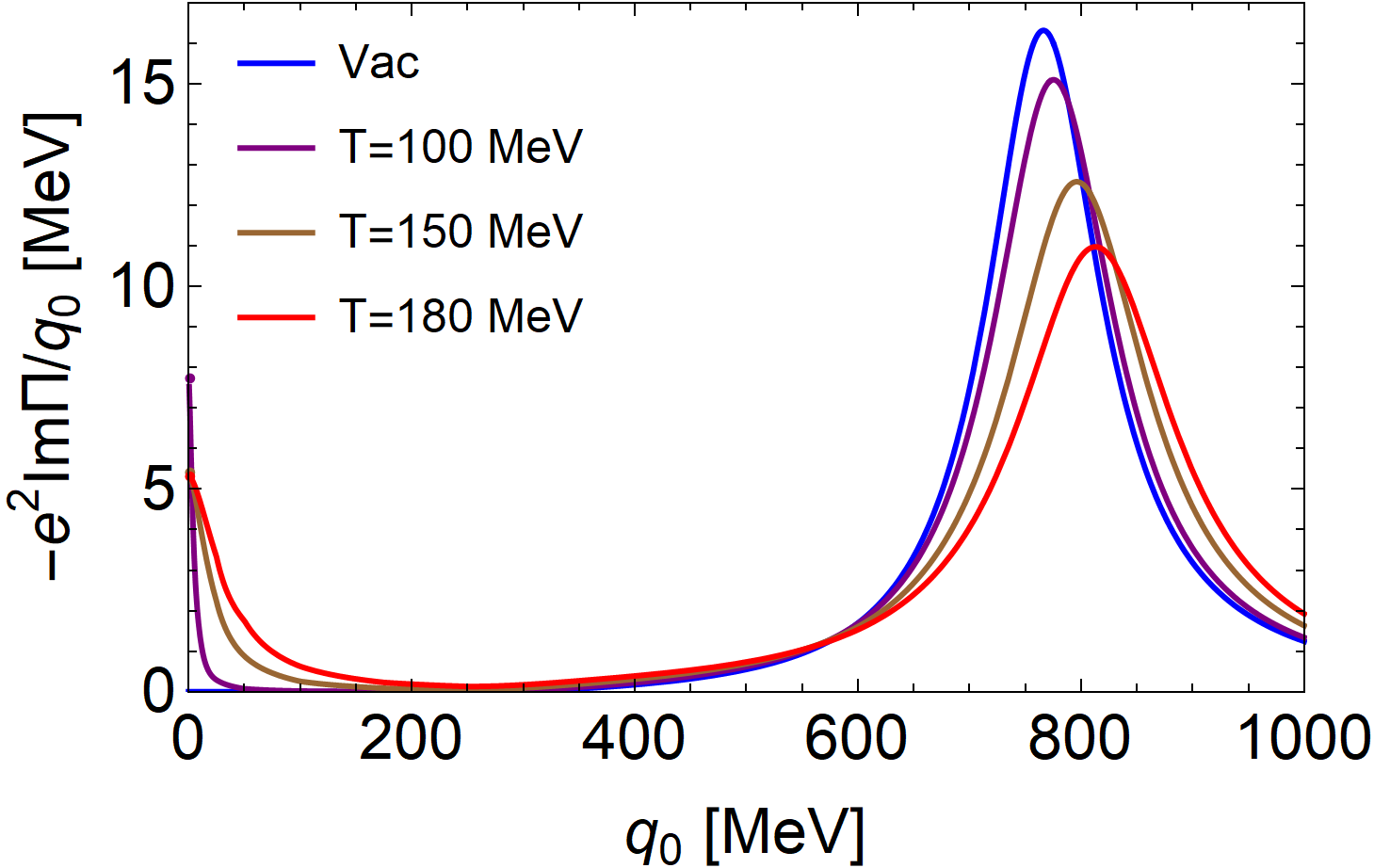}

\vspace{0.3cm}

\includegraphics[width=20pc]{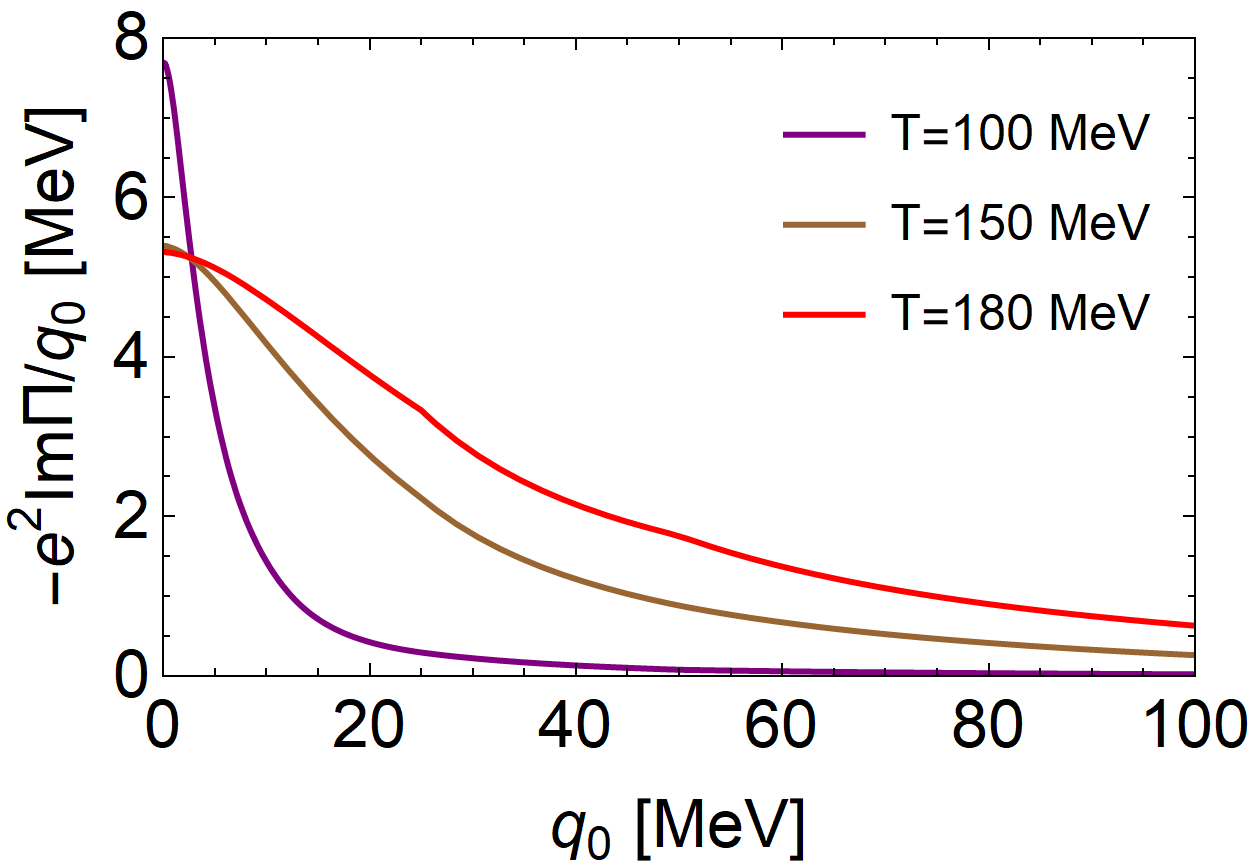}
\caption{Electromagnetic spectral function, scaled such that the zero-energy intercept corresponds to $\sig$, plotted as a function of energy at $\vec{q}=0$
(upper and lower panels differ by the selected energy range). Results excluding vertex corrections are plotted for vacuum (blue line), at $T$=100 MeV (purple lines), $T$=150 MeV (brown lines), and $T$=180 MeV (red lines). }
\label{fig:piem-novcorr}
\end{figure}

Turning to the temperature dependence, but still without vertex corrections, the $\rho$ resonance peak shows the expected broadening
increasing with temperature, cf.~Fig.~\ref{fig:piem-novcorr}. 
At  low energies, the conductivity decreases significantly  from $T$=100 MeV to 150\,MeV, but is almost unchanged  
between 150 and 180\,MeV. This shows that a minimum can develop in the conductivity, caused by a misalignment of thermal pion energies with that of 
the resonant particles towards higher temperatures; in other words, if the thermal energies of the charge carriers are beyond those needed for 
resonance excitations, charge can be transported more freely again.

\begin{figure}
\includegraphics[width=20pc]{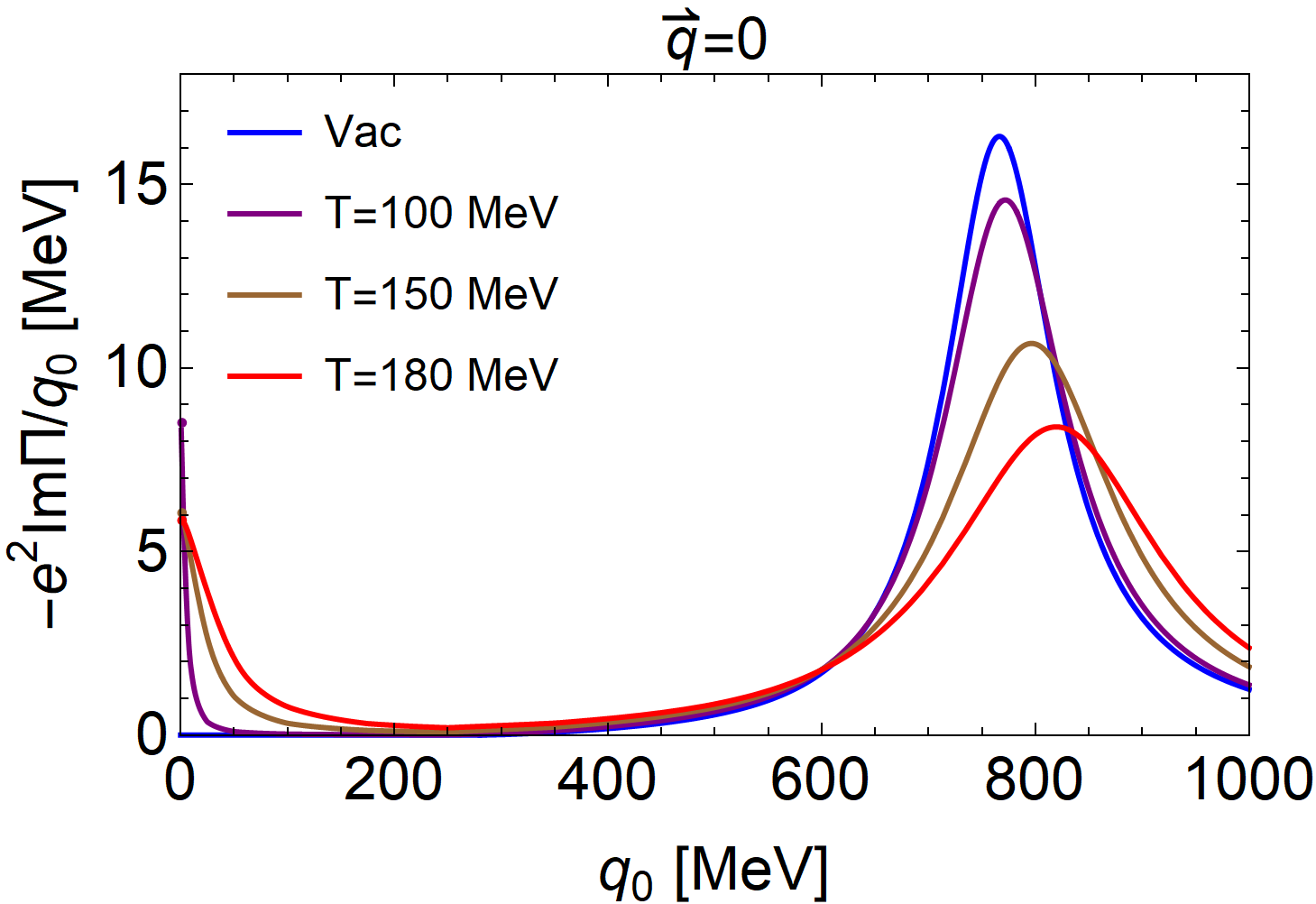}

\vspace{0.5cm}

\includegraphics[width=20pc]{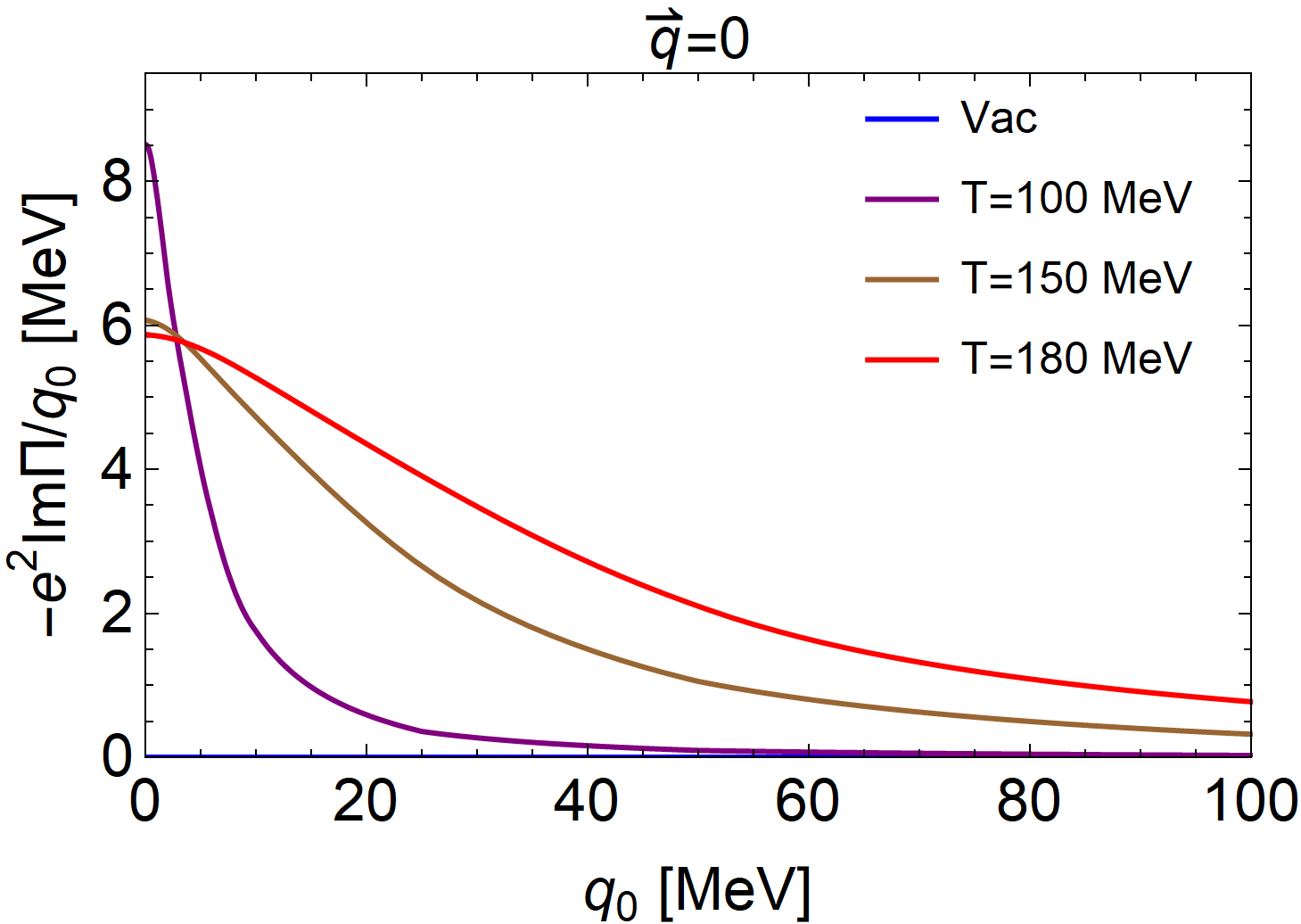}
\caption{Electromagnetic spectral function, scaled such that the zero-energy intercept corresponds to $\sig$, plotted as a function of energy at $\vec{q}=0$
(upper and lower panels differ by the selected energy range). We plot results including vertex corrections for $T$=100 MeV (purple lines), $T$=150 MeV 
(brown lines), and $T$=180 MeV (red lines). The vacuum line is also plotted in blue.}
\label{fig:piem}
\end{figure}

Finally, Fig.~\ref{fig:piem} displays the EM spectral function when vertex corrections are included, for various temperatures. The vertex corrections result in a broadening of the $\rho$ peak that increases with temperature. This is expected because the corrections increase the $\rho$'s interaction with the medium. Furthermore, we see an increase of the transport peak due to the vertex corrections. This is somewhat counter-intuitive, because the transport peak is proportional to the electric conductivity. Therefore, it appears that increasing the medium interaction by including vertex corrections has resulted in a more conductive medium. In order to understand this phenomenon, we note that the $\rho$ meson does not transmit electric charge through the medium. Charge is transmitted by the charged pion states, and any increase in $\textrm{Im}\Sigma_{\pi}$ will reduce the conductivity. On the other hand, the $\rho$ self-energy determines how the external photon couples to the medium. Therefore, the various $\rho$ self-energy diagrams represent different channels through which 
electric charge can travel, \ie, the vertex corrections introduce additional channels, increasing the electric conductivity.

%%%%%%%%%%%%%%%%%%%%%%%%%%%%
\subsection{Electric conductivity over temperature}
\label{ssec:cond}
\begin{figure}
\includegraphics[width=19pc]{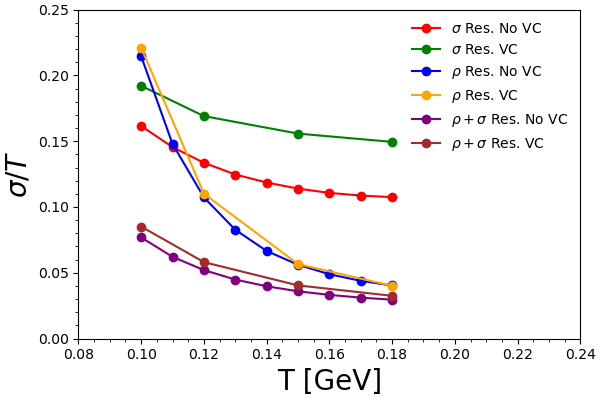}
\caption{Pion gas conductivity over temperature for different $\pi\pi$ scattering resonances, with and without vertex corrections. Results including only the 
$\sigma$ resonance are plotted in red (vertex corrections excluded) and green (vertex corrections included). Results including only the $\rho$ resonance 
are plotted in blue (vertex corrections excluded) and orange (vertex corrections included). Results including both resonances are plotted in purple (vertex corrections excluded) and brown (vertex corrections included).}
\label{fig:sigT}
\end{figure}

%%%%%%%%%%%%%%%%%%%%%%%%%%%%
Let us first summarize in Fig.~\ref{fig:sigT} our pion gas results for the conductivity, divided by temperature, for various scenarios as a function of temperature. 
When only including $P$-wave ($\rho$-resonance) scattering for the pion self-energies the temperature dependence is most pronounced, indicating that
with increasing temperature the typical thermal pion energies provide markedly increasing overlap for resonance formation, thus reducing the conductivity;
vertex corrections have little impact on this result. On the other, the temperature dependence is much less pronounced when only the much broader $\sigma$
resonance is included. The pertinent conductivity is actually smaller than for $\rho$ resonance reactions only, since the relative low thermal energies at
small temperatures are dominated by $S$-wave interactions; however, around  $T\gsim$\,110\,MeV, the $\rho$ resonance interactions take over and 
dominate at still higher $T$. Interestingly, the vertex corrections have a larger effect in the $\sigma$ channels, caused primarily by 
$\Gamma_{\mu \, ab3}^{(3)C\sigma}$ (expressions are given in Appendix~\ref{app:vcorr}).  When both resonances are included, the effect of 
this correction is greatly reduced, because the dominant $\rho$ self-energy diagram from  $\Gamma_{\mu \, ab3}^{(3)C\sigma}$ is inversely 
proportional to the pion width squared, and the $\rho$ resonance significantly increases the pion width. At the same time, other $\rho$ self-energy diagrams 
are only suppressed by the inverse pion width. In particular, the largest corrections to the $\rho$ self-energy, when both resonances are included, turn 
out to arise from $\Gamma_{\mu\nu \, ab33}^{(3)E\rho}$ and $\Gamma_{\mu\nu \, ab33}^{(3)E\sigma}$ (see Appendix~\ref{app:vcorr} for the explicit 
expressions), accounting for about 70\% of the vertex corrections' contribution to the $\rho$ self-energy.
Overall, with both resonances included, the vertex corrections increase the conductivity by approximately 10\% throughout.

\begin{figure}
\includegraphics[width=19pc]{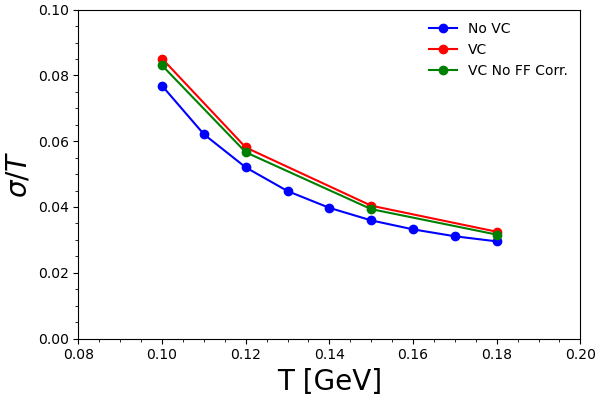}
\caption{Pion gas conductivity over temperature without vertex corrections (blue), with vertex corrections (red), and including all vertex corrections except those induced by the form factor (green). The results include $\pi\pi$ scattering through rho and sigma resonances.}
\label{fig:cond-FF}
\end{figure}

Next, we examine the impact of the additional vertex corrections induced by the form factors. The conductivity with and without these corrections is 
plotted in Fig~\ref{fig:cond-FF}, along with the conductivity without any vertex corrections. Form factors are still included in all results, in order to ensure 
convergence of the calculation. One can see that the form factor correction terms provide only a small increase in the total conductivity, of approximately 
2.5\%. 
%However, because the contribution of the vertex corrections to the conductivity is rather small to begin, the form factor correction amount to an increase the effect of the vertex corrections by about 30\%. This increase is fairly concerning, thus we intend to further investigate the violation of gauge invariance due to the form factor in future works. 

\begin{figure}
\includegraphics[width=20pc]{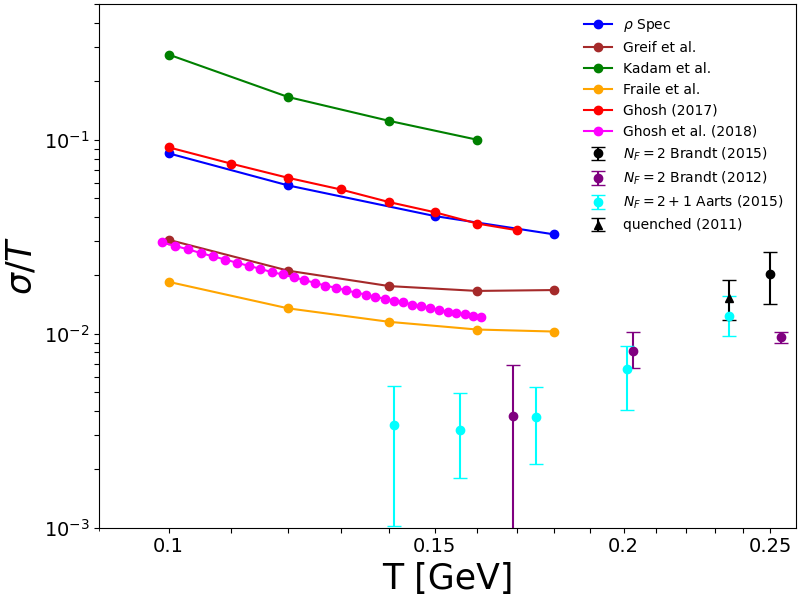}
\caption{Our results for the electric conductivity over temperature (blue line) compared to previous calculations in pion matter and lQCD results: 
brown line:  kinetic-theory using Breit-Wigner cross sections~\cite{Greif}, green line: $K$-matrix approach~\cite{Kadam}, orange line: chiral 
perturbation theory~\cite{Fraile}, red line: real-time thermal field theory~\cite{Ghosh}, magenta line: relaxation time approximation with Chapman-Enskog 
technique~\cite{Ghosh2}, black, purple and cyan dots: lattice QCD for $N_f$=2 light flavors~\cite{Brandt2013,Brandt2016} and for $N_f$=2+1~\cite{Aarts}, respectively. The black dashed line indicates a proposed lower bound from Ref.~\cite{Huot} using a holographic calculation for a supersymmetric 
Yang-Millls plasma.  
}
%{\bf it would be good to enlarge the figure by moving the legend from the side to the bottom, or drop it altogether.}
\label{fig:cond-lit}
\end{figure}

Finally, we compare our results for the hot pion matter conductivity to results from the literature in Fig.~\ref{fig:cond-lit}. Our conductivity is significantly 
larger than the results from kinetic theory using Breit-Wigner cross sections~\cite{Greif}, chiral perturbation theory~\cite{Fraile}, and a relaxation time approximation~\cite{Ghosh2}, but it is smaller than the $K$-matrix results of Ref.~\cite{Kadam}. However, our calculation agrees well with the real-time 
field theory results of Ref.~\cite{Ghosh}. In Refs.~\cite{Greif,Fraile,Kadam,Ghosh}  expressions for the conductivity are provided in terms of either the 
pion width or the relaxation time (sometimes equated to the collisions time), which are similar to our Eq.~\ref{ApproxCond11}; however, the inputs for the 
pion width vary considerably. For example, in Ref.~\cite{Ghosh2}, the momentum averaged charged-pion relaxation time at $T$=150\,MeV (using 
vacuum $\rho$ and $\sigma$ channel cross sections) amounts to ca.~2 fm/$c$ (3\,fm/$c$ for neutral pions), which translates into a reaction rate
of $\sim$100\,MeV,  substantially larger than our optical potential of $\Gamma=-2 {\rm Im}U_\pi\simeq 20-30$\,MeV.  
Figure~\ref{fig:cond-lit} also indicates that the pion gas results are significantly larger than lQCD calculations, with most lQCD results falling below a 
proposed lower bound from a calculation for a strongly coupled supersymmetric Yang-Mills plasma using AdS/CFT duality~\cite{Huot} (which, 
however, depends on the number of degrees of freedom in the calculation and therefore may not be 
appropriate to be compared to pion matter; we will return to this issue Sec.~\ref{ssec:sus} below). Furthermore, in Ref.~\cite{Aarts} it is cautioned 
that the extraction of the conductivity at low temperature from lQCD computations of Euclidean vector-current correlators faces difficulties in extracting 
narrow transport peaks created by hadronic interactions. 

Our results support a pion matter conductivity significantly higher than the lower bound proposed in Ref.~\cite{Huot}. Furthermore, our calculations indicate 
that the effects of the vertex corrections are rather small (at the $\sim10\%$ level), whereas the conductivity is dominated by the Landau cut of the 
$\rho$ self-energy, which is related to the pion's collisional width. As demonstrated in Eq.~\ref{ApproxCond11}, consistent with kinetic theory, 
the conductivity is essentially inversely proportional to the pion's width. Therefore, the conductivity is sensitive to pionic interactions and a robust 
calculation of the pion's width is required in order to reliably extract the conductivity. 

%%%%%%%%%%%%%%%%%%%%%%%%%%%%%%%%%%%%%%%
\section{Applciations of the electromagnetic spectral function}
\label{sec:appl}
%%%%%%%%%%%%%%%%%%%%%%%%%%%%%%%%%%%%%%%
In this section we further analyze the EM spectral function by utilizing it to calculate the charge susceptibility which facilitates a more quantitative context for comparisons to conjectured quantum lower bound, and testing our calculation's  with a current conservation sum rule.

%%%%%%%%%%%%%%%%%%%%%%%%%%%%
\subsection{Charge susceptibility}
\label{ssec:sus}
%%%%%%%%%%%%%%%%%%%%%%%%%%%%
When comparing $\sig/T$ to proposed quantum lower bounds (and lQCD calculations), some care is in order as this quantity, albeit dimensionless, 
depends on the number of degrees of freedom in the theory, which is quite different for a pion gas. To mitigate this difference, it has been 
proposed~\cite{Huot} to rather divide the conductivity by the charge susceptibility. To leading order in $e^{2}$ the susceptibility is related to the 
EM Debye mass squared~\cite{McLerran,Kapusta,Prakash,Blaizot}. Within the VDM one can express the Debye mass in terms of 
of the temporal component of the $\rho$ propagator~\cite{Prakash:2001xm},
\begin{eqnarray}\label{ChargeS3}
\Xi&=&m_{D}^{2}=-\lim_{\vec{q} \to 0}\Big[(m_{\rho}^{0})^{4}/g_{\rho}^{2}D_{\rho}^{00}(q_{0}=0,\vec{q})\nonumber\\
&&-(m_{\rho}^{0})^{4}/g_{\rho}^{2}D_{\rho \, vac}^{00}(0,0)\Big] \ .
\end{eqnarray}
We recall that the conductivity corresponds to the time-like limit of $\textrm{Im}\Pi_{\textrm{EM}}^{\mu\nu}$, while the susceptibility is determined 
by the space-like limit of $\textrm{Re}\Pi_{\textrm{EM}}^{\mu\nu}$. Though our formalism can be extended to finite $\vec{q}$, we have not calculated 
vertex corrections at finite $\vec{q}$ yet, as this presents significant numerical challenges while only producing a $\sim$10\% effect. Therefore, we here 
extract $\Xi$ using from $\rho$ self-energy in $\Pi_{\textrm{EM}}^{00}$ with dressed pion propagators neglecting vertex corrections. 

\begin{figure}
\includegraphics[width=20pc]{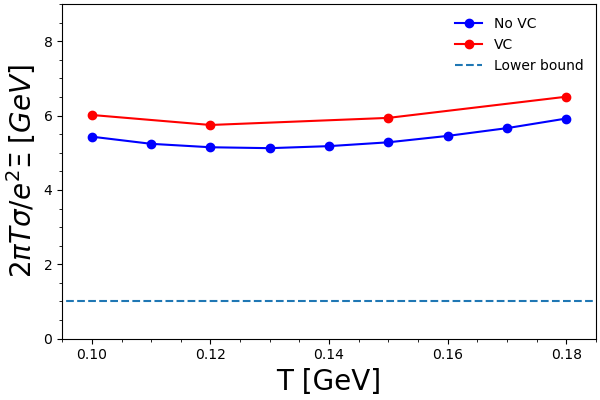}
\caption{Electric conductivity divided by $e^{2}$ times the charge susceptibility, normalized such that the lower bound is one. The susceptibility is calculated without  vertex corrections, but results are shown for the conductivity with (red line) and without (blue line) vertex corrections and compared to the lower bound calculated in Ref.~\cite{Huot} 
(black dashed line). }
%{\bf com: better to put the legend in the plot and enlarge the figure.}
\label{fig:cond-sus}
\end{figure}

%{\bf com RR: why not further use the results of the Huot paper and include a $2\pi T$ factor in the above ratio so that the lower bound simply becomes 1?}\\
%{\bf com JA: I will do this when I create the new plots.}\\
In Fig.~\ref{fig:cond-sus} we plot our result for $(2\pi T\sig)/(e^{2}\Xi)$ and compare it to the proposed lower bound of 1 from 
Ref.~\cite{Huot}. We see that our pion-gas values are approximately a factor of five larger than this lower bound. Furthermore, our results display a minimum between 120 and 140 MeV. Therefore, we see that even when the number of degrees of freedom is accounted for, our result is well above the proposed lower bound.

%%%%%%%%%%%%%%%%%%%%%%%%%%%%%%%%%%
\subsection{\label{sec6:level2:2}Current conservation sum rule}
\label{ssec:sunrule}
%%%%%%%%%%%%%%%%%%%%%%%%%%%%%%%%%%%
As a further test of the internal consistency of our calculation, we insert our EM spectral function into a sum rule derived based on charge 
conservation in Ref.~\cite{Gubler}, 
\begin{eqnarray}\label{eq142}
\int_{0}^{\infty}dw\frac{1}{w}\textrm{Im}\Pi(w,T)&=&\int_{0}^{\infty}dw\frac{1}{w}\textrm{Im}\Pi_{vac}(w) \ .
\end{eqnarray}
In order to quantify the violation of the sum rule, we calculate the percent change of the left-hand side of Eq.~(\ref{eq142}) relative to the vacuum value. 
The violations with and without vertex correction are collected in Tab.~\ref{SumRuleT}. Before including vertex corrections the violation is on 
average less than 0.5\%. When vertex corrections are included the violation increases for $T$=100 MeV and 120 MeV, but is reduced for $T$=150 MeV 
and 180 MeV. Furthermore, adding the corrections induced by the form factors tends to increase the violation. While this increase provides 
further motivation to investigate a form factor that improves on gauge invariance, it is reassuring that the violation is consistently less than 1\%. 
In order to gain precision at this level we would likely also have to refine the numerical computations and calculate further iterations of the 
self-consistency equation induced by the pion self-energy.
\begin{table}[!tb]%The best place to locate the table environment is directly after its first reference in text
\caption{\label{SumRuleT}%
\label{SumRuleT}Violation of the sum rule proposed in ref. \cite{Gubler} for various temperatures.}
\begin{ruledtabular}
\begin{tabular}{lcccc}
\textrm{Calc}&
\textrm{100 MeV}&
\textrm{120 MeV}&
\textrm{150 MeV}&
\textrm{180 MeV}\\
\colrule
No VC & .32\% & .34\% & .36\% & .57\%\\ 
VC & .92\% & .67\% & .33\% & .24\%\\
VC No FF Corr. & .38\% & .39\% & .32\% & .35\%\\
\end{tabular}
\end{ruledtabular}
\end{table}

%%%%%%%%%%%%%%%%%%%%%%%%%%
\section{Summary and future work}
\label{sec:sum}
%%%%%%%%%%%%%%%%%%%%%%%%%%%
We have calculated the electric conductivity of hot pion matter employing a quantum-field theoretic approach that is rooted in successful descriptions 
of thermal-dilepton spectra in heavy-ion collisions~\cite{RappDiL,Rapp:1999us}. Based on the VDM, the EM spectral function is related to the imaginary 
part of the light vector-meson propagators (most notably the $\rho$), and interactions with the thermal medium have been evaluated through the $\rho$ 
self-energy. In pion matter the latter exhibits a transport peak through its Landau cut which corresponds to the scattering of a low-energy $\rho$ (or photon) 
with thermal pions from the heat bath. The width of the transport peak (and with it a finite conductivity) is generated through thermal $\pi\pi$ scattering  
producing finite collisional widths for both in- and outgoing pions. We have included $S$- and $P$-wave scatterings through $\sigma$ and $\rho$ resonances, respectively, to obtain in-medium pion self-energies and resummed those through the pion propagators within the $\rho$ self-energy. We then remedied 
violations of gauge invariance induced by this dressing with suitably constructed corrections to the $\rho\pi\pi$ and $\rho\rho\pi\pi$ vertices guided by
Ward-Takahashi identities. The effect of the vertex corrections on the conductivity turns out to be rather moderate, resulting in an approximately 10\% 
increase due to additional charge-conduction channels provided by the modified vertices. Both $S$- and $P$-wave $\pi\pi$ scattering contribute 
to the conductivity, with the latter (former) being the leading contribution at temperatures above (below) $\sim$100\,MeV. 
%Although, the vertex corrections do not tend to improve the current conservation sum rule, the violations with or without vertex corrections are less than 1\%, and correcting these violations will likely require the calculation of higher order vertex corrections, the dressing of thermal pions within $\Sigma_{\pi}$, and higher numerical precision. The choice of the center of mass momentum in the form factor may also contribute to the violation. 

Our results suggest a conductivity significantly larger than found in previous calculations using kinetic theory, chiral perturbation theory, a Chapman-Enskog 
type relaxation time approximation, or a $K$-matrix formalism~\cite{Greif,Fraile,Ghosh2,Kadam}. Our values agree with the real-time field theory 
calculation in Ref.~\cite{Ghosh}. However, the variation in the conductivity does not appear to be due to the choice of formalism, as the different formalisms
produce similar expressions for the conductivity in terms of the pion width. Rather, the most important differences appear to originate from the model 
for the in-medium self-energy. Furthermore, our conductivity is well above the quantum lower bound proposed in Ref.~\cite{Huot}. This can be better 
quantified when normalizing the $\sig$ values by the pertinent charge susceptibilities, as to remove the dependency on the number of degrees of freedom
in the different systems. In that normalization our result is about a factor of $\sim$5 above the lower bound.

We intend to extend our formalism to study the strong-coupling limit by increasing $g_{\rho}$, as our calculations, rooted in quantum mechanics, may be 
expected to respect a pertinent lower bound. In this limit, the higher-order corrections may not be as heavily suppressed, therefore they will require 
further scrutiny. Another extension concerns the inclusion of additional hadronic species in the confined medium. This will require a more complete 
calculation of the in medium $\rho$ self-energy, for which existing results can be utilized~\cite{Rapp:1999us}. In particular, we will include baryonic 
effects on the self-energy, which will be crucial when comparing our calculation to results from HICs. The effect of nucleons at finite density and 
temperature can be included by combining this work's pion self-energy with that of Ref.~\cite{Urban:1999im} while accounting for additional vertex 
corrections to the $\rho$ self-energy. Furthermore, we will include direct couplings of the $\rho$ to mesons and baryons in the surrounding hot 
hadronic medium which are usually dominated by resonance excitations. 
Both the probing of the strong-coupling limit and the more complete hadronic matter calculation will likely bring the conductivity closer to the lower 
bound. Furthermore, in a more strongly coupled medium an increase in the conductivity due to the vertex corrections may also occur and provide 
a significant contribution. This may be one of the mechanisms that may contribute to a saturation value within a quantum-mechanical 
framework with conserving approximations, and thus enable to probe quantum lower bounds in a controlled way. 
Finally, experimental efforts to measure very-low mass dileptons are underway by HADES at GSI, STAR at BNL, and ALICE-3 for LHC run-5 down to a 
few tens of MeV~\cite{UpStar,HADES,ALICE:2022wwr}. Our calculations suggest that these measurements should be able to access the mass region where
the transport peak is prevalent, and thus may provide a measurement of the conductivity. However, more quantitative predictions of thermal-emission
spectra from heavy-ion collisions in the very-low-mass region require a more complete hadronic matter calculation, as indicated above.

%%%%%%%%%%%
\acknowledgments
This work has been supported by the U.S.~National Science foundation under grants PHY-1913286 and PHY-2209335.
%%%%%%%%%%%%%%%

%%%%%%%%%%%%%%%%%%%%%%%%%%%
\appendix
%%%%%%%%%%%%%%%%%%%
\section{Vertex correction formulation}
\label{app:vcorr}
%%%%%%%%%%%%%%%%%%%%
In this appendix we elaborate on our calculation of the vertex corrections. We give the explicit expressions for the vertex corrections in Sec.~\ref{app:int}, implement form factors in Sec.~\ref{app:FF}, and discuss the elimination of double-counting issues from the $\rho$ self-energy in Sec.~\ref{app:double}.

\begin{widetext}
%%%%%%%%%%%%%%%%%%%%%
\subsection{Vertex correction integrals}
\label{app:int}
%%%%%%%%%%%%%%%%%%%%%
The corrections to the $\rho\pi\pi$ vertex due to the $\rho$ resonance are given by
\begin{eqnarray}
\label{eq51FF}
\Gamma_{\mu \, ab3}^{(3)A\rho}(k,q)&=&\epsilon_{3ab}\frac{3g_{\rho}^{3}}{2}T\sum_{n(even)}\int \frac{d^{3}p}{(2\pi)^{3}} \bigg[ D_{\pi}(p)D_{\rho}^{\nu\lambda}(k+p) g_{\mu\lambda}(k-p)_{\nu}\textrm{FF}_{\rho}[p,k]\textrm{FF}_{\rho}[q+p,-q+k]\bigg]_{p_{0}=iw_{n}},\\
\label{eq52FF}
\Gamma_{\mu \, ab3}^{(3)B\rho}(k,q)&=&\epsilon_{3ab}\frac{3g_{\rho}^{3}}{2}T\sum_{n(even)}\int \frac{d^{3}p}{(2\pi)^{3}}\bigg[ D_{\pi}(p)D_{\rho}^{\nu\lambda}(q+k+p)g_{\mu\lambda}(q+k-p)_{\nu}\nonumber\\
&&\times\textrm{FF}_{\rho}[p,q+k]\textrm{FF}_{\rho}[q+p,k]\bigg]_{p_{0}=iw_{n}},\\
\label{eq53FF}
\Gamma_{\mu \, ab3}^{(3)C\rho}(k,q)&=&\epsilon_{3ab}\frac{g_{\rho}^{3}}{2}T\sum_{n(even)}\int \frac{d^{3}p}{(2\pi)^{3}}\bigg[ D_{\pi}(p)D_{\pi}(q+p)D_{\rho}^{\nu\lambda}(q+k+p)(2p+q)_{\mu}(-p+k+q)_{\lambda}(k-p-q)_{\nu}\nonumber\\
&&\times\textrm{FF}_{\rho}[p,q+k]\textrm{FF}_{\rho}[q+p,k]\bigg]_{p_{0}=iw_{n}} \ .
\nonumber\\
\end{eqnarray}

The corrections to the $\rho\pi\pi$ vertex due to the $\sigma$ resonance are given by
\begin{eqnarray}
\label{eq55FF}
\Gamma_{\mu \, ab3}^{(3)C\sigma}(k,q)&=&\epsilon_{3ab}\frac{g_{\rho} g_{\sigma}^{2}}{2}T\sum_{n(even)}\int \frac{d^{3}p}{(2\pi)^{3}}\bigg[ D_{\pi}(p)D_{\pi}(q+p)D_{\sigma}(q+k+p)\times((q+k+p)^{2}-m_{\pi}^{2})(2p+q)_{\mu}\nonumber\\
&&\textrm{FF}_{\rho}[p,q+k]\textrm{FF}_{\rho}[q+p,k]\bigg]_{p_{0}=iw_{n}} \ . 
\nonumber\\
\end{eqnarray}

The corrections to the $\rho\rho\pi\pi$ vertex due to the $\rho$ resonance are given by
\begin{eqnarray}
\label{eq57FF}
\Gamma_{\mu\nu \, ab33}^{(4)A\rho}(k,q)&=&(5\delta_{ab}-3\delta_{3a}\delta_{3b})\frac{ig_{\rho}^{4}}{2}T\sum_{n(even)}\bigg[\int \frac{d^{3}p}{(2\pi)^{3}} D_{\pi}(p)D_{\rho}^{\mu\nu}(q+k+p)\nonumber\\
&&\times\textrm{FF}_{\rho}[q+p,k]^{2}\bigg]_{p_{0}=iw_{n}}+(k\rightarrow-k),\\
\label{eq58FF}
\Gamma_{\mu\nu \, ab33}^{(4)B_{1}\rho}(k,q)&=&-2\delta_{ab}\frac{ig_{\rho}^{4}}{2}T\sum_{n(even)}\bigg[\int \frac{d^{3}p}{(2\pi)^{3}}D_{\pi}(p)D_{\pi}(p+q)D_{\rho\nu\beta}(k+p)\nonumber\\
&&\times(2p+q)_{\mu}(p-k)^{\beta}\textrm{FF}_{\rho}[p,k]^{2}\bigg]_{p_{0}=iw_{n}}+(k\rightarrow-k)),\\
\label{eq58FFa}
\Gamma_{\mu\nu \, ab33}^{(4)B_{2}\rho}(k,q)&=&\Gamma_{\nu\mu ab}^{(4)B_{1}\rho}(k,q),\\
\label{eq61FF}
\Gamma_{\mu\nu \, ab33}^{(4)E\rho}(k,q)&=&(\delta_{ab}+\delta_{3a}\delta_{3b})\frac{ig_{\rho}^{4}}{2}T\sum_{n(even)}\bigg[\int \frac{d^{3}p}{(2\pi)^{3}} D_{\pi}(p)^{2}D_{\pi}(p+q)D_{\rho}(k+p)_{\alpha\beta}\nonumber\\
&&\times(2p+q)_{\mu}(p-k)^{\alpha}(p-k)^{\beta}(2p+q)_{\nu}\textrm{FF}_{\rho}[p,k]^{2}\bigg]_{p_{0}=iw_{n}}+(k\rightarrow-k),\\
\label{eq63FF}
\Gamma_{\mu\nu \, ab33}^{(4)G\rho}(k,q)&=&-2(\delta_{3a}\delta_{3b}+\delta_{ab})\frac{ig_{\rho}^{4}}{2}T\sum_{n(even)}\bigg[\int \frac{d^{3}p}{(2\pi)^{3}} D_{\pi}(p)^{2}D_{\rho}^{\alpha\beta}(k+p)(k-p)_{\alpha}(k-p)_{\beta}g_{\mu\nu}\nonumber\\
&&\times\textrm{FF}_{\rho}[p,k]^{2}\bigg]_{p_{0}=iw_{n}} \ . 
\end{eqnarray}

Finally the corrections to the $\rho\rho\pi\pi$ vertex due to the $\sigma$ resonance are given by
\begin{eqnarray}
\label{eq65FF}
\Gamma_{\mu\nu \, ab33}^{(4)E\sigma}(k,q)&=&(\delta_{ab}-\delta_{3a}\delta_{3b})\frac{ig_{\rho}^{2}g_{\sigma}^{2}}{2}T\sum_{n(even)}\bigg[\int \frac{d^{3}p}{(2\pi)^{3}} D_{\pi}(p)^{2}D_{\pi}(p+q)D_{\sigma}(k+p)((k+p)^{2}-m_{\pi}^{2})\nonumber\\
&&\times(2p+q)_{\mu}(2p+q)_{\nu}\textrm{FF}_{\sigma}[p,k]^{2}\bigg]_{p_{0}=iw_{n}}+(k\rightarrow-k),\\
\label{eq66FF}
\Gamma_{\mu\nu \, ab33}^{(4)G\sigma}(k,q)&=&2(\delta_{3a}\delta_{3b}-\delta_{ab})\frac{ig_{\rho}^{2}g_{\sigma}^{2}}{2}T\sum_{n(even)}\bigg[\int \frac{d^{3}p}{(2\pi)^{3}}D_{\pi}(p)^{2}D_{\sigma}(k+p)((k+p)^{2}-m_{\pi}^{2})g_{\mu\nu}\textrm{FF}_{\sigma}[p,k]^{2}\bigg]_{p_{0}=iw_{n}} \ ,\nonumber\\
\end{eqnarray}
where we have excluded all corrections including $\rho\rho\rho$ or $\rho\rho\rho\rho$ vertices.

%%%%%%%%%%%%%%%%%%%%%%%%%%%%%
\subsection{Vertex correction form factor correction}
\label{app:FF}
%%%%%%%%%%%%%%%%%%%%%%%%%%%%%
Here we give the expressions for the total regularized vertex corrections, including terms where the $\rho$ meson couples to a heavy-pion propagator.
\begin{eqnarray}
\label{UrbanFFVC3rsX}
\Gamma'^{(3)}_{\mu \, ab3}(k,q)&=&\Gamma^{(3)\rho}_{\mu \, ab3}(k,q)+\Gamma^{(3)\sigma}_{\mu \, ab3}(k,q)+g_{\rho}\epsilon_{3ab}(2k+q)_{i}\Big(\frac{\Sigma_{\pi}^{(\rho)}(q+k)}{A_{\rho}^{2}+\vec{k}^{2}}+\frac{\Sigma_{\pi}^{(\rho)}(k)}{A_{\rho}^{2}+(\vec{q}+\vec{k})^{2}}\Big)\nonumber\\
&&+g_{\rho}\epsilon_{3ab}(2k+q)_{i}\Big(\frac{\Sigma_{\pi}^{(\sigma)}(q+k)}{A_{\sigma}^{2}+\vec{k}^{2}}+\frac{\Sigma_{\pi}^{(\sigma)}(k)}{A_{\sigma}^{2}+(\vec{q}+\vec{k})^{2}}\Big),
\end{eqnarray}
\begin{eqnarray}
\label{UrbanFFVC4rsX}
\Gamma'^{(4)}_{\mu\nu \, ab33}(k,q)&=&\Gamma^{(4)\rho}_{\mu\nu \, ab33}(k,q)+\Gamma^{(4)\sigma}_{\mu\nu \, ab33}(k,q)-\frac{ig_{\rho}\epsilon_{3ca}}{A_{\rho}^{2}+\vec{k}^{2}}\Big[(2k-q)_{i}\Gamma^{(3)\rho}_{\nu \, bc3}(k,-q)+(2k+q)_{j}\Gamma^{(3)\rho}_{\mu \, cb3}(-q-k,q)\Big]\nonumber\\
&&-\frac{ig_{\rho}\epsilon_{3ca}}{A_{\sigma}^{2}+\vec{k}^{2}}\Big[(2k-q)_{i}\Gamma^{(3)\sigma}_{\nu \, bc3}(k,-q)+(2k+q)_{j}\Gamma^{(3)\sigma}_{\mu \, cb3}(-q-k,q)\Big]\nonumber\\
&&-\frac{ig_{\rho}\epsilon_{3bc}}{A_{\rho}^{2}+\vec{k}^{2}}\Big[(2k+q)_{i}\Gamma^{(3)\rho}_{\nu \, ac3}(-k,-q)+(2k-q)_{j}\Gamma^{(3)\rho}_{\mu \, ca3}(-q+k,q)\Big]\nonumber\\
&&-\frac{ig_{\rho}\epsilon_{3bc}}{A_{\sigma}^{2}+\vec{k}^{2}}\Big[(2k+q)_{i}\Gamma^{(3)\sigma}_{\nu \, ac3}(-k,-q)+(2k-q)_{j}\Gamma^{(3)\sigma}_{\mu \, ca3}(-q+k,q)\Big]\nonumber\\
&&-ig_{\rho}^{2}(\delta_{ab}-\delta_{3a}\delta_{3b})\Bigg\{(2k-q)_{i}(2k-q)_{j}\Bigg[\frac{\Sigma_{\pi}^{(\rho)}(-q+k)}{(A_{\rho}^{2}+\vec{k}^{2})^{2}}+\frac{2\Sigma_{\pi}^{(\rho)}(k)}{(A_{\rho}^{2}+(-\vec{q}+\vec{k})^{2})(A_{\rho}^{2}+\vec{k}^{2})}\nonumber\\
&&+\frac{\Sigma_{\pi}^{(\sigma)}(-q+k)}{(A_{\sigma}^{2}+\vec{k}^{2})^{2}}+\frac{2\Sigma_{\pi}^{(\sigma)}(k)}{(A_{\sigma}^{2}+(-\vec{q}+\vec{k})^{2})(A_{\sigma}^{2}+\vec{k}^{2})}\Bigg]+(2k+q)_{i}(2k+q)_{j}\Bigg[\frac{\Sigma_{\pi}^{(\rho)}(q+k)}{(A_{\rho}^{2}+\vec{k}^{2})^{2}}\nonumber\\
&&+\frac{2\Sigma_{\pi}^{(\rho)}(k)}{(A_{\rho}^{2}+(\vec{q}+\vec{k})^{2})(A_{\rho}^{2}+\vec{k}^{2})}+\frac{\Sigma_{\pi}^{(\sigma)}(q+k)}{(A_{\sigma}^{2}+\vec{k}^{2})^{2}}+\frac{2\Sigma_{\pi}^{(\sigma)}(k)}{(A_{\sigma}^{2}+(\vec{q}+\vec{k})^{2})(A_{\sigma}^{2}+\vec{k}^{2})}\Bigg]\nonumber\\
&&-4\delta_{ij}\frac{\Sigma_{\pi}^{(\rho)}(k)}{A_{\rho}^{2}+\vec{k}^{2}}-4\delta_{ij}\frac{\Sigma_{\pi}^{(\sigma)}(k)}{A_{\sigma}^{2}+\vec{k}^{2}}\Bigg\} \ ,
\end{eqnarray}
where $\Gamma^{(3)(\rho/\sigma)}_{\mu \, ab3}$ and $\Gamma^{(4)(\rho/\sigma)}_{\mu\nu \, ab33}$ are the total regularized vertex corrections, $A_{R}=\sqrt{\Lambda_{R}^{2}/4+m_{\pi}^{2}}$, and the indices $i$ and $j$
designate the spatial components of $\mu$ and $\nu$. For $i=0$ or $j=0$ the corresponding term should be dropped. 
%I updated the constants in these equations to better reflect the updated form factor.
%%%%%%%%%%%%%%%%%%%%%%
\end{widetext}

\subsection{Removing double counting}
\label{app:double}
%%%%%%%%%%%%%%%%%%%
We must take care to avoid double-counting of self-energy diagrams when calculating $\Sigma_{\rho}^{\mu\nu}$. Double-counting occurs due to the 
self-consistent treatment of the $\rho\pi\pi$ vertex and the pion propagator. For the corrections to the $\rho\pi\pi$ vertex, double-counting is encountered 
due to the presence of the Landau cut within vertex corrections, \ie, thermal $\pi\pi$ scattering with the external $\rho$. Furthermore, double-counting is generated in all the corrections to the $\rho\rho\pi\pi$ vertex.

For the corrections to the $\rho\pi\pi$ vertex, double-counting occurs due to $\Gamma_{\mu \, ab3}^{(3)C\rho}$ and 
$\Gamma_{\mu \, ab3}^{(3)C\sigma}$. We see that these vertex corrections produce the same self-energy diagram when used to dress the right 
or left hand vertex of the $\pi\pi$-loop. Furthermore, consider the higher-order corrections obtained by dressing $\Gamma_{\mu \, ab3}^{(3)C\rho}$ or 
$\Gamma_{\mu \, ab3}^{(3)C\sigma}$ with three-point vertex corrections. Such higher-order vertex corrections do not produce unique contributions 
to the $\rho$ self-energy, because these self-energies are also produced by dressing one vertex of the $\pi\pi$-loop with $\Gamma_{\mu \, ab3}^{(3)C\rho}$ 
or $\Gamma_{\mu \, ab3}^{(3)C\sigma}$ and the other vertex with a three-point vertex correction. Double-counting is encountered because the 
thermal particles in the vertex corrections are identical to those in the $\rho$ self-energy. The vertex corrections are defined such that they do not include 
vacuum particles, therefore we only encounter double-counting when dressing the Landau cut of $\Sigma_{\rho}^{\mu\nu}$. In this work we avoid 
double-counting in the $\pi\pi$-loop by only dressing the right-hand $\rho\pi\pi$ vertex with the self-consistently calculated $\Gamma_{\mu \, ab3}^{(3)C\rho}$ or $\Gamma_{\mu \, ab3}^{(3)C\sigma}$ vertex correction.

Next, we address double-counting in the corrections to the $\rho\rho\pi\pi$ vertex. As we have seen for the unitarity cut of the $\pi\pi$-loop, double-counting 
is avoided when we dress the vacuum tadpole loop and its Bose enhancement. However, when the pion in the tadpole loop is thermal, all of the four-point corrections generate double-counting. The double counting is generated by one of two scenarios: First, the $\rho$ self-energy contribution derived from the 
correction is equivalent to a self-energy obtained from dressing the $\pi\pi$-loop. We see that $\Gamma_{\mu\nu \, ab33}^{(4)B_{1}\rho}$ and
 $\Gamma_{\mu\nu \, ab33}^{(4)B_{2}\rho}$ can be generated by dressing the $\pi\pi$-loop with $\Gamma_{\mu \, ab3}^{(3)A\rho}$. Alternatively, 
$\Gamma_{\mu\nu \, ab33}^{(4)E\rho}$, $\Gamma_{\mu\nu \, ab33}^{(4)G\rho}$, $\Gamma_{\mu\nu \, ab33}^{(4)E\sigma}$, and $
\Gamma_{\mu\nu \, ab33}^{(4)G\sigma}$ simply dress a pion propagator in $\Sigma_{\rho}^{\mu\nu}$ with a thermal $\pi\rho$ or $\pi\sigma$ loop.
However, these diagrams where already included by resumming the pion propagators with $\Sigma_{\pi}$.
The second scenario occurs in $\Gamma_{\mu\nu \, ab33}^{(4)A\rho}$. In this case a unique diagram is generated by the vertex corrections, however, the diagram includes a $\pi\pi$-loop in which the two pions have 
identical 4-momentum. This configuration generates double-counting, because the pions are indistinguishable. Therefore, an additional symmetry factor of 
$\frac{1}{2}$ needs to be added to $\Gamma_{\mu\nu \, ab33}^{(4)A\rho}$ when the tadpole loop contains a thermal pion. 

We have verified analytically that if all double counting is dropped gauge invariance is not violated.

%%%%%%%%%%%%%
\section{Matsubara sums}
\label{app:Matsubara}
%%%%%%%%%%%%%%%%%%
In this appendix we establish the spectral representations used to perform the Matsubara sums appearing in the $\rho$ self-energy, and provide the 
results of the summations. We begin by presenting the corrections to the $\pi\pi$-loop from the three-point vertex corrections in Sec.~\ref{app:pipi}. In Sec.~\ref{app:tadpole} we 
present the corrections to the tadpole loop from the four-point vertex corrections, and in Sec.~\ref{app:total} we give the total correction to the $\rho$ self-energy.
%%%%%%%%%%%%%%%%%%%%%%%%%%
\subsection{$\pi\pi$-loop corrections}
\label{app:pipi}
%%%%%%%%%%%%%%%%%%%%%%%%%%
Here we write the imaginary parts of $\Gamma_{\mu \, ab3}^{(3)A\rho}$, $\Gamma_{\mu \, ab3}^{(3)B\rho}$, $\Gamma_{\mu \, ab3}^{(3)C\rho}$, 
and $\Gamma_{\mu \, ab3}^{(3)C\sigma}$, relevant for maintaining gauge invariance.
\begin{widetext}
\begin{eqnarray}
\label{eq74}
\textrm{Im}\Gamma_{\mu \, ab3}^{(3)A\rho}(k,q)&=&-\epsilon_{3ab}\frac{3g_{\rho}^{3}}{2}\int \frac{d^{3}p}{(2\pi)^{3}}\int_{0}^{\infty}\frac{dw}{-\pi}\bigg[\textrm{Im}[D_{\pi}(p)]\textrm{Im}[D_{\rho}^{\nu\lambda}(p+k)] g_{\mu\lambda}(k-p)_{\nu}\nonumber\\
&&\times\textrm{FF}_{\rho}[-q+p,q+k]\textrm{FF}_{\rho}[p,k]\Theta(k_{0}+w)(f(w)-f(k_{0}+w))\bigg]_{p_{0}=w}\nonumber\\
&&+\epsilon_{3ab}\frac{3g_{\rho}^{3}}{2}\int \frac{d^{3}p}{(2\pi)^{3}}\int_{0}^{\infty}\frac{dw}{-\pi}\bigg[ \textrm{Im}[D_{\pi}(p)]\textrm{Im}[D_{\rho}^{\nu\lambda}(p+k)] g_{\mu\lambda}(k-p)_{\nu}\nonumber\\
&&\times\textrm{FF}_{\rho}[-q+p,q+k]\textrm{FF}_{\rho}[p,k]\Theta(-k_{0}+w)(f(w)-f(-k_{0}+w))\bigg]_{p_{0}=-w},\\
\label{eq74B}
\textrm{Im}\Gamma_{\mu \, ab3}^{(3)B\rho}(k,q)&=&-\epsilon_{3ab}\frac{3g_{\rho}^{3}}{2}\int \frac{d^{3}p}{(2\pi)^{3}}\int_{0}^{\infty}\frac{dw}{-\pi}\bigg[\textrm{Im}[D_{\pi}(p)]\textrm{Im}[D_{\rho}^{\nu\lambda}(q+k+p)] g_{\mu\lambda}(q+k-p)_{\nu}\nonumber\\
&&\times\textrm{FF}_{\rho}[q+p,k]\textrm{FF}_{\rho}[p,q+k]\Theta(q_{0}+k_{0}+w)(f(w)-f(q_{0}+k_{0}+w))\bigg]_{p_{0}=w}\nonumber\\
&&+\epsilon_{3ab}\frac{3g_{\rho}^{3}}{2}\int \frac{d^{3}p}{(2\pi)^{3}}\int_{0}^{\infty}\frac{dw}{-\pi}\bigg[\textrm{Im}[D_{\pi}(p)]\textrm{Im}[D_{\rho}^{\nu\lambda}(q+k+p)]\times g_{\mu\lambda}(q+k-p)_{\nu}\nonumber\\
&&\times \textrm{FF}_{\rho}[q+p,k]\textrm{FF}_{\rho}[p,q+k]\Theta(-q_{0}-k_{0}+w)(f(w)-f(-q_{0}-k_{0}+w))\bigg]_{p_{0}=-w},
\end{eqnarray}
In order to help group $\Gamma_{\mu \, ab3}^{(3)C\rho}$ and $\Gamma_{\mu \, ab3}^{(3)C\sigma}$ into spectral representations we break each term 
into four functions:
\begin{eqnarray}
\label{eq116}
\textrm{Im}\Gamma_{\mu \, ab3}^{(3)C\rho k}(k,q)&=&\epsilon_{3ab}\frac{g_{\rho}^{3}}{2}\int \frac{d^{3}p}{(2\pi)^{3}}\int_{0}^{\infty}\frac{dw}{-\pi}\bigg[\textrm{Im}[D_{\pi}(p)]\textrm{Re}[D_{\pi}(q-p)]\textrm{Im}[D_{\rho}^{\nu\lambda}(p+k)](-2p+q)_{\mu}(k+2q-p)_{\lambda}\nonumber\\
&&\times(k-p)_{\nu}\textrm{FF}_{\rho}[p,k]\textrm{FF}_{\rho}[-q+p,q+k]\Theta(k_{0}+w)(f(w)-f(k_{0}+w))\bigg]_{p_{0}=w}\nonumber\\
&&-\epsilon_{3ab}\frac{g_{\rho}^{3}}{2}\int \frac{d^{3}p}{(2\pi)^{3}}\int_{0}^{\infty}\frac{dw}{-\pi}\bigg[\textrm{Im}[D_{\pi}(p)]\textrm{Re}[D_{\pi}(q-p)]\textrm{Im}[D_{\rho}^{\nu\lambda}(p+k)](-2p+q)_{\mu}(-p+k+2q)_{\lambda}\nonumber\\
&&\times(k-p)_{\nu}\textrm{FF}_{\rho}[p,k]\textrm{FF}_{\rho}[-q+p,q+k]\Theta(-k_{0}+w)(f(w)-f(-k_{0}+w))\bigg]_{p_{0}=-w},\\
\label{eq117}
\textrm{Im}\Gamma_{\mu \, ab3}^{(3)C\rho q }(k,q)&=&\epsilon_{3ab}\frac{g_{\rho}^{3}}{2}\int \frac{d^{3}p}{(2\pi)^{3}}\int_{0}^{\infty}\frac{dw}{-\pi}\bigg[\textrm{Im}[D_{\pi}(p)]\textrm{Re}[D_{\pi}(q+p)]\textrm{Im}[D_{\rho}^{\nu\lambda}(q+k+p)](-2p-q)_{\mu}(-p+k+q)_{\lambda}\nonumber\\
&&\times(k-q-p)_{\nu}\textrm{FF}_{\rho}[q+p,k]\textrm{FF}_{\rho}[p,q+k]\Theta(q_{0}+k_{0}+w)(f(w)-f(q_{0}+k_{0}+w))\bigg]_{p_{0}=w}\nonumber\\
&&-\epsilon_{3ab}\frac{g_{\rho}^{3}}{2}\int \frac{d^{3}p}{(2\pi)^{3}}\int_{0}^{\infty}\frac{dw}{-\pi}\bigg[\textrm{Im}[D_{\pi}(p)]\textrm{Re}[D_{\pi}(q+p)]\textrm{Im}[D_{\rho}^{\nu\lambda}(q+k+p)](-2p-q)_{\mu}(-p+k+q)_{\lambda}\nonumber\\
&&\times(k-q-p)_{\nu}\textrm{FF}_{\rho}[q+p,k]\textrm{FF}_{\rho}[p,q+k]\Theta(-q_{0}-k_{0}+w)(f(w)-f(-q_{0}-k_{0}+w))\bigg]_{p_{0}=-w},\\
\label{eq116pipik}
\textrm{Im}\Gamma_{\mu \, ab3}^{(3)C\rho k\pi}(k,q)&=&-\epsilon_{3ab}\frac{g_{\rho}^{3}}{2}\int \frac{d^{3}p}{(2\pi)^{3}}\int_{0}^{\infty}\frac{dw}{-\pi}\bigg[\textrm{Im}[D_{\pi}(p)]\textrm{Im}[D_{\pi}(q-p)]\textrm{Re}[D_{\rho}^{\nu\lambda}(p+k)](-2p+q)_{\mu}(-p+k+2q)_{\lambda}\nonumber\\
&&\times(k-p)_{\nu}\textrm{FF}_{\rho}[p,k]\textrm{FF}_{\rho}[-q+p,q+k](f(w)-f(q_{0}+w))\bigg]_{p_{0}=-w},\\
\label{eq116pipiq}
\textrm{Im}\Gamma_{\mu \, ab3}^{(3)C\rho q\pi}(k,q)&=&\epsilon_{3ab}\frac{g_{\rho}^{3}}{2}\int \frac{d^{3}p}{(2\pi)^{3}}\int_{0}^{\infty}\frac{dw}{-\pi}\bigg[\textrm{Im}[D_{\pi}(p)]\textrm{Im}[D_{\pi}(q+p)]\textrm{Re}[D_{\rho}^{\nu\lambda}(q+k+p)](-2p-q)_{\mu}(-p+k+q)_{\lambda}\nonumber\\
&&\times(k-q-p)_{\nu}\textrm{FF}_{\rho}[q+p,k]\textrm{FF}_{\rho}[p,q+k](f(w)-f(q_{0}+w))\bigg]_{p_{0}=w},\\
\label{eq118}
\textrm{Im}\Gamma_{\mu \, ab3}^{(3)C\sigma k}(k,q)&=&\epsilon_{3ab}\frac{g_{\rho} g_{\sigma}^{2}}{2}\int \frac{d^{3}p}{(2\pi)^{3}}\int_{0}^{\infty}\frac{dw}{-\pi}\bigg[\textrm{Im}[D_{\pi}(p)]\textrm{Re}[D_{\pi}(q-p)]\textrm{Im}[D_{\sigma}(p+k)](p+k)^{2}-m_{\pi}^{2})\nonumber\\
&&\times(-2p+q)_{\mu}\textrm{FF}_{\rho}[p,k]\textrm{FF}_{\rho}[-q+p,q+k]\Theta(k_{0}+w)(f(w)-f(k_{0}+w))\bigg]_{p_{0}=w}\nonumber\\
&&-\epsilon_{3ab}\frac{g_{\rho} g_{\sigma}^{2}}{2}\int \frac{d^{3}p}{(2\pi)^{3}}\int_{0}^{\infty}\frac{dw}{-\pi}\bigg[\textrm{Im}[D_{\pi}(p)]\textrm{Re}[D_{\pi}(q-p)]\textrm{Im}[D_{\sigma}(p+k)]((p+k)^{2}-m_{\pi}^{2})\nonumber\\
&&\times(-2p+q)_{\mu}\textrm{FF}_{\rho}[p,k]\textrm{FF}_{\rho}[-q+p,q+k]\Theta(-k_{0}+w)(f(w)-f(-k_{0}+w))\bigg]_{p_{0}=-w},
\end{eqnarray}
\begin{eqnarray}
\label{eq119}
\textrm{Im}\Gamma_{\mu \, ab3}^{(3)C\sigma q}(k,q)&=&\epsilon_{3ab}\frac{g_{\rho} g_{\sigma}^{2}}{2}\int \frac{d^{3}p}{(2\pi)^{3}}\int_{0}^{\infty}\frac{dw}{-\pi}\bigg[\textrm{Im}[D_{\pi}(p)]\textrm{Re}[D_{\pi}(q+p)]\textrm{Im}[D_{\sigma}(q+k+p)]((q+k+p)^{2}-m_{\pi}^{2})\nonumber\\
&&\times(-2p-q)_{\mu}\textrm{FF}_{\rho}[q+p,k]\textrm{FF}_{\rho}[p,q+k]\Theta(q_{0}+k_{0}+w)(f(w)-f(q_{0}+k_{0}+w))\bigg]_{p_{0}=w}\nonumber\\
&&-\epsilon_{3ab}\frac{g_{\rho} g_{\sigma}^{2}}{2}\int \frac{d^{3}p}{(2\pi)^{3}}\int_{0}^{\infty}\frac{dw}{-\pi}\bigg[\textrm{Im}[D_{\pi}(p)]\textrm{Re}[D_{\pi}(q+p)]\textrm{Im}[D_{\sigma}(q+k+p)]((q+k+p)^{2}-m_{\pi}^{2})\nonumber\\
&&\times(-2p-q)_{\mu}\textrm{FF}_{\rho}[q+p,k]\textrm{FF}_{\rho}[p,q+k]\Theta(-q_{0}-k_{0}+w)(f(w)-f(-q_{0}-k_{0}+w))\bigg]_{p_{0}=-w},\\
\label{eq118k}
\textrm{Im}\Gamma_{\mu \, ab3}^{(3)C\sigma k\pi}(k,q)&=&-\epsilon_{3ab}\frac{g_{\rho} g_{\sigma}^{2}}{2}\int \frac{d^{3}p}{(2\pi)^{3}}\int_{0}^{\infty}\frac{dw}{-\pi}\bigg[\textrm{Im}[D_{\pi}(p)]\textrm{Im}[D_{\pi}(q-p)]\textrm{Re}[D_{\sigma}(p+k)]((p+k)^{2}-m_{\pi}^{2})\nonumber\\
&&\times(-2p+q)_{\mu}\textrm{FF}_{\sigma}[p,k]\textrm{FF}_{\sigma}[-q+p,q+k](f(w)-f(q_{0}+w))\bigg]_{p_{0}=w},
\end{eqnarray}

\begin{eqnarray}
\label{eq118q}
\textrm{Im}\Gamma_{\mu \, ab3}^{(3)C\sigma q\pi}(k,q)&=&\epsilon_{3ab}\frac{g_{\rho} g_{\sigma}^{2}}{2}\int \frac{d^{3}p}{(2\pi)^{3}}\int_{0}^{\infty}\frac{dw}{-\pi}\bigg[\textrm{Im}[D_{\pi}(p)]\textrm{Im}[D_{\pi}(q+p)]\textrm{Re}[D_{\sigma}(q+k+p)]((q+k+p)^{2}-m_{\pi}^{2})\nonumber\\
&&\times(-2p-q)_{\mu}\textrm{FF}_{\sigma}[q+p,k]\textrm{FF}_{\sigma}[p,q+k](f(w)-f(q_{0}+w))\bigg]_{p_{0}=w}.
\end{eqnarray}
Here the heaviside step function $(\Theta)$ is introduced, so that we only include vertex corrections generated by the Landau cut of the pion self-energy. 

Next, we address the real parts of the vertex corrections. In the calculation of $\text{Re}\Sigma_{\pi}$ we never explicitly utilized $\textrm{Re}D_{(\rho/\sigma)}$, opting instead to calculate $\text{Re}\Sigma_{\pi}$ with subtracted dispersion relations. In order to maintain gauge invariance, we must calculate the vertex corrections with the same $\textrm{Re}D_{(\rho/\sigma)}$ as was used in $\text{Re}\Sigma_{\pi}$. Thus the real parts of the vertex corrections are also calculated with subtracted dispersion relations:
\begin{eqnarray}
\label{eq84A}
\textrm{Re}\Gamma_{\mu \, ab3}^{(3)A\rho}(k,q)&=&\frac{-1}{\pi}\int_{-\infty}^{\infty}dw\Bigg(\frac{\textrm{Im}\Gamma_{\mu \, ab3}^{(3)A\rho}(\{w,\vec{k}\},q)}{k_{0}-w}-\frac{\textrm{Im}\Gamma_{\mu \, ab3}^{(3)A\rho}\big(\{w,\vec{k}\},q\big)}{-w}\Bigg),\\
\label{eq84B}
\textrm{Re}\Gamma_{\mu \, ab3}^{(3)B\rho}(k,q)&=&\frac{-1}{\pi}\int_{-\infty}^{\infty}dw\Bigg(\frac{\textrm{Im}\Gamma_{\mu \, ab3}^{(3)B\rho}(\{w,\vec{k}\},q)}{k_{0}-w}-\frac{\textrm{Im}\Gamma_{\mu \, ab3}^{(3)B\rho}\big(\{w,\vec{k}\},q\big)}{-q_{0}-w}\Bigg),\\
\label{eq84Ck}
\textrm{Re}\Gamma_{\mu \, ab3}^{(3)C(\rho/\sigma) k}(k,q)&=&\frac{-1}{\pi}\int_{-\infty}^{\infty}dw\Bigg(\frac{\textrm{Im}\Gamma_{\mu \, ab3}^{(3)C(\rho/\sigma) k}(\{w,\vec{k}\},q)}{k_{0}-w}-\frac{\textrm{Im}\Gamma_{\mu \, ab3}^{(3)C(\rho/\sigma) k}\big(\{w,\vec{k}\},q\big)}{-w}\Bigg),\\
\label{eq84Cq}
\textrm{Re}\Gamma_{\mu \, ab3}^{(3)C(\rho/\sigma) q}(k,q)&=&\frac{-1}{\pi}\int_{-\infty}^{\infty}dw\Bigg(\frac{\textrm{Im}\Gamma_{\mu \, ab3}^{(3)C(\rho/\sigma) q}(\{w,\vec{k}\},q)}{k_{0}-w}-\frac{\textrm{Im}\Gamma_{\mu \, ab3}^{(3)C(\rho/\sigma) q}\big(\{w,\vec{k}\},q\big)}{-q_{0}-w}\Bigg),\\
\label{eq84Cpk}
\textrm{Re}\Gamma_{\mu \, ab3}^{(3)C(\rho/\sigma) \pi k}(k,q)&=&\frac{-1}{\pi}\int_{-\infty}^{\infty}dw\frac{\textrm{Im}\Gamma_{\mu \, ab3}^{(3)C(\rho/\sigma)\pi k}(\{w,\vec{k}\},q)}{k_{0}-w},\\
\label{eq84Cpq}
\textrm{Re}\Gamma_{\mu \, ab3}^{(3)C(\rho/\sigma)\pi q}(k,q)&=&\frac{-1}{\pi}\int_{-\infty}^{\infty}dw\frac{\textrm{Im}\Gamma_{\mu \, ab3}^{(3)C(\rho/\sigma) \pi q}(\{w,\vec{k}\},q)}{k_{0}-w}  \ .
\end{eqnarray}
The subtractions are determined by the argument of the $\rho$ or $\sigma$ propagator in the vertex correction, and are fixed by the subtractions used in the pion self-energy. Notably, $\textrm{Re}\Gamma_{\mu \, ab3}^{(3)C(\rho/\sigma) \pi (k/q)}$ does not include a subtraction, because it is not proportional to $\textrm{Re}D_{(\rho/\sigma)}$. Instead, $\textrm{Re}\Gamma_{\mu \, ab3}^{(3)C(\rho/\sigma) \pi (k/q)}$ is proportional to $\textrm{Im}D_{(\rho/\sigma)}$ and $\textrm{Im}\Gamma_{\mu \, ab3}^{(3)C(\rho/\sigma) \pi (k/q)}$ is proportional to $\textrm{Re}D_{(\rho/\sigma)}$. Thus, to maintain consistent $\rho$ and $\sigma$ propagators we perform a subtraction on $\textrm{Im}\Gamma_{\mu \, ab3}^{(3)C(\rho/\sigma) \pi (k/q)}$, rather than $\textrm{Re}\Gamma_{\mu \, ab3}^{(3)C(\rho/\sigma) \pi (k/q)}$. 
\begin{eqnarray}
\label{eq84ImCpkR}
\textrm{Im}\tilde{\Gamma}_{\mu \, ab3}^{(3)C\rho \pi k}(k,q)&=&\textrm{Im}\Gamma_{\mu \, ab3}^{(3)C(\rho/\sigma) \pi k}(k,q)-\textrm{Im}\Gamma_{\mu \, ab3}^{(3)C\rho \pi k}(\{k_{0}=0,\vec{k}\},q),\\
\label{eq84ImCpqR}
\textrm{Im}\tilde{\Gamma}_{\mu \, ab3}^{(3)C\rho\pi q}(k,q)&=&\textrm{Im}\Gamma_{\mu \, ab3}^{(3)C(\rho/\sigma) \pi q}(k,q)-\textrm{Im}\Gamma_{\mu \, ab3}^{(3)C\rho \pi q}(\{k_{0}=-q_{0},\vec{k}\},q),\\
\label{eq84ImCpkS}
\textrm{Im}\tilde{\Gamma}_{\mu \, ab3}^{(3)C\sigma \pi k}(k,q)&=&\textrm{Im}\Gamma_{\mu \, ab3}^{(3)C(\rho/\sigma) \pi k}(k,q)-\textrm{Im}\Gamma_{\mu \, ab3}^{(3)C\sigma \pi k}(\{k_{0}=0,\vec{k}\},q),\\
\label{eq84ImCpqS}
\textrm{Im}\tilde{\Gamma}_{\mu \, ab3}^{(3)C\sigma\pi q}(k,q)&=&\textrm{Im}\Gamma_{\mu \, ab3}^{(3)C(\rho/\sigma) \pi q}(k,q)-\textrm{Im}\Gamma_{\mu \, ab3}^{(3)C\sigma \pi q}(\{k_{0}=-q_{0},\vec{k}\},q)
.
\end{eqnarray}
The subtractions allow for a straightforward implementation of spectral representations, without introducing an additional violation of gauge invariance. The spectral representations are given by:
\begin{eqnarray}
\label{VC3Speck}
D_{\pi}(k)\Gamma_{\mu \, ab3}^{(3)k}(k,q)&=&\frac{-1}{\pi}\int_{-\infty}^{\infty}dv\frac{\textrm{Im}\big[D_{\pi}(v,\vec{k})\Gamma_{\mu \, ab3}^{(3)k}(\{v,\vec{k}\},q)\big]}{k_{0}-v+i\epsilon},\\
\label{VC3Specq}
D_{\pi}(q+k)\Gamma_{\mu \, ab3}^{(3)q}(q+k,q)&=&\frac{-1}{\pi}\int_{-\infty}^{\infty}dv'\frac{\textrm{Im}\big[D_{\pi}(v',\vec{k})\Gamma_{\mu \, ab3}^{(3)q}(\{v',\vec{k}\},q)\big]}{q_{0}+k_{0}-v'+i\epsilon},\\
\label{VC3Speckk}
D_{\pi}(k)\Gamma_{\mu \, ab3}^{(3)k}(k,q)^{2}&=&\frac{-1}{\pi}\int_{-\infty}^{\infty}dv\frac{\textrm{Im}\big[D_{\pi}(v,\vec{k})\Gamma_{\mu \, ab3}^{(3)k}(\{v,\vec{k}\},q)^{2}\big]}{k_{0}-v+i\epsilon},\\
\label{VC3Specqq}
D_{\pi}(q+k)\Gamma_{\mu \, ab3}^{(3)q}(q+k,q)^{2}&=&\frac{-1}{\pi}\int_{-\infty}^{\infty}dv'\frac{\textrm{Im}\big[D_{\pi}(v',\vec{k})\Gamma_{\mu \, ab3}^{(3)q}(\{v',\vec{k}\},q)^{2}\big]}{q_{0}+k_{0}-v'+i\epsilon}  \ ,
\end{eqnarray}
where we define the functions:
\begin{eqnarray}
\label{VC3Gk}
\Gamma_{\mu \, ab3}^{(3)k}(k,q)&=&\Gamma_{\mu \, ab3}^{(3)A\rho}(k,q)+\Gamma_{\mu \, ab3}^{(3)C\rho k}(k,q)+\tilde{\Gamma}_{\mu \, ab3}^{(3)C\rho k\pi}(k,q)+\Gamma_{\mu \, ab3}^{(3)C\sigma k}(k,q)+\tilde{\Gamma}_{\mu \, ab3}^{(3)C\sigma k\pi}(k,q)\nonumber\\
&&+g_{\rho}\epsilon_{3ab}(2k+q)_{i}\Big(\frac{\Sigma_{\pi(\rho)}(k)}{\Lambda_{2\rho}^{2}+(\vec{q}+\vec{k})^{2}}+\frac{\Sigma_{\pi(\sigma)}(k)}{\Lambda_{2\sigma}^{2}+(\vec{q}+\vec{k})^{2}}\Big),\\
\label{VC3Gq}
\Gamma_{\mu \, ab3}^{(3)q}(k,q)&=&\Gamma_{\mu \, ab3}^{(3)B\rho}(-q+k,q)+\Gamma_{\mu \, ab3}^{(3)C\rho q}(-q+k,q)+\tilde{\Gamma}_{\mu \, ab3}^{(3)C\rho q\pi}(-q+k,q)+\Gamma_{\mu \, ab3}^{(3)C\sigma q}(-q+k,q)\nonumber\\
&&+\tilde{\Gamma}_{\mu \, ab3}^{(3)C\sigma q\pi}(-q+k,q)+g_{\rho}\epsilon_{3ab}(2k-q)_{i}\Big(\frac{\Sigma_{\pi(\rho)}(k)}{\Lambda_{2\rho}^{2}+(\vec{k})^{2}}+\frac{\Sigma_{\pi(\sigma)}(k)}{\Lambda_{2\sigma}^{2}+(\vec{k})^{2}}\Big)  \ .
\end{eqnarray}
The real part of the integrals in eqs. \ref{VC3Speck} through \ref{VC3Specqq} are calculated from the principal values of the integrals.

We are now in place to calculate the transverse projection of the $\pi\pi$-loop diagrams containing vertex corrections, at $\vec{q}=0$. In terms of 
$\Gamma_{\mu \, ab3}^{(3)k}$ and $\Gamma_{\mu \, ab3}^{(3)q}$ one can write the transverse projection of the vertex corrections to the 
$\pi\pi$-loop as 
\begin{eqnarray}
\label{SERTVC3}
\Sigma_{\rho1}(q_{0},\vec{q}=0)&=&\frac{2\pi}{3}T\sum_{n(even)}\int\frac{d|\vec{k}|\vec{k}^{2}}{(2\pi)^3}\int_{-\infty}^{\infty}\frac{dvdv'}{\pi^{2}}\frac{1}{(k_{0}-v+i\epsilon)(q_{0}+k_{0}-v'+i\epsilon)}\nonumber\\
&&\Big[2g_{\rho}\epsilon_{3ab}(2|\vec{k}|)\textrm{Im}[D_{\pi}(v,\vec{k})\Gamma_{3 \, ba3}^{(3)k}(\{v,\vec{k}\},q)]\textrm{Im}[D_{\pi}(v',\vec{k})]\nonumber\\
&&+2g_{\rho}\epsilon_{3ab}(2|\vec{k}|)\textrm{Im}[D_{\pi}(v,\vec{k})]\textrm{Im}[D_{\pi}(v',\vec{k})\Gamma_{3 \, ba3}^{(3)q}(\{v',\vec{k}\},q)]\nonumber\\
&&+2\textrm{Im}[D_{\pi}(v,\vec{k})\Gamma_{3 \, ab3}^{(3)k}(\{v,\vec{k}\},q)]\textrm{Im}[D_{\pi}(v',\vec{k})\Gamma_{3 \, ba3}^{(3)q}(\{v',\vec{k}\},q)]\nonumber\\
&&-\textrm{Im}\big[D_{\pi}(v,\vec{k})\big(\Gamma_{3 \, ab3}^{(3)k}(\{v,\vec{k}\},q)\big)^{2}\big]\textrm{Im}[D_{\pi}(v',\vec{k})]\nonumber\\
&&-\textrm{Im}[D_{\pi}(v,\vec{k})]\textrm{Im}\big[D_{\pi}(v',\vec{k})\big(\Gamma_{3 \, ab3}^{(3)q}(\{v',\vec{k}\},q)\big)^{2}\big]\Big]_{k_{0}=i\omega_{n}}+\textrm{PV} \ .
\end{eqnarray}
One can now perform the Matsubara sum to obtain
\begin{eqnarray}
\label{SERTVC3MS}
\Sigma_{\rho1}(q_{0},\vec{q}=0)&=&\frac{-2\pi}{3}\int\frac{d|\vec{k}|\vec{k}^{2}}{(2\pi)^3}\int_{-\infty}^{\infty}\frac{dvdv'}{\pi^{2}}\frac{(f(v)-f(v'))}{q_{0}+v-v'}\nonumber\\
&&\Big[2g_{\rho}\epsilon_{3ab}(2|\vec{k}|)\textrm{Im}[D_{\pi}(v,\vec{k})\Gamma_{3 \, ba3}^{(3)k}(\{v,\vec{k}\},q)]\textrm{Im}[D_{\pi}(v',\vec{k})]\nonumber\\
&&+2g_{\rho}\epsilon_{3ab}(2|\vec{k}|)\textrm{Im}[D_{\pi}(v,\vec{k})]\textrm{Im}[D_{\pi}(v',\vec{k})\Gamma_{3 \, ba3}^{(3)q}(\{v',\vec{k}\},q)]\nonumber\\
&&+2\textrm{Im}[D_{\pi}(v,\vec{k})\Gamma_{3 \, ab3}^{(3)k}(\{v,\vec{k}\},q)]\textrm{Im}[D_{\pi}(v',\vec{k})\Gamma_{3 \, ba3}^{(3)q}(\{v',\vec{k}\},q)]\nonumber\\
&&-\textrm{Im}\big[D_{\pi}(v,\vec{k})\big(\Gamma_{3 \, ab3}^{(3)k}(\{v,\vec{k}\},q)\big)^{2}\big]\textrm{Im}[D_{\pi}(v',\vec{k})]\nonumber\\
&&-\textrm{Im}[D_{\pi}(v,\vec{k})]\textrm{Im}\big[D_{\pi}(v',\vec{k})\big(\Gamma_{3 \, ab3}^{(3)q}(\{v',\vec{k}\},q)\big)^{2}\big]\Big]+\textrm{PV}  \ ,
\end{eqnarray}
where PV denotes the Paulli-Villars regularization terms, defined by dressing the vacuum Pauli-Villars terms with thermal-pion self-energies and 
vertex corrections. Finally, we remove double-counting introduced by $\Gamma_{\mu \, ab3}^{(3)C(\rho/\sigma)}$ from the Landau cut of $\Sigma_{\rho1}$. This is achieved by introducing a function, 
$\tilde{\Sigma}_{\rho1}$, which subtracts the double-counting, 
\begin{eqnarray}
\label{SERTVC3DC}
\tilde{\Sigma}_{\rho1}(q_{0},\vec{q}=0)&=&\frac{2\pi}{3}\int\frac{d|\vec{k}|\vec{k}^{2}}{(2\pi)^3}\int_{-\infty}^{\infty}\frac{dvdv'}{\pi^{2}}\frac{\Theta(vv')(f(v)-f(v'))}{q_{0}+v-v'}\nonumber\\
&&\Big[g_{\rho}\epsilon_{3ab}(2|\vec{k}|)\textrm{Im}[D_{\pi}(v,\vec{k})\chi^{k}(\{v,\vec{k}\},q)]\textrm{Im}[D_{\pi}(v',\vec{k})]\nonumber\\
&&+g_{\rho}\epsilon_{3ab}(2|\vec{k}|)\textrm{Im}[D_{\pi}(v,\vec{k})]\textrm{Im}[D_{\pi}(v',\vec{k})\chi^{q}(\{v',\vec{k}\},q)]] \ ,
\end{eqnarray}
where
\begin{eqnarray}
\label{Chik}
\chi^{k}(k,q)&=&\Gamma_{3 \, ba3}^{(3)C\rho k}(k,q)+\tilde{\Gamma}_{3 \, ba3}^{(3)C\rho k\pi}(k,q),\\
\label{Chiq}
\chi^{q}(k,q)&=&\Gamma_{3 \, ba3}^{(3)C\rho q}(-q+k,q)+\tilde{\Gamma}_{3 \, ba3}^{(3)C\rho q\pi}(-q+k,q)  \ .
\end{eqnarray}

%%%%%%%%%%%%%%%%%%%%%%%%
\subsection{Tadpole loop corrections}
\label{app:tadpole}
%%%%%%%%%%%%%%%%%%%%%%%%%%
In this section we give the corrections to the tadpole loop from the $\rho\rho\pi\pi$ vertex corrections. To begin, we consider 
$\Gamma_{\mu\nu \, ab33}^{(4)A\rho}$, which we rewrite in terms of a function $\tilde{\Gamma}_{\mu\nu \, ab33}^{(4)A\rho}$ such that
\begin{eqnarray}
\label{ATilde1}
\Gamma_{\mu\nu \, ab33}^{(4)A\rho}(k,q)&=&i\big(\tilde{\Gamma}_{\mu\nu \, ab33}^{(4)A\rho}(k,q)+\tilde{\Gamma}_{\mu\nu \, ab33}^{(4)A\rho}(-k,q)\big) \ .
\end{eqnarray}
We have factored a complex $i$ out of the vertex correction, so that it more closely resembles the vacuum $\pi\pi\rho\rho$ vertex. Additionally, it is
more convenient to define spectral representations in terms of $\tilde{\Gamma}_{\mu\nu \, ab33}^{(4)A\rho}$. 
The imaginary part of $\tilde{\Gamma}_{\mu\nu \, ab33}^{(4)A\rho}$ is given by
\begin{eqnarray}
\label{eq78}
\textrm{Im}\tilde{\Gamma}_{\mu\nu \, ab33}^{(4)A\rho}(k,q)&=&(3\delta_{3a}\delta_{3b}-5\delta_{ab})\frac{g_{\rho}^{4}}{2}\int \frac{d^{3}p}{(2\pi)^{3}}\bigg[\int_{0}^{\infty}\frac{dw}{-\pi}\textrm{Im}[D_{\pi}(p)]\textrm{Im}[D_{\rho}^{\mu\nu}(q+k+p)]\nonumber\\
&&\times(f(w)-f(q_{0}+k_{0}+w))\Theta(q_{0}+k_{0}+w)\textrm{FF}_{\rho}[q+p,k]^{2}\bigg]_{p_{0}=w}\nonumber\\
&&-(3\delta_{3a}\delta_{3b}-5\delta_{ab})\frac{g_{\rho}^{4}}{2}\int \frac{d^{3}p}{(2\pi)^{3}}\bigg[\int_{0}^{\infty}\frac{dw}{-\pi}\textrm{Im}[D_{\pi}(p)]\textrm{Im}[D_{\rho}^{\mu\nu}(q+k+p)]\nonumber\\
&&\times(f(w)-f(-q_{0}-k_{0}+w))\Theta(-q_{0}-k_{0}+w)\textrm{FF}_{\rho}[q+p,k]^{2}\bigg]_{p_{0}=-w} \ ,
\end{eqnarray}
where we only write the cuts necessary to maintain gauge invariance. The real part of $\Gamma_{\mu\nu \, ab33}^{(4)A\rho}$ can be calculated with 
a subtracted dispersion relation,
\begin{eqnarray}
\label{eq84b}
\textrm{Re}\tilde{\Gamma}_{\mu\nu \, ab33}^{(4)A\rho}&=&\frac{-1}{\pi}\int_{-\infty}^{\infty}dw\Big(\frac{\textrm{Im}\tilde{\Gamma}_{\mu\nu \, ab33}^{(4)A\rho}(\{w,\vec{k}\},q)}{k_{0}-w}-\frac{\textrm{Im}\tilde{\Gamma}_{\mu\nu \, ab33}^{(4)A\rho}\big(\{w,\vec{k}\},q\big)}{-q_{0}-w}\Big) \ .
\end{eqnarray}
The transverse projection of the $\rho$ self-energy corrections derived from $\Gamma_{\mu\nu \, ab33}^{(4)A\rho}$ at $\vec{q}=0$ are given by
\begin{eqnarray}
\label{RhoSE4A}
\Sigma_{\rho2}(q_{0},0)&=&\frac{4\pi}{3}T\sum_{n(even)}\int\frac{d|\vec{k}|\vec{k}^{2}}{(2\pi)^3}\int_{-\infty}^{\infty}\frac{dvdv'}{\pi^{2}}\textrm{Im}[D_{\pi}(v,\vec{k})]\textrm{Im}[\tilde{\Gamma}_{ii \, aa33}^{(4)A\rho}(\{-q_{0}+v',\vec{k}\},q)]\nonumber\\
&&\times\frac{1}{(i\omega_{n}-v+i\epsilon)(q_{0}+i\omega_{n}-v'+i\epsilon)}+\textrm{PV}\nonumber\\
&=&\frac{-4\pi}{3}\int\frac{d|\vec{k}|\vec{k}^{2}}{(2\pi)^3}\int_{-\infty}^{\infty}\frac{dvdv'}{\pi^{2}}\textrm{Im}[D_{\pi}(v,\vec{k})]\textrm{Im}[\tilde{\Gamma}_{ii \, aa33}^{(4)A\rho}(\{-q_{0}+v',\vec{k}\},q)]\nonumber\\
&&\times\frac{1}{q_{0}+v-v'+i\epsilon}(1-\frac{1}{2}\Theta(v v'))(f(v)-f(v'))+\textrm{PV},\\
\label{RhoSE4Ar}
\Sigma_{\rho2}^{0}&=&\frac{4\pi}{3}T\sum_{n(even)}\int\frac{d|\vec{k}|\vec{k}^{2}}{(2\pi)^3}\int_{-\infty}^{\infty}\frac{dv}{-\pi}\textrm{Re}[\tilde{\Gamma}_{ii \, aa33}^{(4)A\rho}(\{0,\vec{k}\},0)]\frac{\textrm{Im}[D_{\pi}(v,\vec{k})]}{(i\omega_{n}-v+i\epsilon)}+\textrm{PV}\nonumber\\
&=&-\frac{4\pi}{3}\int\frac{d|\vec{k}|\vec{k}^{2}}{(2\pi)^3}\int_{-\infty}^{\infty}\frac{dv}{-\pi}\textrm{Re}[\tilde{\Gamma}_{ii \, aa33}^{(4)A\rho}(\{0,\vec{k}\},0)]\textrm{Im}[D_{\pi}(v,\vec{k})](1-\frac{1}{2}\Theta(v))f(v)+\textrm{PV} \ ,
\end{eqnarray}
where the theta functions are added in the last equalities to remove double-counting from the Landau cut, as was described in appendix~\ref{app:double}. Unlike the corrections to the $\rho\pi\pi$ vertex, $\tilde{\Gamma}_{ii \, aa33}^{(4)A\rho}$ introduces a nondispersive constant into the rho self-energy. Thus we explicitly calculated this constant with $\Sigma_{\rho2}^{0}$.

The corrections to the tadpole diagram due to the addition of a form factor can be expressed in a similar form. We write these contributions in terms 
of the following functions
\begin{eqnarray}
\label{FFVCk}
Y_{k}(k,\{q_{0},\vec{q}=0\})&=&-\frac{4g_{\rho}\epsilon_{3ca}k_{3}}{\Lambda_{2\rho}^{2}+\vec{k}^{2}}\Big[\Gamma_{3 \, ac3}^{(3)A\rho}(k,q)+\Gamma_{3 \, ac3}^{(3)C\rho k}(k,q)+\Gamma_{3 \, ac3}''^{(3)C\rho k\pi}(k,q)\Big]\nonumber\\
&&-\frac{4g_{\rho}\epsilon_{3ca}k_{3}}{\Lambda_{2\sigma}^{2}+\vec{k}^{2}}\Big[\Gamma_{3 \, ac3}^{(3)C\sigma k}(k,q)+\Gamma_{\mu \, ac3}^{(3)C\sigma k\pi}(k,q)\Big]\nonumber\\
&&-4g_{\rho}^{2}\Bigg\{4 k_{3}^{2}\Bigg[\frac{\Sigma_{\pi(\rho)}(k)}{(\Lambda_{2\rho}^{2}+\vec{k}^{2})^{2}}+\frac{\Sigma_{\pi(\sigma)}(k)}{(\Lambda_{2\sigma}^{2}+\vec{k}^{2})^{2}}\Bigg]-3\frac{\Sigma_{\pi(\rho)}(k)}{\Lambda_{2\rho}^{2}+\vec{k}^{2}}-3\frac{\Sigma_{\pi(\sigma)}(k)}{\Lambda_{2\sigma}^{2}+\vec{k}^{2}}\Bigg\},\\
\label{FFVCq}
Y_{q}(k,\{q_{0},\vec{q}=0\})&=&-\frac{4g_{\rho}\epsilon_{3ca}k_{3}}{\Lambda_{2\rho}^{2}+\vec{k}^{2}}\Big[\Gamma_{3 \, ab3}^{(3)B\rho}(-q+k,q)+\Gamma_{3 \, ab3}^{(3)C\rho q}(-q+k,q)+\Gamma_{3 \, ab3}^{(3)C\rho q\pi}(-q+k,q)\Big]\nonumber\\
&&-\frac{4g_{\rho}\epsilon_{3ca}k_{3}}{\Lambda_{2\sigma}^{2}+\vec{k}^{2}}\Big[\Gamma_{3 \, ab3}^{(3)C\sigma q}(-q+k,q)+\Gamma_{3 \, ab3}^{(3)C\sigma q\pi}(-q+k,q)\Big]\nonumber\\
&&-8g_{\rho}^{2}k_{3}^{2}\Bigg[\frac{\Sigma_{\pi(\rho)}(k)}{(\Lambda_{2\rho}^{2}+\vec{k}^{2})^{2}}+\frac{\Sigma_{\pi(\sigma)}(k)}{(\Lambda_{2\sigma}^{2}+\vec{k}^{2})^{2}}\Bigg] \ .
\end{eqnarray}
The transverse projection of the corresponding $\rho$ self-energy contributions at $\vec{q}=0$ is given by
\begin{eqnarray}
\label{RhoSE4FF}
\Sigma_{\rho3}(q_{0},0)&=&\frac{2\pi}{3}T\sum_{n(even)}\int\frac{d|\vec{k}|\vec{k}^{2}}{(2\pi)^3}\Big[D_{\pi}(k)\big(Y_{q}(q+k,q)+Y_{q}(q-k,q)+Y_{k}(k,q)+Y_{k}(-k,q)\big)\Big]_{k_{0}=i\omega_{n}}\nonumber\\
&=&\frac{-4\pi}{3}\int\frac{d|\vec{k}|\vec{k}^{2}}{(2\pi)^3}\int_{-\infty}^{\infty}\frac{dvdv'}{\pi^{2}}\textrm{Im}[Y_{q}(\{v',\vec{k}\},q)]\textrm{Im}[D_{\pi}(v,\vec{k})]\nonumber\\
&&\times\frac{1}{q_{0}+v-v'+i\epsilon}(f(v)-f(v'))+\textrm{PV},\\
\label{RhoSE4FFr}
\Sigma_{\rho3}^{0}&=&\frac{4\pi}{3}T\sum_{n(even)}\int\frac{d|\vec{k}|\vec{k}^{2}}{(2\pi)^3}\int_{-\infty}^{\infty}\frac{dv}{-\pi}\textrm{Re}[Y_{q}(\{0,\vec{k}\},0)]\frac{\textrm{Im}[D_{\pi}(v,\vec{k})]}{(i\omega_{n}-v+i\epsilon)}\nonumber\\
&&+\frac{4\pi}{3}T\sum_{n(even)}\int\frac{d|\vec{k}|\vec{k}^{2}}{(2\pi)^3}\int_{-\infty}^{\infty}\frac{dv}{-\pi}\frac{\textrm{Im}\big[Y_{k}(\{v,\vec{k}\},0)D_{\pi}(v,\vec{k})\big]}{(i\omega_{n}-v+i\epsilon)}\nonumber\\
&=&-\frac{4\pi}{3}\int\frac{d|\vec{k}|\vec{k}^{2}}{(2\pi)^3}\int_{-\infty}^{\infty}\frac{dv}{-\pi}\textrm{Re}[Y_{q}(\{0,\vec{k}\},0)]\textrm{Im}[D_{\pi}(v,\vec{k})]f(v)\nonumber\\
&&-\frac{4\pi}{3}\int\frac{d|\vec{k}|\vec{k}^{2}}{(2\pi)^3}\int_{-\infty}^{\infty}\frac{dv}{-\pi}\textrm{Im}\big[Y_{k}(\{v,\vec{k}\},0)D_{\pi}(v,\vec{k})\big]f(v)+\textrm{PV} \ ,
\end{eqnarray}
where we again explicitly calculate a nondispersive constant with $\Sigma_{\rho3}^{0}$.

Next, we calculate corrections to the tadpole that can be expressed in terms of unitarity cuts of $\rho\pi$ or $\sigma\pi$-loops, 
$\Gamma_{\mu\nu \, ab33}^{(4)B_{1}\rho}$, $\Gamma_{\mu\nu \, ab33}^{(4)B_{2}\rho}$, $\Gamma_{\mu\nu \, ab33}^{(4)E\rho}$, 
$\Gamma_{\mu\nu \, ab33}^{(4)G\rho}$, $\Gamma_{\mu\nu \, ab33}^{(4)E\sigma}$, and $\Gamma_{\mu\nu \, ab33}^{(4)G\sigma}$. These $\rho$ 
self-energies dress a pion propagator with the unitarity cut of the pion self-energy $(\Sigma_{\pi}^{U})$. For positive energy $\text{Im}\Sigma_{\pi}^{U(R)}$ is given by
\begin{eqnarray}
\label{eq34U1}
\textrm{Im}\Sigma_{\pi \rho}^{U(R)}(k,T)&=&g_R^2\int \frac{d^3p}{(2\pi)^3}\int_{0}^{\infty}\frac{dw}{-\pi}\Big[-\Theta(k_{0}-w)\textrm{Im}[D_{\pi}(p)]\textrm{Im}[D_{R}(k+p)]\nonumber\\
&&\times N_{R\pi\pi}[p,k,p_{0}+k_{0}]\textrm{FF}_{R}[p,k]^{2}(1+f(w)+f(k_{0}-w))\Big]_{p_{0}=-w} \ ,
\end{eqnarray}
For negative energies, $\Sigma_{\pi}^{U}$ is determined by forcing the imaginary part of the pion self-energy to be odd, that is enforcing the retarded property of the pion self-energy. The real part of $\Sigma_{\pi}^{U}$ can then be evaluated with the dispersion relation
\begin{eqnarray}
\label{ReSEPiU}
\textrm{Re}\Sigma_{\pi}^{U}(k)&=&\frac{-1}{\pi}\textrm{p.v.}\int_{0}^{\infty}dv^{2}\frac{\textrm{Im}\Sigma_{\pi}^{U}(v,\vec{k})}{k_{0}^{2}-v^{2}}-\frac{\textrm{Im}\Sigma_{\pi}^{U}(v,\vec{k})}{-v^{2}} \ .
\end{eqnarray}
However, this dispersion relations converges quite slowly, resulting in an unphysical shift of the pion mass by approximately 800 MeV. The shift is produced by vacuum $\pi\to\rho\pi$ and $\pi\to\sigma\pi$ decays and in principal should be absorbed into the pion mass; however, modifying the real part of the vacuum loop in only these diagrams would violate gauge invariance. In order to remove the shift systematically, while preserving gauge invariance, we resum the pion propagators in the $\rho$ self-energy with $\Sigma_{\pi}^{U}$ yielding
\begin{eqnarray}
\label{SERhoU}
\Sigma_{\rho \, U}^{\mu\nu}(q)&=&g_\rho^2 \int \frac{d^3k}{(2\pi)^3} \int_{-\infty}^{\infty}\frac{dvdv'}{\pi^{2}}\frac{(2k+q)^{\mu} (2k+q)^{\nu}}{q_{0}+v-v'+i\epsilon}(f(v)-f(v'))\Theta(vv')
\nonumber\\
&&\times\textrm{Im}[\frac{1}{v^{2}-\vec{k}^{2}-m_{\pi}^{2}-\Sigma_{\pi}(v,\vec{k},T)-\Sigma_{\pi}^{U}(v,\vec{k},T)-\textrm{Re}\Sigma_{\pi}^{U}(m_{\pi},0,0)}]
\nonumber\\
&&\times\textrm{Im}[\frac{1}{(v')^{2}-(\vec{q}+\vec{k})^{2}-m_{\pi}^{2}-\Sigma_{\pi}(v',\vec{q}+\vec{k},T)-\Sigma_{\pi}^{U}(v',\vec{q}+\vec{k},T)-\textrm{Re}\Sigma_{\pi}^{U}(m_{\pi},0,0)}]
\nonumber\\
&&-2g_\rho^2g^{\mu\nu} \int  \frac{d^3k}{(2\pi)^3}\int_{-\infty}^{\infty}\frac{dv}{-\pi} f(v)\Theta(v)
\nonumber\\
&&\times\textrm{Im}[\frac{1}{v^{2}-\vec{k}^{2}-m_{\pi}^{2}-\Sigma_{\pi}(v,\vec{k},T)-\Sigma_{\pi}^{U}(v,\vec{k},T)-\textrm{Re}\Sigma_{\pi}^{U}(m_{\pi},0,0)}] \ ,
\end{eqnarray}
where we have limited the integration with theta functions to ensure that we do not introduce new vacuum $\rho$ self-energy diagrams. Finally, we 
apply Pauli-Villars regularization to the vacuum $\Sigma_{\pi}^{U}$ loop, such that
\begin{eqnarray}
\label{ImSEPiUPV}
\textrm{Im}\Sigma_{\pi}^{U}(k,m_{\pi},T)&\to&\textrm{Im}\Sigma_{\pi}^{U}(k,m_{\pi},T)-2\textrm{Im}\Sigma_{\pi}^{U}(k,\sqrt{m_{\pi}^{2}+\Lambda_{0}^{2}},T=0)\\
&&+\textrm{Im}\Sigma_{\pi}^{U}(k,\sqrt{m_{\pi}^{2}+2\Lambda_{0}^{2}},T=0) \ .
\end{eqnarray}

Equation~(\ref{SERhoU}) contains the tadpole corrections derived from $\Gamma_{\mu\nu \, ab33}^{(4)E\rho}$, $\Gamma_{\mu\nu \, ab33}^{(4)G\rho}$, 
$\Gamma_{\mu\nu \, ab33}^{(4)E\sigma}$, and $\Gamma_{\mu\nu \, ab33}^{(4)G\sigma}$ in the resummation of the pion propagator. However, in 
Eq.~(\ref{SERhoU}) we are free to treat $\Sigma_{\pi}^{U}$ equivalently to $\Sigma_{\pi}$, performing a zero-energy subtraction on 
$\textrm{Re}\Sigma_{\pi}^{U}$. Furthermore, we add a constant shift $\textrm{Re}\Sigma_{\pi}^{U}(k_{0}=m_{\pi},\vec{k}=0,T=0)$ to ensure that 
the vacuum pion mass is 140 MeV at $\vec{p}=0$. We are able add this constant without violating gauge invariance, because it simply amounts to 
a redefinition of the bare pion mass. It should be noted that because we have dressed the pion propagators with $\Sigma_{\pi}^{U}$ the Ward identities imply that 
additional vertex corrections must be calculated to maintain gauge invariance. These additional corrections correspond to the unitarity cuts of the previously calculated thermal 
vertex corrections. However, we have already encountered these corrections by dressing the unitarity cut of the $\rho$ self-energy with thermal 
vertex corrections. This symmetry is precisely why double-counting was encountered in the Landau cut. In fact, the first-order corrections to the 
resummation in Eq.~\ref{SERhoU} are already included in our formalism, and attempting to explicitly calculate first-order vertex corrections due to 
$\Sigma_{\pi}^{U}$ would only introduce double-counting into the unitarity cut of $\Sigma_{\rho}^{\mu\nu}$.

We now take the transverse projections of $\Sigma_{\rho \, U}^{\mu\nu}$, at $\vec{q}=0$ to obtain
\begin{eqnarray}
\label{SERhoUPt}
\Sigma_{\rho 4}(q_{0},0)&=&\frac{4\pi g_\rho^{2}}{3} \int \frac{dk\vec{k}^{2}}{(2\pi)^3} \int_{-\infty}^{\infty}\frac{dvdv'}{\pi^{2}}\frac{4\vec{k}^{2}\Theta(vv')(f(v)-f(v'))}{q_{0}+v-v'+i\epsilon}
\nonumber\\
&&\times\textrm{Im}[\frac{1}{v^{2}-\vec{k}^{2}-m_{\pi}^{2}-\Sigma_{\pi}(v,\vec{k},T)-\Sigma_{\pi}^{U}(v,\vec{k},T)-\textrm{Re}\Sigma_{\pi}^{U}(m_{\pi},0,0)}]
\nonumber\\
&&\times\textrm{Im}[\frac{1}{(v')^{2}-(\vec{q}+\vec{k})^{2}-m_{\pi}^{2}-\Sigma_{\pi}(v',\vec{q}+\vec{k},T)-\Sigma_{\pi}^{U}(v',\vec{q}+\vec{k},T)-\textrm{Re}\Sigma_{\pi}^{U}(m_{\pi},0,0)}]
\\
\label{SERhoTadUPt}
\Sigma_{\rho 4}^{0}(q_{0},0)&=&-\frac{8g_\rho^2}{3} \int  \frac{dk\vec{k}^{2}}{(2\pi)^3}\int_{-\infty}^{\infty}\frac{dv}{-\pi}\textrm{Im}[\frac{f(v)\Theta(v)}{v^{2}-\vec{k}^{2}-m_{\pi}^{2}-\Sigma_{\pi}(v,\vec{k},T)-\Sigma_{\pi}^{U}(v\vec{k},T)-\textrm{Re}\Sigma_{\pi}^{U}(m_{\pi},0,0)}] \ ,
\end{eqnarray}
where $\Sigma_{\rho4}^{0}$ again calculates a nondispersive constant.

Finally, we calculate the $\rho$ self-energy corrections due to $\Gamma_{\mu\nu \, ab33}^{(4)B_{1}\rho}$ and $\Gamma_{\mu\nu \, ab33}^{(4)B_{2}\rho}$. These corrections can be expressed in terms of the unitarity cut of $\Gamma_{\mu \, ab3}^{(3)A\rho}$ as 
\begin{eqnarray}
\label{eqA3U}
\textrm{Im}\Gamma_{\mu \, ab3}^{(3)AU\rho}(k,q)&=&\epsilon_{3ab}\frac{3g_{\rho}^{3}}{2}\int \frac{d^{3}p}{(2\pi)^{3}}\int_{0}^{\infty}\frac{dw}{-\pi}\bigg[\textrm{Im}[D_{\pi}(p)]\textrm{Im}[D_{\rho}^{\nu\lambda}(p+k)]g_{\mu\lambda}(k-p)_{\nu}
\nonumber\\
&&\times\textrm{FF}_{\rho}[-q+p,q+k]\textrm{FF}_{\rho}[p,k]\Theta(k_{0}-w)(1+f(w)+f(k_{0}-w))\bigg]_{p_{0}=-w}
\nonumber\\
&&-\epsilon_{3ab}\frac{3g_{\rho}^{3}}{2}\int \frac{d^{3}p}{(2\pi)^{3}}\int_{0}^{\infty}\frac{dw}{-\pi}\bigg[\textrm{Im}[D_{\pi}(p)]\textrm{Im}[D_{\rho}^{\nu\lambda}(p+k)]g_{\mu\lambda}(k-p)_{\nu}
\nonumber\\
&&\times\textrm{FF}_{\rho}[-q+p,q+k]\textrm{FF}_{\rho}[p,k]\Theta(-k_{0}-w)(1+f(w)+f(-k_{0}-w))\bigg]_{p_{0}=w}
\nonumber\\
&&-2\Big[m_{\pi}\to\sqrt{m_{\pi}^{2}+\Lambda_{0}^{2}}\Big]_{T=0}+\Big[m_{\pi}\to\sqrt{m_{\pi}^{2}+2\Lambda_{0}^{2}}\Big]_{T=0} \ ,
\end{eqnarray}
where the last line implements the Pauli-Villars regularization on the vacuum loop. The real part of $\Gamma_{\mu \, ab3}^{(3)AU\rho}$ is given by the subtracted dispersion relation
\begin{eqnarray}
\label{eqRA3U}
\textrm{Re}\Gamma_{\mu \, ab3}^{(3)AU\rho}(k,q)&=&\frac{-1}{\pi}\textrm{p.v.}\int_{-\infty}^{\infty}dv\frac{\textrm{Im}\Gamma_{\mu \, ab3}^{(3)AU\rho}(\{v,\vec{k}\},q)}{k_{0}-v}-\frac{\textrm{Im}\Gamma_{\mu \, ab3}^{(3)AU\rho}(\{v,\vec{k}\},q)}{-v} \ .
\end{eqnarray}
The self-energy correction arising from $\Gamma_{\mu\nu \, ab33}^{(4)B_{1}\rho}$ and $\Gamma_{\mu\nu \, ab33}^{(4)B_{2}\rho}$ is then given by
\begin{eqnarray}
\label{RhoSEU3}
\Sigma_{\rho5}(q_{0},0)&=&\frac{-8\pi}{3}\int\frac{d|\vec{k}|\vec{k}^{2}}{(2\pi)^3}\int_{-\infty}^{\infty}\frac{dvdv'}{\pi^{2}}\textrm{Im}[D_{\pi}(v',\vec{k})]\textrm{Im}\big[D_{\pi}(v,\vec{k})\Gamma_{3 ba}^{(3)AU\rho}(v,\vec{k})\big]
\nonumber\\
&&\times(2g_{\rho}|\vec{k}|\epsilon_{3ab})\frac{(f(v)-f(v'))}{(q_{0}+v-v'+i\epsilon)}\Theta(vv') \ ,
\end{eqnarray}
where we again limit our calculation to the Landau cut of $\Sigma_{\rho}^{\mu\nu}$.

%%%%%%%%%%%%%%%%%%%%%%
\subsection{Total corrections}
\label{app:total}
%%%%%%%%%%%%%%%%%%%%%%%%%
The total contribution of the vertex corrections to the transverse projections of the $\rho$ self-energy, at $\vec{q}=0$, is given by
\begin{eqnarray}\label{VCRhoTot}
\Sigma_{\rho}^{VC}(q_{0},0)&=&\Sigma_{\rho1}(q_{0},0)+\tilde{\Sigma}_{\rho1}(q_{0},0)+\Sigma_{\rho2}(q_{0},0)+\Sigma_{\rho2}^{0}+\Sigma_{\rho3}(q_{0},0)+\Sigma_{\rho3}^{0}
\nonumber\\
&&+\Sigma_{\rho4}(q_{0},0)+\Sigma_{\rho4}^{0}+\Sigma_{\rho5}(q_{0},0) \ .
\end{eqnarray}
The imaginary part can be calculated from $\Sigma_{\rho1}$, $\tilde{\Sigma}_{\rho1}$, $\Sigma_{\rho2}$, $\Sigma_{\rho3}$, $\Sigma_{\rho4}$, and 
$\Sigma_{\rho5}$, by converting $\frac{1}{q_{0}+v-v'+i\epsilon}$ into a delta-function and performing the remaining integrations. The real part can then 
be calculated through a dispersion relation,
\begin{eqnarray}
\label{ReDiss}
\textrm{Re}\Sigma_{\rho}^{VC}(q)&=&\frac{-1}{\pi}\textrm{p.v.}\int_{0}^{\infty}\frac{dv^{2}\textrm{Im}\Sigma_{\rho}^{VC}(v,\vec{q})}{q_{0}^{2}-v^{2}}+\Sigma_{\rho2}^{0}+\Sigma_{\rho3}^{0}+\Sigma_{\rho4}^{0} \ ,
\end{eqnarray}
where we do not perform a subtraction, because we have explicitly calculated the nondispersive constant with $\Sigma_{\rho2}^{0}$, 
$\Sigma_{\rho3}^{0}$, and $\Sigma_{\rho4}^{0}$.

\end{widetext}

\bibliography{cond-pigas-v2}

\end{document}